\documentclass[traditabstract]{aa} 

\usepackage{graphicx, times, txfonts, float, color, url}

\newcommand{\less}{\raisebox{-1.1mm}{$\stackrel{<}{\sim}$}} 
\newcommand{\more}{\raisebox{-1.1mm}{$\stackrel{>}{\sim}$}}

\begin{document}

\title{
The VMC Survey – XXXIII. The tip of the red giant branch in the Magellanic Clouds
\thanks{
Based on observations made with VISTA at ESO under programme ID 179.B-2003.
} 
}  
 
\author{ 
M.~A.~T.~Groenewegen\inst{1}
\and
M.-R.~L.~Cioni\inst{2}
\and
L.~Girardi\inst{3}
\and 
R.~de~Grijs\inst{4,5,6} 
\and
V.~D.~Ivanov\inst{7}
\and
M.~Marconi\inst{8}
\and
T.~Muraveva\inst{9}
\and
V.~Ripepi\inst{8}
\and
J.~Th.~van Loon\inst{10}
}

\institute{ 
Koninklijke Sterrenwacht van Belgi\"e, Ringlaan 3, B--1180 Brussels, Belgium \\ \email{martin.groenewegen@oma.be}
\and
Leibniz-Institut f\"ur Astrophysik Potsdam, An der Sternwarte 16, D--14482 Potsdam, Germany 
\and
Dipartimento di Fisica e Astronomia, Universit\`a di Padova, Vicolo dell’Osservatorio 2, I--35122 Padova, Italy 
\and
Department of Physics and Astronomy, Macquarie University, Balaclava Road, Sydney, NSW 2109, Australia
\and
Research Centre for Astronomy, Astrophysics and Astrophotonics, Macquarie University, Balaclava Road, Sydney, NSW 2109, Australia
\and
International Space Science Institute--Beijing, 1 Nanertiao, Zhongguancun, Hai Dian District, Beijing 100190, China
\and
European Southern Observatory, Karl-Schwarzschild-Str. 2, D--85748 Garching bei M\"unchen, Germany
\and
INAF -- Osservatorio Astronomico di Capodimonte, via Moiariello 16, I--80131, Naples, Italy  
\and
INAF -- Osservatorio di Astrofisica e Scienza dello Spazio di Bologna, via Piero Gobetti 93/3, I--40129, Bologna, Italy 
\and
Lennard-Jones Laboratories, Keele University, ST5 5BG, UK 
} 
 
\date{received: July 2018, accepted: 2018} 
 
\offprints{Martin Groenewegen}

\titlerunning{The Tip of the Red Giant Branch in the Magellanic Clouds}

\abstract {
In this paper $JK_{\rm s}$-band data from the VISTA Magellanic Cloud (VMC) survey are used to investigate the tip of the red giant branch (TRGB)
as a distance indicator. A linear fit to recent theoretical models is used as the basis for the absolute calibration which reads
$M_{K_{\rm s}} =  -4.196 -2.013 \; (J-K_{\rm s})$, valid in the colour range $0.75 < (J-K_{\rm s}) < 1.3$ mag and in the 2MASS system. 

The observed TRGB is found based on a classical first-order derivative filter and a second-order derivative filter applied to 
the binned luminosity function using the "sharpened" magnitude that takes the colour term into account. 
Extensive simulations are carried out to investigate any biases and errors in the 
derived distance modulus (DM).
Based on these simulations criteria are established related to the number of stars per bin in the 0.5~magnitude range below the TRGB 
and related to the significance  with which the peak in the filter response curve is determined such that the derived distances are unbiased.
The DMs based on the second-order derivative filter are found to be more stable and are therefore adopted, although 
this requires twice as many stars per bin.
Given the surface density of TRGB stars in the Magellanic Clouds (MCs), areas of $\sim 0.5\deg^2$ in the densest parts 
to $\sim 10\deg^2$ in the outskirts of the MCs need to be considered to obtain accurate and reliable values for the DMs.

The TRGB method is applied to specific lines-of-sight where independent distance estimates exist, based on
detached eclipsing binaries in the Large and Small Magellanic Clouds (LMC, SMC), classical Cepheids in the LMC,
RR Lyrae stars in the SMC, and fields in the SMC where the star formation history (together with reddening and distance) 
has been derived from deep VMC data.

The analysis shows that the theoretical calibration is consistent with the data, that the systematic error on the DM is approximately 
0.045 mag (about evenly split between the theoretical calibration and the method), and that random errors of 0.015 mag are achievable.

Reddening is an important element in deriving the distance: we derive
mean DMs ranging from 18.92 mag (for a typical $E(B-V)$ of 0.15 mag) to 19.07 mag ($E(B-V) \sim$ 0.04 mag) for the SMC, 
     and ranging from 18.48 mag ($E(B-V) \sim$ 0.12 mag) to 18.57 mag ($E(B-V) \sim$ 0.05 mag) for the LMC.
}


\maketitle

\section{Introduction} 

The  VISTA Magellanic Cloud (VMC) ESO public survey is a photometric survey in the three filters $Y$, $J$,
and $K_{\rm s}$ \citep{Cioni11} performed with the Visible and Infrared Survey Telescope for Astronomy (VISTA) telescope 
using the VISTA InfraRed CAMera (VIRCAM) camera \citep{Sutherland15}. The latter provides a spatial
resolution of 0.34\arcsec\ per pixel and a non-contiguous field-of-view of 1.65$\degr$ in diameter sampled by 16 detectors. 
To homogeneously cover the field-of-view it is necessary to fill the gaps between individual detectors using a six-point mosaic. 
This unit area of VISTA surveys is called a tile and covers 1.77~deg$^2$ of which the central area 
of 1.475$\degr$$\times$1.017$\degr$ is covered by at least two of the six pointins in the mosaic.

The VMC survey covers an area of approximately 170~deg$^2$ (110 tiles) of the Magellanic Cloud (MC) system and 
includes stars as faint as 22 mag in $K_{\rm s}$ (5$\sigma$, Vega mag); see \citet{Cioni11} for a description of the survey.

The main scientific goals of the VMC survey are to derive the spatially resolved star formation history (SFH) 
across the Magellanic system \citep{Rubele12, Rubele15, Rubele18} and to measure its three-dimensional geometry 
(e.g. \citealt{Ripepi17, Subramanian17, Muraveva2018}, see below), 
which drive, respectively, the depth and the monitoring strategy of the survey. 
There is much additional science that has been done using VMC data, for example on background galaxies 
(including quasars), asymptotic giant branch (AGB) stars, planetary nebulae, eclipsing binaries, stellar clusters, variable stars, 
and the proper motion of the MCs (see \citealt{Cioni16} for some recent science highlights).

The study of the 3D structure of the MCs relies on the use of different stellar distance indicators available
in the MCs. The VMC team has addressed this in various papers using the data 
available, in particular, using 
Type-II Cepheids (T2Cs; \citealt{Ripepi15}, 13 tiles in the Large MC, LMC), 
Classical Cepheids (CCs; \citealt{Ripepi12}, two tiles in the LMC centred on the south ecliptic pole and 30 Doradus; 
\citealt{Ripepi16, Ripepi17}, analysing almost 4800 CCs detected in the OGLE-IV survey across the entire SMC), 
RR Lyrae (RRL; \citealt{Muraveva2018}, all 27 tiles in the Small MC, SMC), and
the Red Clump (RC; \citealt{Tatton13}, one tile centred on 30 Doradus;
\citealt{Subramanian17}, 13 tiles covering the central part of the SMC).

In this paper we investigate and use yet another distance indicator, the tip of the red giant branch (TRGB), and apply it to VMC data in the MCs.
Over the years the TRGB distance has become an important rung of the distance ladder as distances can be routinely obtained 
with the {\it Hubble Space Telescope} (HST) with moderate effort out to $\sim 10$ Mpc (see for example \citealt{McQuinn17} using two orbits of HST)
or
 $\sim 15$ Mpc (see for example \citealt{Hatt2018} using six orbits of HST). 
The {\it Extragalactic Distance Database}\footnote{ \url{http://edd.ifa.hawaii.edu/} } \citep{Jacobs09} currently contains $400+$ galaxies with TRGB distances.

The classical paper on the subject is \citet{Lee93} which introduced the method of using an edge-detection algorithm 
to determine the tip (the TRGB was recognised and used as a distance indicator before, but more in a qualitative way; 
see references in \citealt{Lee93}). \citet{Lee93} also introduced the classical method of using the $I$-band for absolute calibration.
Later it was recognised that the absolute magnitude in $I$ (or $K_{\rm s}$, see later) of the tip is not constant but is a shallow function 
of metallicity, or, in the observational plane, colour (see \citealt{Salaris05} for a theoretical point of view).

\citet{Madore09} took this into consideration and introduced the idea of “sharpening” the tip by colour-correcting the $I$-band data 
before producing the luminosity function. The function marginalized for the tip detection had the form $T = I - \beta \cdot (V-I)$, 
where $\beta$ is the slope of the tip magnitude as a function of colour, thereby correcting for the metallicity sensitivity of the TRGB.

The TRGB method can also be applied in the near-infrared (NIR), where reddening is lower than in the optical, and TRGB stars 
are intrinsically brighter, $M_{K_{\rm s}} \approx -6.5$ (see later) versus $M_{I} \approx -4.0$ mag (see e.g. \citealt{Serenelli17} 
and references therein).

\citet{Cioni2000} appear to have been the first to investigate the TRGB in the NIR, 
using $I, J, K_{\rm s}$ data
from the Deep Near Infrared Survey of the Southern Sky (DENIS,  \citealt{Epchtein1999}) for the MCs. 
They also introduced a new method to detect the tip, based on the second-order derivative
of the luminosity function (LF), rather than the traditional Sobel filter \citep{Sobel70} which is a first-order derivative 
filter (see Sect.~\ref{S-Model}).
They found that the TRGB is located at a dereddened magnitude (in the DENIS system) of $K_{\rm s}= 11.94 \pm 0.04$ (LMC) 
and $12.58 \pm 0.04$ mag (SMC).
In that paper the distance to the MCs is not actually derived from the TRGB in the infrared, but from the TRGB in 
bolometric magnitude, calculated from $J, K_{\rm s}$, a bolometric correction, and a theoretical calibration. 
They found distance moduli (DM) of $18.55 \pm 0.04 \pm 0.08$ mag for the LMC 
and $18.99 \pm 0.03 \pm 0.08$ mag for the SMC (where the two error bars indicate formal and systematic errors, respectively), which 
imply (in the DENIS system) $M_{K_{\rm s}} = -6.61 \pm 0.09$ mag and $M_{K_{\rm s}} = -6.41 \pm 0.09$ mag for the LMC and SMC, respectively.

\citet{Macri15} presented the results of the LMC Near-Infrared Synoptic Survey (LMCNISS) 
covering 18~deg$^2$ down to $K_{\rm s} \sim 16.5$ mag.
They found the TRGB to be located at (observed magnitudes, calibrated in the 2MASS system) $J= 13.23 \pm 0.03$, 
$H= 12.35 \pm 0.02$, and $K_{\rm s}= 12.11 \pm 0.01$ mag. They used a typical reddening of $E(V-I)= 0.08$ mag (from \citealt{HaschkeLMC}),
and the distance to the LMC based on detached eclipsing binaries (dEBs; DM= $18.493 \pm 0.048$ mag, \citealt{Pietrzynski13}) 
to find $M_{K_{\rm s}} = -6.41 \pm 0.05$ mag. 
Taking into account the difference in adopted DM, the remaining difference with \citet{Cioni2000} is explained by 
the difference in the photometric passbands. According to \citet{Delmotte02}, $K_{\rm s}$(DENIS)= $K_{\rm s}$(2MASS) $-(0.14 \pm 0.05)$ mag.

\citet{Gorski16} investigated the TRGB in the MCs using the $I$-band (from OGLE), $J$, $K_{\rm s}$ (from a survey with 
the InfraRed Survey Facility, IRSF, see \citealt{Kato07}, and bolometric magnitudes.
They considered 17 fields in the LMC and 5 in the SMC, each 35\arcmin\ $\times$ 35\arcmin, 
selected to have a reddening of $E(V-I)< 0.1$ mag according to \citet{Haschke11}.
They used a kernel of the form [$-2$ $-1$ 0 +1 +2] and then calculated the Gaussian-smoothed LF introduced by \citet{Sakai1996} to 
detect the edge.
The mean magnitudes of the measured TRGB in the LMC and SMC are $K_{\rm s}= 12.13 \pm 0.04$ mag, and $12.91 \pm 0.04$ mag, respectively,
with mean $K$-band reddening values of 0.05, and 0.02 mag, respectively, in agreement with the estimates above.
They appear to assume that the IRSF magnitudes are effectively in the 2MASS system but \citet{Kato07} indicate 
differences of 0.01 mag in $J$ and 0.04 mag $K_{\rm s}$, 
and then reach the conclusion that the DM
to the LMC and SMC is about 0.2 mag longer than the values based on dEBs \citep{Pietrzynski13, Graczyk14}.
For the absolute calibration (see Sect.~\ref{S-Calib}) they used the relation of \citet{Valenti04} adopting 
metallicities of [Fe/H]= $-0.6$ and $-1.0$ dex for the LMC and the SMC, respectively.
In their latest paper \citet{Gorski18} credit this difference of 0.2 mag in DM to population effects and advocate 
the use of colour-dependent calibration relations rather then metallicity-dependent ones. 

The TRGB method in the $K$-band has been applied to galaxies other than the MCs, namely
Fornax \citep{Gullieuszik2007,Pietrzynski2009,Whitelock09}, Carina \citep{Pietrzynski2009}, Sculptor \citep{Menzies11}, 
NGC 205 \citep{Jung12} and IC 1613 \citep{Chun15}. 
The latter two papers use the method introduced by \citet{Cioni2000} to detect 
the edge using the second-order derivative of the LF\footnote{Neither paper discusses the correction one needs to apply to
the edge magnitude to obtain the true TRGB magnitude when using \citet{Cioni2000}'s original method.
}.
The TRGB method has been applied to 23 nearby galaxies (\less 4 Mpc) by \citet{Dalcanton12} using the HST {\it F110W} and {\it F160W} filters.
%
Most recently, \cite{Madore18} and \cite{Hoyt18} 
discuss the TRGB in the $JHK$ band in IC 1613 
and the LMC. A more detailed comparison to their work is done in Sect.~\ref{S-Calib}.

In the present paper we apply the TRGB method in the $K_{\rm s}$-band across the SMC and LMC using VMC data. 
In Sect.~\ref{S-Sample} the selection of the sample is discussed.
In Sect.~\ref{S-Calib} the absolute magnitude of the TRGB in the infrared is discussed, while
Sect.~\ref{S-Model} discusses the model, which includes a classical (first-order derivative) edge-detection, 
and an extension and improvement of the second-order derivative method of \citet{Cioni2000}.

\section{Data overview and sample selection} 
\label{S-Sample}

\begin{figure}
\centering
\resizebox{\hsize}{!}{\includegraphics{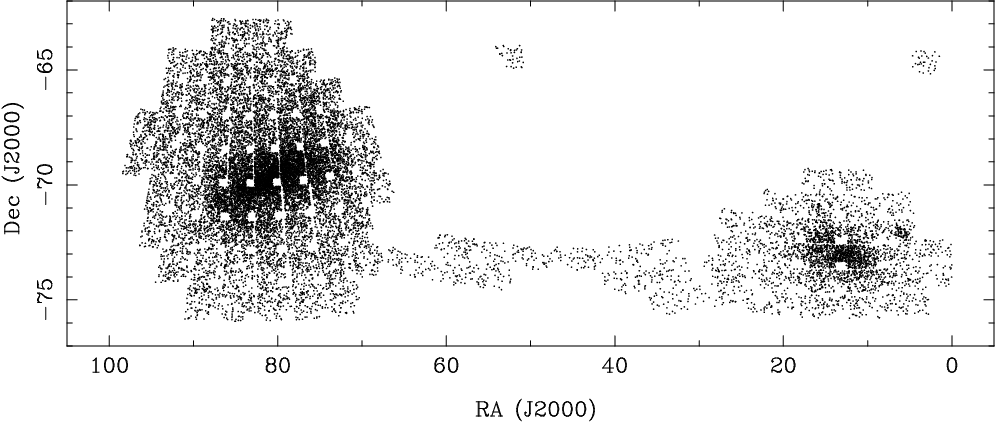}}
\caption[]{ 
Position on the sky of the selected VMC sources. For clarity only every 40th object is plotted.
The LMC, the SMC, the two tiles in the MS, and the MB are apparent.
The small regions missing in the corner of every tile correspond to detector 16 which are excluded by 
enforcing the constraint on {\it ksppErrBits}.
} 
\label{Fig-RaDec} 
\end{figure}

From the VISTA Science Archive (VSA; \citealt{Cross12}) all sources\footnote{Containing data processed until September 2016.} 
brighter than $K_{\rm s}= 15$ mag are selected, with a photometric error of $<0.1$ mag and a quality bit flag indicating at best minor warnings.
This query results in 885~558 sources. 
There are several magnitudes listed in the source tables. 
The recommended {\it aperMag3} is taken, which is based on a 2\arcsec\ aperture in diameter and includes an aperture correction 
and a saturation correction for the brightest stars (not relevant here).
%
%
Only likely and probable point sources are selected reducing the number of objects to 851~658\footnote{Selecting stars 
with {\it mergedClass} of $-1$ or $-2$.}.
The sky distribution is shown in Fig.~\ref{Fig-RaDec}.
The LMC, the SMC, the two tiles in the Magellanic Stream (MS), and the Magellanic Bridge (MB) are apparent.
The small regions missing in the corner of every tile correspond to detector 16 which are excluded by 
selecting on the quality bit flag\footnote{Selecting objects with {\it ksppErrBits} $<$ 256.}.

The data are dereddened based on the reddening law of \citet{Cardelli1989} 
for $R_{V}= 3.1$ which in the VISTA passbands leads to $A_{J}/A_{V} = 0.283$ and $A_{K_{\rm s}}/A_{V} = 0.114$ \citep{Rubele15}.
The dereddened data are then transformed from the VISTA system to the 2MASS system, which will
be the reference photometric system in this paper.
Transformation formulae from 2MASS to VISTA are given by \citet{VISTACal}\footnote{In their Appendix~C1 for software version 1.3.} 
which can be inverted to give:
\begin{equation}
J = J_{\rm VISTA} + 0.0703 \; (J-K_{\rm s})_{\rm VISTA}
\end{equation}
\begin{displaymath}
K_{\rm s} = K_{\rm s, VISTA} - 0.0108 \; (J-K_{\rm s})_{\rm VISTA},
\end{displaymath}
with the subscript "VISTA" indicating magnitudes in the VISTA system.

\begin{figure}
\centering

\begin{minipage}{0.24\textwidth}
\resizebox{\hsize}{!}{\includegraphics{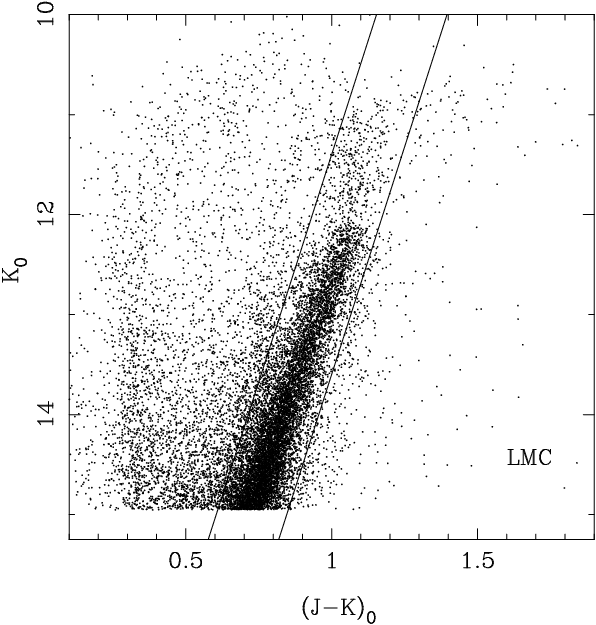}}
\end{minipage}
\begin{minipage}{0.24\textwidth}
\resizebox{\hsize}{!}{\includegraphics{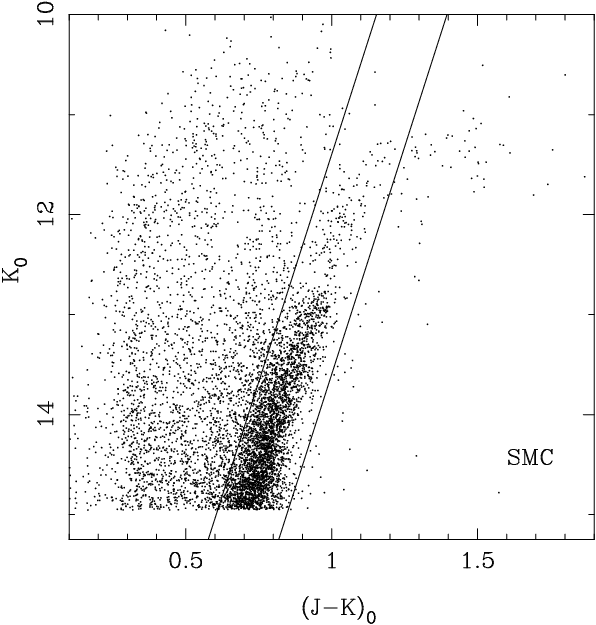}}
\end{minipage}

\begin{minipage}{0.24\textwidth}
\resizebox{\hsize}{!}{\includegraphics{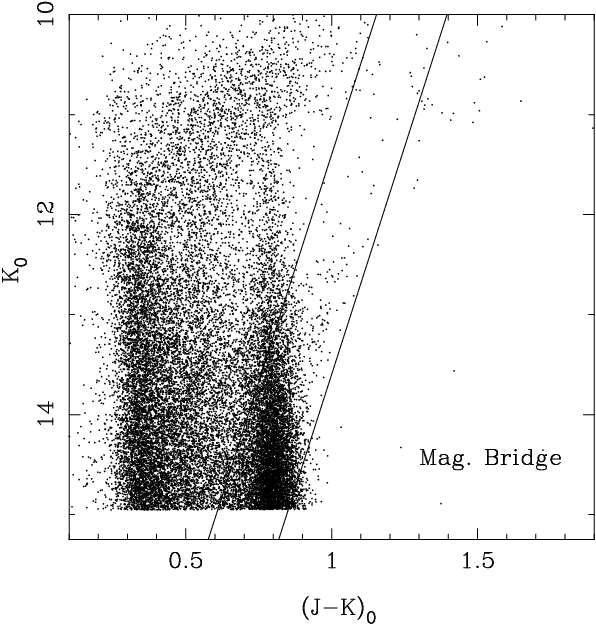}}
\end{minipage}
\begin{minipage}{0.24\textwidth}
\resizebox{\hsize}{!}{\includegraphics{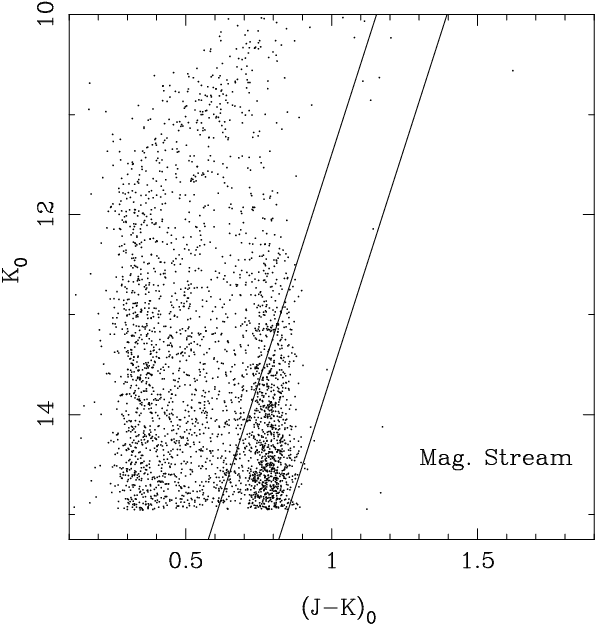}}
\end{minipage}

\caption[]{ 
Colour--magnitude diagrams of the LMC, SMC, MS, and the MB. 
For clarity only every 40th (LMC), or 20th (SMC) point is shown, and all points for the MS and MB.
The solid lines (see text) indicate the adopted borders to select RGB stars, independent of spatial location (see Eq.~\ref{eq-limits}).
} 
\label{Fig-CMD} 
\end{figure}

Figure~\ref{Fig-CMD} shows the colour--magnitude diagram (CMD) for the LMC, SMC, MS and MB.
For this figure, a constant $E(B-V)$ of 0.12 (LMC) and 0.075 mag (SMC, MS, MB) are adopted for simplicity, 
the average of the reddening towards the known dEBs in the LMC and the SMC (see Table~\ref{Tab-EB}).
The RGB is very well developed in the LMC and the SMC, but there are only a few RGB stars in the MS and MB.
The figure also includes lines which are used to select stars for further analysis.
The TRGB method is applied to stars with
\begin{displaymath}
K_0 > -9.1 \; (J-K_{\rm s})_0 + 20.50 {\rm \; (mag), and}
\end{displaymath}
\begin{equation}
K_0 < -9.1 \; (J-K_{\rm s})_0 + 22.70 {\rm \; (mag).}  
\label{eq-limits}
\end{equation}
These relations are determined by eye to select predominantly RGB stars and minimise AGB/foreground contaminants.
As Figure~\ref{Fig-CMD} shows the same relations are effective in making this selection for SMC and LMC alike.
When the method outlined below is applied to another stellar system a different set of equations should be 
determined to take into account differences in DM and colour of the RGB.
We note that photometric uncertainties are very small in the VMC data, at $K_{\rm s}= 12, 13, 14$ mag, and 
the typical photometric errors are 1.5, 2.0 and 4.2 millimags, respectively. 

The model to detect the TRGB is introduced in Sect.~\ref{S-Model}, but we first discuss the
absolute calibration of the TRGB in the infrared as this also enters into the method.

\section{Absolute calibration of the TRGB in the $K_{\rm s}$-band} 
\label{S-Calib}

The default calibration for the brightness of the TRGB in the present paper is based on the 
theoretical calculations of \citet{Serenelli17} which provide the 
absolute magnitude in several filters ($V$ and $I$, $J$ and $K_{\rm s}$ in the 2MASS system, and HST {\it F110W} and {\it F160W} filters) 
based on stellar evolution models, using bolometric corrections to convert luminosity, effective temperature 
and metallicity to the observational plane.
In their Table~1 they provide second-order polynomial
fits to  $M_{K_{\rm s}}$ for two ranges in $(J-K_{\rm s})$. Here we use a subset of their dataset (kindly provided by M.~Salaris) to fit
a linear equation in the colour range of interest.

\begin{figure}
\centering
\resizebox{\hsize}{!}{\includegraphics{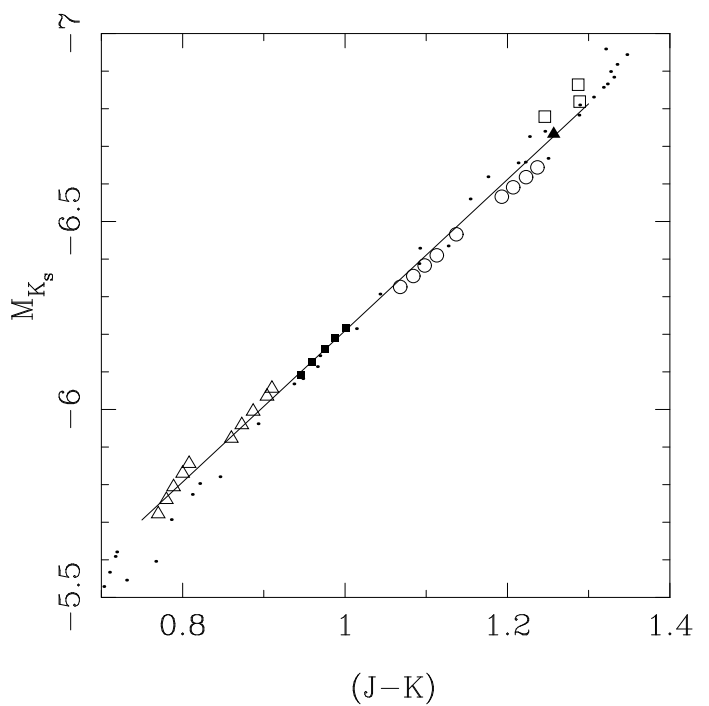}}
\caption[]{ 
Calibration of the $K_{\rm s}$-band absolute magnitude of the TRGB as a function of $(J-K_{\rm s})$ colour,
based on the data of \citet{Serenelli17}.
The solid line indicates the fit to models in the colour range 0.75 $< (J-K_{\rm s}) < 1.3$ mag and ages between 4 and 14 Gyr.
Sets of different metallicities are indicated by 
open triangles  ([Fe/H]= $-1.49$ dex), 
filled squares  ([Fe/H]= $-1.27$ dex), 
open circles    ([Fe/H]= $-0.96$ and $-0.66$ dex), 
filled triangle ([Fe/H]= $-0.35$ dex), and
open squares    ([Fe/H]= $-0.25$, $-0.01$ and $+0.06$ dex).
Models outside these criteria 
are indicated by the small dots.
} 
\label{Fig-Calib} 
\end{figure}

Restricting the fit to the colour range $0.75 < (J-K_{\rm s}) < 1.3$ mag to broadly match the colour range 
of the SMC and LMC TRGBs, model ages older than 4 Gyr (see the discussion in \citealt{Serenelli17}), 
and model ages younger than 14 Gyr, the bi-sector fit is: 
\begin{equation}
M_{K_{\rm s}} =  (-4.196 \pm 0.030) - (2.013 \pm 0.042) \; (J-K_{\rm s})
\label{Eq-AC}
\end{equation}
with an rms of 0.030 mag ($N= 28$). The fit is shown as the solid line in Fig.~\ref{Fig-Calib}.
In Sect.~\ref{SS-Calib} the sensitivity of the results to this calibration is investigated.
An alternative calibration, restricting the colour range to specifically match that of the SMC and LMC TRGBs 
makes the relation shallower,  
$M_{K_{\rm s}} =  (-4.331 \pm 0.025) - (1.873 \pm 0.023)\, (J-K_{\rm s})$ for $0.82 < (J-K_{\rm s}) < 1.2$ mag 
with an rms of 0.009 mag ($N= 16$).

When the current paper was near completion \citet{Madore18} and \citet{Hoyt18} discussed the absolute calibration of 
the TRGB in $JHK$\footnote{Also see \citet{Gorski18} which appeared when this paper was under review.}.
They derived the slope from data in IC 1613, and found $\beta= -1.85 \pm 0.27$, consistent with \citet{Serenelli17} in general 
and the specific values from our fits.
Using NIR data in the bar of the LMC, adopting the distance to the LMC from the dEBs in \citet{Pietrzynski13}, 
$\beta= -1.85$ from the work on IC 1613, and a low reddening to the LMC of $E(B-V)= 0.03 \pm 0.03$ mag, they derived 
a zero point (ZP) of $-6.14$ mag (at $(J-K_{\rm s})=1.0$ mag). 
The error in the ZP they claimed is 0.01 mag (statistical) and 0.06 (systematic), of which 0.02 is due to the uncertainty 
in the reddening, and 0.05 mag to the adopted LMC distance. 

The reddening \citet{Hoyt18} adopted is quite low, but is also inconsistent with the (mean) reddening towards the dEBs in the LMC, 
the (mean) distance of which is used to calibrate the ZP. Adopting $E(B-V)= 0.12$ mag (see earlier, and Table~\ref{Tab-EB}) 
their ZP would become $-6.17$ mag  (at $(J-K_{\rm s})=1.0$ mag).
This ZP compares to $-6.21$ and $-6.20$ mag  (at $(J-K_{\rm s})=1.0$ mag) that we derive from the data in \citet{Serenelli17}.

\section{Model} 
\label{S-Model}

The calculations are carried out using a numerical program, which reads in the VMC data.
Other inputs are the right ascension (RA) and declination (Dec) of the line-of-sight (los) of interest, 
the radius, $r$, of the circle centred on (RA, Dec) to select the data from the VMC input, 
the adopted reddening $E(B-V)$ for that los, and
the adopted width of the bin, $w$, for the binning of the LF. 

The VISTA $J, K_{\rm s}$ magnitudes are de-reddened and transformed to the 2MASS system as outlined in Sect.~\ref{S-Sample}.
If the absolute calibration relation is $M_{K_{\rm s}} = \alpha + \beta \cdot (J-K_{\rm s})$, the "sharpened" magnitude 
$T = K_0 - \beta \cdot (J-K_{\rm s})_0$ is constructed with $\beta = -2.013$ as standard value following Sect.~\ref{S-Calib}.
The error in $T$ is calculated from the  propagation of the errors in $J, K$, and $\beta$.
We also keep track of $(J-K_{\rm s})_0$ and its error.
Stars in the region defined by Eq.~2 are selected and the LF in $T$ is constructed using the adopted bin size.

Two edge-detection algorithms are run on the binned LF, based on the first-order and second-order derivative of the LF. 
The derivatives are calculated using Savitzky-Golay coefficients as implemented in Fortran in "Numerical Recipes" \citep{Press1992}. 
At a point $i$ the function $f$ is replaced by a linear combination $g$, of itself and $n_{\rm L}$ "left" and $n_{\rm R}$ "right" neighbouring values:
\begin{equation}
g_i = \sum_{n= -n_{\rm L}}^{n_{\rm R}} c_n \; f_{i+n}  
\end{equation}
The Savitzky-Golay coefficients are determined in such a way that the filter fits a polynomial of degree $M$ to 
the moving window, and then evaluates the derivative of chosen order $L$.
\citet{Cioni2000} performed extensive tests and used $M= 2$ and $n_{\rm L} = n_{\rm R} = 3$ for their second-order derivative 
filter which we adopt here as well\footnote{Within the implementation in "Numerical Recipes" the functional call is 
savgol(SG, nSG, 3, 3, 2, 2), where SG is an array of size nSG, and leads to the (approximate) kernel [+0.60 0.0 $-0.36$ $-0.48$ $-0.36$ 0.0 +0.60].
The convolution is performed with the routine {\it convlv}.}.
For the first-order derivative we use $M= 1$ and $n_{\rm L} = n_{\rm R} = 2$, resulting in the kernel used by \citet{Sakai1996}\footnote{The functional call is 
savgol(SG, nSG, 2, 2, 1, 1) and leads to the kernel [$-2$ $-1$ 0 +1 +2]. 
The call savgol(SG, nSG, 1, 1, 1, 1) would lead to the classical kernel [$-1$ 0 +1], as first introduced by \citet{Lee93}. 
Note that \citet{MF95} use yet another kernel,  [$-1$ $-2$ 0 +2 +1] to determine the first derivative.}.

The filter response of the LF to the first-order derivative kernel is fitted with a single Gaussian (SG) plus a constant:
\begin{equation}
F(m)=  a_1 + a_2 \; \exp( -(m - a_3)^2 / (2 a_4^2) ),
\end{equation}
where the TRGB magnitude is given by the peak of the Gaussian.

\citet{Cioni2000} also fitted a SG to the response function of the LF to the second-order derivative filter and then applied a correction which depends
on the width of the Gaussian fit (see Figure~A2 in \citealt{Cioni2000}), which can be a few tenths of a magnitude.
Here we find (Appendix~\ref{App-Simul}) that the response function to the second-order derivative filter can be well fitted 
by a double Gaussian (DG) of the form:
\begin{eqnarray}
 F(m) &=&  a_1     + a_2 \; \exp( -(m - a_3 + a_5)^2 / (2 a_4^2) )  \nonumber \\
      & & \quad {} - a_2 \; \exp( -(m - a_3 - a_5)^2 / (2 a_4^2) ).
\end{eqnarray}
Compared to the SG it has one additional free parameter, the distance between the positive and negative peaks of the Gaussians, $a_5$, 
and where the TRGB magnitude is given by the magnitude in between the peaks.
%
%
For both the SG and DG fits the DM for a given los is then $a_3 + \alpha$.

In Appendix~\ref{App-Simul} the numerical details of the method are discussed extensively, 
including simulations to estimate any biases in the method, the influence of the bin size, and error estimates.
It is found that both the first- and the second-order derivative methods can be applied with negligible bias (a few millimag) 
if certain criteria are met that concern the significance with which the peak in the response function is detected 
(SNpk= $a_2/\sigma_{\rm a_2}$), the average number of stars per bin ($N$/bin) in the 0.5 mag below the tip of the RGB, and 
the error in the magnitude of the peak ($\sigma_{a_3}$) relative to the width of the bin.
The second-order derivative method is more stable to noise in the data but needs more stars per bin. 
\cite{Cioni2000} also prefer the second-order derivative (as mentioned before however, their implementation differs from the current one) 
over the first-order derivative method.

In the applications discussed below the code is run for a given los for all combinations 
of 18 radii\footnote{
Radii $r$= 0.45\degr\ in steps of 0.05 to 1.0, 1.25 to 2.0\degr\ in steps of 0.25, 2.5 and 3.0\degr.}
and bin widths\footnote{Twenty bin widths $w$= 0.033 in steps of 0.001 to 0.048, 0.05, 0.06, 0.07, and 0.08 mag    for the second-order filter, and 
                            19 bin widths $w$= 0.016 in steps of 0.001 to 0.030, 0.035, 0.040, 0.045, and 0.050 mag for the first-order filter.}.
The best model is adopted to be the one with the lowest reduced $\chi^2$ ($\chi_{\rm r,min}^2$) that meets the criteria on SNpk, $N$/bin and $a_3/w$.
Below, we also investigate the range in the parameters for models with $\chi_{\rm r}^2 < 2 \cdot \chi_{\rm r, min}^2$ to have an independent
estimate of the errors on the derived distances.

\section{Applications} 
\label{S-Appl}

\subsection{TRGB distances towards dEBs in the MCs}
\label{SS-EB}

In a first application we considered the TRGB in the los towards nine dEBs in the LMC and five in the SMC.
In particular for the LMC, the eight systems in \citet{Pietrzynski13} give a DM to the LMC barycentre
of 18.493 $\pm$ 0.008 (statistical) $\pm$ 0.047 (systematic) mag which has become the de-facto value adopted
after 2013 for the DM to the LMC in most papers.
For the SMC, \citet{Graczyk14} give a mean DM based on five dEBs of 18.965 $\pm$ 0.025 (statistical) $\pm$ 0.048 (systematic) mag.
For comparison, based on a careful, statistical analysis of a large number of recent distance estimates, 
grouped by main stellar population tracers, \citet{deGB2014} and \citet{deGB2015} recommend DMs of 
18.49 $\pm$ 0.09 to the LMC, and
18.96 $\pm$ 0.02 mag (formal errors), with additional systematic uncertainties possibly exceeding 0.15--0.20 mag, for the SMC.

Table~\ref{Tab-EB} lists the identifier, DM and error, and the reddening (the error is given on the second line) 
given by the references listed in the fourth column. 
Columns 5--11 contain the results of our analysis:
The DM and error, the estimated $(J-K_{\rm s})_0$ magnitude at the TRGB and error (see Appendix~\ref{App-Simul} on how they are derived), 
the radius of the circle used, the bin width, the average number of RGB stars per bin in the 0.5 mag below the TRGB, 
the significance with which the peak in the response function is detected, and the reduced $\chi^2$. 
The errors quoted are the formal errors.

\begin{figure}
\centering

\includegraphics[width=0.72\hsize]{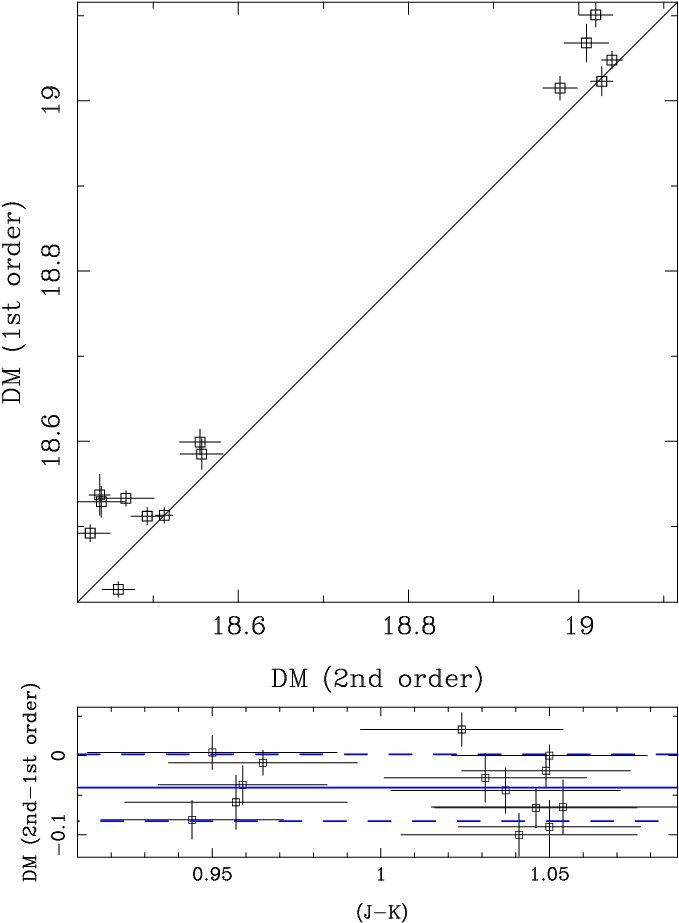}
\bigskip

\includegraphics[width=0.83\hsize]{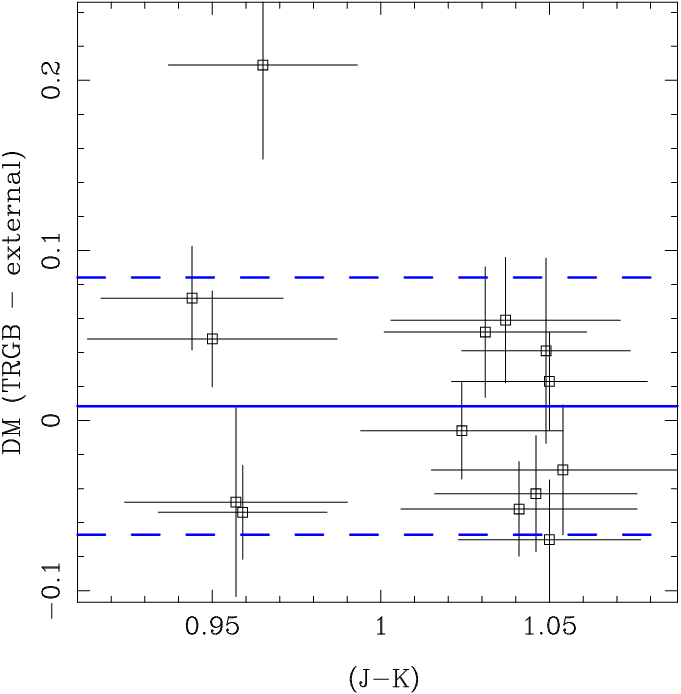}

\caption[]{ 
Comparison of the TRGB DM based on the first- and second-order derivatives (top panel),
and the difference plotted against $(J-K_{\rm s})$ colour (middle panel) towards the 14 los containing dEBs.
The one-to-one relation is shown in the top  panel.
In the middle panel in blue are indicated the median of the difference (solid line) and plus-minus the 
dispersion (taken as 1.48 $\cdot$ MAD; dashed lines).
The bottom panel shows the difference between the second-order-derivative-based TRGB distance and the 
DM of the dEB systems against colour.
The median of the difference (solid line) and plus-minus 1.48 $\cdot$ MAD (dashed line) are shown as the blue lines.
} 
\label{Fig-TRGB-EB} 
\end{figure}

Figure~\ref{Fig-TRGB-EB} shows the comparison between the first- and second-order-derivative-based DM and the difference
plotted against $(J-K_{\rm s})$ colour of the TRGB (left-hand panel), and the comparison of the second-order-derivative-based DM with
the published values of the DM for the dEBs.

Interestingly, an offset between the second- and first-order-derivative-based DM is observed that is
not predicted by the simulations. The difference is small (median offset of $-0.040$, a weighted mean offset of $-0.026$ mag)
and insignificant (the error in this offset is 0.042 mag). It is observed in other applications as well, and we return to this 
in Sect.~\ref{S-Sum}. 
The simulations in Appendix~\ref{App-Simul} do suggest that the second-order-derivative-based DM 
is the more reliable and stable of the two methods in reproducing the input DM, and therefore we choose this option in the 
comparisons to external catalogues. 
The simulations show that this method requires approximately twice as many stars per magnitude bin than the first-order derivative filter.
Inspection of Tables~\ref{Tab-EB}-\ref{Tab-VMC} indeed shows that for the best fits, when the resulting areas on the sky are similar 
for the second- and first-order derivative results, the bin size in the former case is almost always larger than for the latter.

The bottom panel of Fig.~\ref{Fig-TRGB-EB} compares the second-order-derivative-based DM with the published values for the dEBs systems.
There is excellent agreement with a difference of 0.009 $\pm$ 0.075 mag. There is no trend of the offset with colour.
Part of the scatter could be due to the depth along the los. The TRGB distance is based on the RGB stars in a 
field of $\sim0.4-2$\degr\ radius spread along the los while the DM to each dEB is that to a single object.

\begin{table*}
\setlength{\tabcolsep}{1.25mm}

\caption{TRGB distances to MC fields surrounding dEBs.}
\label{Tab-EB}
\centering
  \begin{tabular}{lcccccccrrr}
  \hline\hline
   System ID  &         DM$_{\rm EB}$    & $E(B-V)$ &  Ref &    DM$_{\rm TRGB}$          &   $(J-K_{\rm s})_0$ @ TRGB   & Rlim    & bin width & $N$/bin & SNpk & $\chi^2_{\rm r}$ \\
OGLE-         &        (mag)      & (mag)    &      &       (mag)        &       (mag)          & (\degr) &   (mag)   &       &      &                 \\
\hline
LMC-ECL-01866 & 18.496 $\pm$ 0.028 & 0.115       &  3 & 18.555 $\pm$ 0.024 & 1.037 $\pm$ 0.034 & 0.85 & 0.070 & 175 & 5.2 & 1.0 \\
              &                    & $\pm$ 0.020 &    & 18.599 $\pm$ 0.015 & 1.032 $\pm$ 0.017 & 0.75 & 0.045 &  92 &  6.6  &  1.5  \\
LMC-ECL-03160 & 18.505 $\pm$ 0.029 & 0.123 &  3 & 18.557 $\pm$ 0.025 & 1.031 $\pm$ 0.030 & 0.80 & 0.060 & 179 & 5.1 & 1.4 \\
              &        &  $\pm$ 0.020        &        & 18.585  $\pm$  0.018  &  1.026  $\pm$ 0.015  & 0.75  & 0.030  &  82  &   5.5  &  0.8  \\
LMC-ECL-06575 & 18.497 $\pm$ 0.019 & 0.107 &  3 & 18.468 $\pm$ 0.033 & 1.054 $\pm$ 0.039 & 0.45 & 0.046 & 92 & 5.0 & 2.4 \\
              &        &  $\pm$ 0.020        &        & 18.533  $\pm$  0.009  &  1.055  $\pm$ 0.010  & 0.75  & 0.018  & 105  &   5.1  &  1.2  \\
LMC-ECL-09114 & 18.465 $\pm$ 0.021 & 0.160 &  3 & 18.459 $\pm$ 0.019 & 1.024 $\pm$ 0.030 & 0.50 & 0.043 & 152 & 7.0 & 10.8 \\
              &        &  $\pm$ 0.020        &        & 18.426  $\pm$  0.009  &  1.034  $\pm$ 0.009  & 0.80  & 0.018  & 144  &   5.1  &  1.0  \\
LMC-ECL-09660 & 18.489 $\pm$ 0.025 & 0.127 &  3 & 18.437 $\pm$ 0.012 & 1.041 $\pm$ 0.035 & 0.85 & 0.042 & 90 & 5.5 & 1.8 \\
              &        &  $\pm$ 0.020        &        & 18.537  $\pm$  0.024  &  1.027  $\pm$ 0.017  & 0.80  & 0.040  &  86  &   5.2  &  1.1  \\
LMC-ECL-10567 & 18.490 $\pm$ 0.027 & 0.102 &  3 & 18.513 $\pm$ 0.010 & 1.050 $\pm$ 0.029 & 0.60 & 0.046 & 193 & 5.4 & 5.0 \\
              &        &  $\pm$ 0.020        &        & 18.513  $\pm$  0.009  &  1.055  $\pm$ 0.011  & 0.70  & 0.024  & 128  &   6.3  &  1.4  \\
LMC-ECL-15260 & 18.509 $\pm$ 0.021 & 0.100 &  3 & 18.439 $\pm$ 0.028 & 1.050 $\pm$ 0.027 & 0.45 & 0.044 & 191 & 17.6 & 2.1 \\
              &        &  $\pm$ 0.020        &        & 18.529  $\pm$  0.018  &  1.041  $\pm$ 0.012  & 0.45  & 0.030  & 146  &   5.0  &  2.1  \\
LMC-ECL-25658 & 18.452 $\pm$ 0.051 & 0.091 &  4 & 18.493 $\pm$ 0.019 & 1.049 $\pm$ 0.025 & 2.00 & 0.040 & 189 & 5.5 & 1.8 \\
              &        &  $\pm$ 0.030        &        & 18.512  $\pm$  0.010  &  1.047  $\pm$ 0.011  & 2.00  & 0.025  & 121  &   7.4  &  2.5  \\
LMC-ECL-26122 & 18.469 $\pm$ 0.025 & 0.140 &  3 & 18.426 $\pm$ 0.023 & 1.046 $\pm$ 0.030 & 0.45 & 0.040 & 147 & 5.6 & 1.5 \\
              &        &  $\pm$ 0.020        &        & 18.492  $\pm$  0.010  &  1.045  $\pm$ 0.011  & 0.50  & 0.023  & 111  &   5.2  &  0.8  \\
SMC-ECL-0195 & 18.948 $\pm$ 0.023 & 0.079 &  1 & 19.020 $\pm$ 0.020 & 0.944 $\pm$ 0.027 & 0.85 & 0.046 & 138 & 6.9 & 1.6 \\
              &        &  $\pm$ 0.020        &        & 19.101  $\pm$  0.014  &  0.923  $\pm$ 0.014  & 0.80  & 0.045  & 137  &   8.3  &  1.1  \\
SMC-ECL-0708 & 18.979 $\pm$ 0.025 & 0.080 &  1 & 19.027 $\pm$ 0.013 & 0.950 $\pm$ 0.037 & 0.45 & 0.070 & 145 & 6.2 & 2.0 \\
              &        &  $\pm$ 0.020        &        & 19.023  $\pm$  0.017  &  0.948  $\pm$ 0.016  & 0.50  & 0.040  & 100  &   5.2  &  0.9  \\
SMC-ECL-1421 & 19.057 $\pm$ 0.049 & 0.067 &  1 & 19.009 $\pm$ 0.026 & 0.957 $\pm$ 0.033 & 0.50 & 0.060 & 157 & 6.7 & 4.3 \\
              &        &  $\pm$ 0.020        &        & 19.068  $\pm$  0.022  &  0.943  $\pm$ 0.015  & 0.50  & 0.050  & 141  &   5.7  &  0.8  \\
SMC-ECL-4152 & 19.032 $\pm$ 0.019 & 0.093 &  1 & 18.978 $\pm$ 0.020 & 0.959 $\pm$ 0.025 & 0.80 & 0.045 & 186 & 5.1 & 1.2 \\
              &        &  $\pm$ 0.020        &        & 19.015  $\pm$  0.014  &  0.950  $\pm$ 0.011  & 0.85  & 0.029  & 144  &   5.5  &  1.6  \\
SMC-ECL-5123 & 18.830 $\pm$ 0.054 & 0.060 &  2 & 19.039 $\pm$ 0.012 & 0.965 $\pm$ 0.028 & 0.95 & 0.048 & 188 & 6.9 & 1.1 \\
              &        &  $\pm$ 0.030        &        & 19.048  $\pm$  0.010  &  0.956  $\pm$ 0.011  & 1.25  & 0.023  & 147  &   6.4  &  1.2  \\
\hline
\end{tabular}
\tablebib{
(1)~\citet{Graczyk14};
(2)~\citet{Graczyk12};
(3)~\citet{Pietrzynski13};
(4)~\citet{Elgueta16}.
}
\tablefoot{Column~1 gives the OGLE identifier, with the DM (Col.~2) and reddening (Col.~3) as given by the references listed in Col.~4.
Columns~5--11 contain the parameters derived in the present paper: The DM, the $(J-K_{\rm s})_0$ colour at the TRGB, 
the radius of the circle used to select the stars in that direction,
the bin width, 
the average number of stars per bin in the 0.5 mag below the tip of the RGB, 
the significance in the detection of the peak in the response function, and
the reduced $\chi^2$.
The first line for each object has the results for the second-order derivative filter response, and
the second line those for the first-order derivative filter.
}
\end{table*}

\subsection{TRGB distances towards LMC Cepheids}
\label{SS-CEP}

A second application concerns the TRGB distances towards CCs in the LMC.
\citet{Inno16} presented DM and reddening estimates for 2504 CCs in the LMC, derived by 
simultaneously fitting $V, I, J, H, K$ and WISE W1 magnitudes (when available) to corresponding period-luminosity ($PL$)-relations.
In the procedure discussed below 16 stars with very negative reddenings ($E(B-V) < -0.07$ mag) and 22 stars with very large $\chi^2$ \
($> 600$, compared to the median of 20) have been excluded from the sample of \citet{Inno16}.

Some scatter in DM is expected due to the finite width of the instability strips and depth effects.
Therefore we average DM and reddening values of Cepheids located close together on the sky in the following way:
starting from the first Cepheid in the list\footnote{We verified that the starting order is irrelevant.} 
in \citet{Inno16} its distance to all neighbours not already marked to belong to another los is calculated.
The number, NN, of nearest neighbours is identified (with NN at least 35).
If the distance to the NN-th nearest neighbour is less than 0.4\degr\ NN is increased by 2, and this is repeated if necessary.
The NN Cepheids are marked as belonging to this los, and one proceeds to the next Cepheid in the list.
This is repeated until no more Cepheids can be assigned to a los (the distance to the NN-th nearest neighbour should be less than 1.5\degr).
The minimum number of Cepheids and the minimum distance are chosen after some testing, using the results of the dEBs that show that the 
radius needed for the TRGB to have reliable results is of order 0.45 to 2\degr\ (see Table~\ref{Tab-EB}).

In this way, 56 independent los were identified containing 2182 CCs. 
For each los the median and standard deviation (calculated as 1.48 times the median-absolute-deviation, 
MAD\footnote{The MAD is robust to outliers, and in the case of a Gaussian distribution 1.48 $\cdot$ MAD is equivalent to $\sigma$ of 
a Gaussian distribution.}) of the DM and reddening were calculated.

The results of the calculations are listed in Table~\ref{Tab-CEP}, which 
lists the identifier (the name of the CC at the centre of each los), the median DM of the CCs in that los, 
the median of the error in the DM of each CC in that los, 
the median of the reddening of the CCs in that los 
(the error, calculated as 1.48 $\cdot$ MAD of the reddening values around the median, is given on the second line).
The radius used to calculate these averages is listed in column~4.
Columns 5--11 in Table~\ref{Tab-CEP} contain the results of our analysis following Table~\ref{Tab-EB}.
The first line for each object contains the results for the second-order derivative filter response, and
the second line those for the first-order derivative filter.

\begin{figure}
\centering

\includegraphics[width=0.72\hsize]{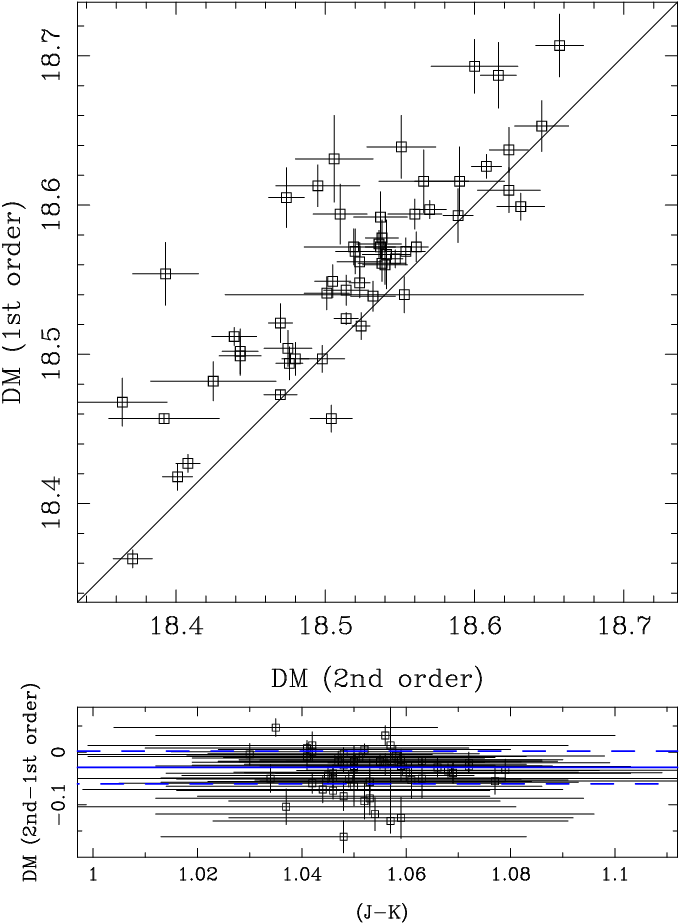}
\bigskip

\includegraphics[width=0.72\hsize]{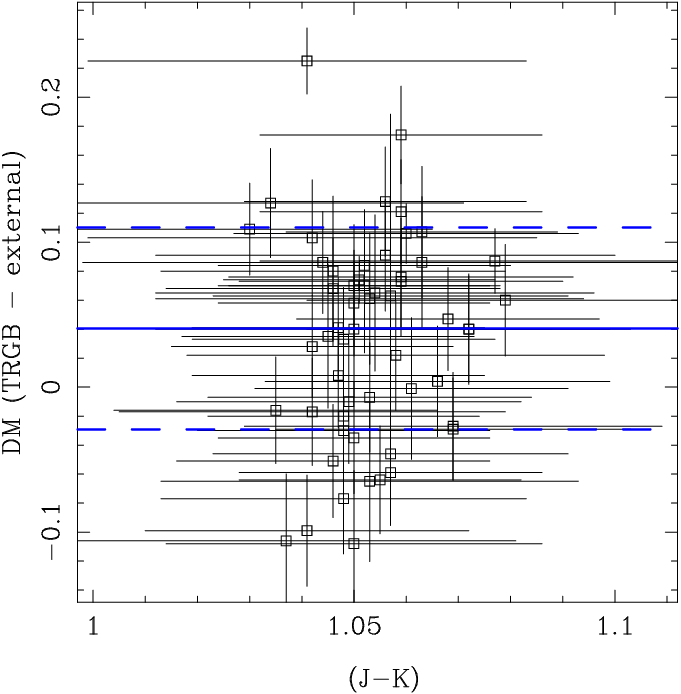}

\includegraphics[width=0.72\hsize]{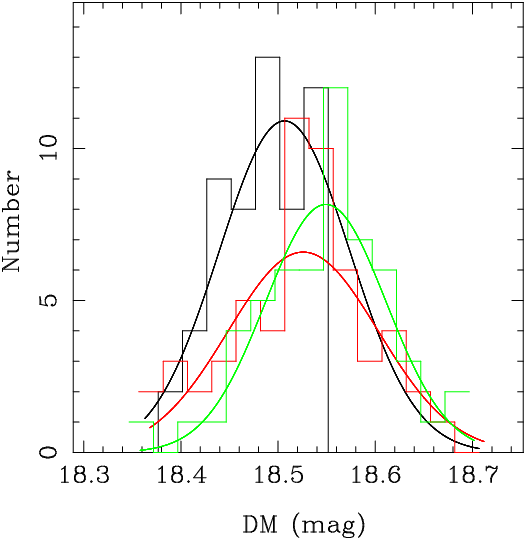}

\caption[]{ 
Same as Fig.~\ref{Fig-TRGB-EB} for 56 los towards CCs in the LMC.
Additionally the bottom panel shows the distribution of the DM for the CCs (in black), and
the first- (green) and second-order-based-derivative TRGB distance (red), and Gaussian fits to these distributions.
For clarity, the green and red histograms have been offset by $-0.005$ and $+0.005$ mag from the black one.
} 
\label{Fig-TRGB-CEP} 
\end{figure}

Figure~\ref{Fig-TRGB-CEP} compares the second-r and first-order-derivative-based DM, and a similar observation is made as in the 
previous section. The difference between the two estimates is $-0.029 \pm 0.031$ mag.
The comparison between the second-order-derivative-based TRGB distance and the median DM for the CCs in that los is good with 
a  negligible difference of $0.041 \pm 0.070$ mag.

With a large number of los spread across the LMC one can also discuss the distribution of the distances and the mean distance 
to the LMC. This is illustrated in the bottom-right panel of Fig.~\ref{Fig-TRGB-CEP}, which  
shows histograms of the DM of the 56 los for the CCs (black), the second-order-derivative-based TRGB distance (red), 
and the first-order-derivative-based TRGB distance (green), and Gaussian fits to these distributions.
As the error bar in an individual DM estimate is non-negligible compared with the width of the distribution 
we also performed Monte Carlo simulations. A new DM for each los was drawn from a Gaussian distributed based 
on its derived value and error. A new histogram based on these new DM was created and a new Gaussian fit was performed.

For the CCs a median DM of 18.491 mag is found with an error on the mean of 0.005 mag. The $\sigma$ of the Gaussian distribution is 0.052 mag.
For the second-order-based-derivative we find $18.521 \pm 0.007$, $\sigma= 0.074$ mag and 
for the first-order-based-derivative          $18.567 \pm 0.006$, $\sigma= 0.078$ mag.
As expected, the value for the CCs is in excellent agreement with the 18.48 $\pm$ 0.10 mag (stat. plus syst.)  quoted by \citet{Inno16} for 
their entire sample.

\longtab{
\setlength{\tabcolsep}{1.4mm}
\begin{longtable}{lcccccccrrr}
  \caption{\label{Tab-CEP} TRGB distances to LMC fields surrounding CCs.} \\
  \hline  \hline
 Name         &         DM        & $E(B-V)$ &     &         DM          &  $(J-K_{\rm s})_0$ @ TRGB  & Rlim    & bin width & $N$/bin & SNpk & $\chi^2_{\rm r}$ \\
              &        (mag)      & (mag)  &  &         (mag)        &      (mag)         & (\degr) &   (mag)   &       &      &                 \\
\hline
\endfirsthead
\caption{continued.}\\
\hline\hline
 Name          &         DM        & $E(B-V)$ &    &       DM          &  $(J-K_{\rm s})_0$ @ TRGB & Rlim    & bin width & $N$/bin & SNpk & $\chi^2_{\rm r}$ \\
               &        (mag)      & (mag)  &  &         (mag)        &      (mag)        & (\degr) &   (mag)   &       &      &                 \\
\hline
\endhead
\hline
\endfoot
HV 955 & 18.430 $\pm$ 0.031 & 0.061 &   &  18.506 $\pm$ 0.026 & 1.059 $\pm$ 0.033 & 1.25 & 0.048 & 181 & 5.0 & 2.4 \\
              &        &  $\pm$ 0.068        &        & 18.631  $\pm$  0.029  &  1.033  $\pm$ 0.021  & 0.80  & 0.050  &  95  &   6.1  &  0.6  \\
HV 6098 & 18.480 $\pm$ 0.042 & 0.070 &   &  18.470 $\pm$ 0.008 & 1.049 $\pm$ 0.033 & 1.75 & 0.034 & 96 & 5.5 & 2.2 \\
              &        &  $\pm$ 0.031        &        & 18.521  $\pm$  0.013  &  1.036  $\pm$ 0.015  & 1.75  & 0.022  &  68  &   6.4  &  3.2  \\
HV 1002 & 18.410 $\pm$ 0.034 & 0.058 &   &  18.538 $\pm$ 0.016 & 1.056 $\pm$ 0.027 & 1.50 & 0.041 & 174 & 5.1 & 0.6 \\
              &        &  $\pm$ 0.077        &        & 18.561  $\pm$  0.012  &  1.050  $\pm$ 0.013  & 1.50  & 0.030  & 131  &   6.5  &  0.6  \\
HV 2827 & 18.430 $\pm$ 0.014 & 0.042 &   &  18.536 $\pm$ 0.015 & 1.060 $\pm$ 0.033 & 1.25 & 0.060 & 150 & 6.8 & 1.1 \\
              &        &  $\pm$ 0.043        &        & 18.574  $\pm$  0.009  &  1.050  $\pm$ 0.011  & 1.75  & 0.024  & 123  &   5.1  &  0.7  \\
LMC-CEP-3568 & 18.490 $\pm$ 0.039 & 0.046 &   &  18.551 $\pm$ 0.023 & 1.053 $\pm$ 0.041 & 1.00 & 0.045 & 90 & 7.5 & 1.8 \\
              &        &  $\pm$ 0.030        &        & 18.639  $\pm$  0.021  &  1.035  $\pm$ 0.015  & 1.25  & 0.029  & 109  &   5.0  &  0.7  \\
LMC-CEP-3506 & 18.530 $\pm$ 0.033 & 0.025 &   &  18.616 $\pm$ 0.012 & 1.044 $\pm$ 0.046 & 0.95 & 0.060 & 110 & 6.1 & 2.5 \\
              &        &  $\pm$ 0.061        &        & 18.687  $\pm$  0.022  &  1.034  $\pm$ 0.014  & 1.50  & 0.029  & 138  &   5.2  &  0.7  \\
LMC-CEP-3649 & 18.530 $\pm$ 0.035 & 0.054 &  &  18.570 $\pm$ 0.011 & 1.072 $\pm$ 0.031 & 1.00 & 0.046 & 147 & 5.6 & 1.3 \\
              &        &  $\pm$ 0.047        &        & 18.597  $\pm$  0.006  &  1.068  $\pm$ 0.009  & 1.50  & 0.017  & 145  &   7.1  &  2.1  \\
LMC-CEP-3320 & 18.420 $\pm$ 0.032 & 0.063 &  &  18.541 $\pm$ 0.016 & 1.059 $\pm$ 0.027 & 1.75 & 0.041 & 193 & 5.3 & 1.5 \\
              &        &  $\pm$ 0.050        &        & 18.567  $\pm$  0.023  &  1.054  $\pm$ 0.020  & 1.00  & 0.030  &  50  &   5.1  &  1.1  \\
LMC-CEP-3258 & 18.460 $\pm$ 0.035 & 0.140 &  &  18.443 $\pm$ 0.012 & 1.042 $\pm$ 0.037 & 0.90 & 0.050 & 127 & 5.8 & 5.9 \\
              &        &  $\pm$ 0.113        &        & 18.502  $\pm$  0.015  &  1.039  $\pm$ 0.014  & 1.00  & 0.029  & 103  &   5.4  &  2.4  \\
LMC-CEP-1128 & 18.510 $\pm$ 0.035 & 0.081 &  &  18.532 $\pm$ 0.015 & 1.058 $\pm$ 0.040 & 0.60 & 0.047 & 97 & 5.1 & 1.2 \\
              &        &  $\pm$ 0.043        &        & 18.539  $\pm$  0.010  &  1.054  $\pm$ 0.011  & 1.00  & 0.020  & 127  &   5.6  &  0.9  \\
LMC-CEP-4544 & 18.430 $\pm$ 0.046 & 0.047 &  &  18.495 $\pm$ 0.028 & 1.054 $\pm$ 0.042 & 1.00 & 0.080 & 112 & 5.7 & 1.6 \\
              &        &  $\pm$ 0.043        &        & 18.613  $\pm$  0.014  &  1.040  $\pm$ 0.012  & 1.75  & 0.030  & 140  &   6.2  &  0.5  \\
LMC-CEP-0107 & 18.520 $\pm$ 0.034 & 0.160 &  &  18.504 $\pm$ 0.014 & 1.035 $\pm$ 0.031 & 0.95 & 0.070 & 187 & 6.8 & 1.4 \\
              &        &  $\pm$ 0.074        &        & 18.457  $\pm$  0.009  &  1.034  $\pm$ 0.008  & 1.50  & 0.016  & 121  &   5.3  &  1.0  \\
LMC-CEP-0046 & 18.520 $\pm$ 0.034 & 0.120 &  &  18.561 $\pm$ 0.008 & 1.047 $\pm$ 0.028 & 1.25 & 0.048 & 160 & 6.4 & 2.9 \\
              &        &  $\pm$ 0.059       &        & 18.572  $\pm$  0.010  &  1.042  $\pm$ 0.011  & 1.50  & 0.027  & 139  &   8.2  &  0.8  \\
LMC-CEP-4064 & 18.440 $\pm$ 0.038 & 0.065 &  &  18.524 $\pm$ 0.006 & 1.052 $\pm$ 0.028 & 1.50 & 0.035 & 116 & 5.3 & 3.7 \\
              &        &  $\pm$ 0.046        &        & 18.519  $\pm$  0.009  &  1.052  $\pm$ 0.012  & 1.75  & 0.029  & 133  &   6.1  &  1.2  \\
LMC-CEP-1538 & 18.490 $\pm$ 0.036 & 0.120 &  &  18.425 $\pm$ 0.042 & 1.053 $\pm$ 0.040 & 0.65 & 0.050 & 120 & 5.2 & 3.1 \\
              &        &  $\pm$ 0.089        &        & 18.482  $\pm$  0.013  &  1.049  $\pm$ 0.011  & 0.90  & 0.023  & 130  &   5.4  &  1.4  \\
LMC-CEP-1954 & 18.520 $\pm$ 0.036 & 0.077 &  &  18.519 $\pm$ 0.033 & 1.061 $\pm$ 0.030 & 0.70 & 0.036 & 128 & 5.8 & 0.4 \\
              &        &  $\pm$ 0.034        &        & 18.572  $\pm$  0.012  &  1.056  $\pm$ 0.011  & 0.80  & 0.028  & 146  &   6.7  &  0.4  \\
LMC-CEP-2337 & 18.470 $\pm$ 0.033 & 0.130 &  &  18.538 $\pm$ 0.011 & 1.046 $\pm$ 0.032 & 0.85 & 0.060 & 181 & 5.9 & 3.6 \\
              &        &  $\pm$ 0.073        &        & 18.578  $\pm$  0.012  &  1.037  $\pm$ 0.017  & 0.75  & 0.025  &  59  &   5.4  &  2.2  \\
LMC-CEP-1100 & 18.530 $\pm$ 0.032 & 0.070 &  &  18.501 $\pm$ 0.015 & 1.069 $\pm$ 0.034 & 0.65 & 0.045 & 125 & 5.2 & 1.3 \\
              &        &  $\pm$ 0.056        &        & 18.541  $\pm$  0.011  &  1.070  $\pm$ 0.010  & 0.90  & 0.025  & 149  &   6.1  &  0.8  \\
LMC-CEP-0545 & 18.500 $\pm$ 0.034 & 0.047 &   &  18.560 $\pm$ 0.018 & 1.079 $\pm$ 0.038 & 0.75 & 0.060 & 146 & 5.7 & 1.8 \\
              &        &  $\pm$ 0.049        &        & 18.594  $\pm$  0.010  &  1.073  $\pm$ 0.012  & 1.00  & 0.022  & 104  &   5.7  &  0.7  \\
LMC-CEP-4357 & 18.380 $\pm$ 0.031 & 0.046 &  &  18.554 $\pm$ 0.013 & 1.059 $\pm$ 0.027 & 1.50 & 0.046 & 205 & 7.1 & 1.0 \\
              &        &  $\pm$ 0.067        &        & 18.569  $\pm$  0.009  &  1.055  $\pm$ 0.013  & 1.50  & 0.021  &  95  &   5.3  &  0.9  \\
LMC-CEP-2534 & 18.470 $\pm$ 0.031 & 0.160 &   &  18.393 $\pm$ 0.022 & 1.048 $\pm$ 0.035 & 0.85 & 0.036 & 122 & 5.9 & 3.4 \\
              &        &  $\pm$ 0.104        &        & 18.554  $\pm$  0.021  &  1.037  $\pm$ 0.018  & 0.65  & 0.045  & 105  &   7.6  &  0.7  \\
LMC-CEP-0249 & 18.520 $\pm$ 0.034 & 0.090 &   &  18.623 $\pm$ 0.021 & 1.042 $\pm$ 0.043 & 0.65 & 0.060 & 95 & 6.9 & 1.0 \\
              &        &  $\pm$ 0.058        &        & 18.610  $\pm$  0.015  &  1.043  $\pm$ 0.013  & 1.00  & 0.030  & 117  &   5.7  &  0.9  \\
LMC-CEP-0467 & 18.530 $\pm$ 0.034 & 0.055 &   &  18.657 $\pm$ 0.016 & 1.034 $\pm$ 0.037 & 0.85 & 0.070 & 165 & 6.5 & 2.1 \\
              &        &  $\pm$ 0.036        &        & 18.707  $\pm$  0.021  &  1.023  $\pm$ 0.018  & 0.80  & 0.050  & 113  &   6.6  &  1.1  \\
LMC-CEP-0068 & 18.550 $\pm$ 0.029 & 0.084 &  &  18.623 $\pm$ 0.013 & 1.059 $\pm$ 0.031 & 1.25 & 0.048 & 165 & 5.5 & 1.6 \\
              &        &  $\pm$ 0.039        &        & 18.637  $\pm$  0.015  &  1.056  $\pm$ 0.011  & 1.50  & 0.022  & 121  &   5.8  &  1.4  \\
LMC-CEP-1290 & 18.480 $\pm$ 0.034 & 0.069 &  &  18.566 $\pm$ 0.030 & 1.063 $\pm$ 0.049 & 0.65 & 0.080 & 95 & 7.7 & 1.4 \\
              &        &  $\pm$ 0.074        &        & 18.616  $\pm$  0.021  &  1.046  $\pm$ 0.017  & 0.90  & 0.040  & 103  &   5.8  &  1.3  \\
LMC-CEP-2226 & 18.500 $\pm$ 0.034 & 0.099 &  &  18.392 $\pm$ 0.037 & 1.050 $\pm$ 0.036 & 0.50 & 0.043 & 134 & 5.4 & 2.6 \\
              &        &  $\pm$ 0.076        &        & 18.457  $\pm$  0.003  &  1.054  $\pm$ 0.010  & 0.75  & 0.018  & 142  &   5.9  &  3.9  \\
LMC-CEP-2244 & 18.490 $\pm$ 0.035 & 0.095 &  &  18.470 $\pm$ 0.011 & 1.048 $\pm$ 0.026 & 0.50 & 0.045 & 194 & 5.2 & 1.8 \\
              &        &  $\pm$ 0.079        &        & 18.473  $\pm$  0.003  &  1.049  $\pm$ 0.010  & 0.65  & 0.016  & 113  &   5.1  &  2.7  \\
LMC-CEP-2492 & 18.470 $\pm$ 0.035 & 0.190 &  &  18.364 $\pm$ 0.030 & 1.037 $\pm$ 0.044 & 0.50 & 0.050 & 103 & 5.6 & 1.2 \\
              &        &  $\pm$ 0.104        &        & 18.468  $\pm$  0.016  &  1.029  $\pm$ 0.018  & 0.50  & 0.040  &  97  &   7.3  &  1.9  \\
LMC-CEP-2831 & 18.470 $\pm$ 0.036 & 0.160 &  &  18.371 $\pm$ 0.013 & 1.041 $\pm$ 0.031 & 0.75 & 0.036 & 149 & 5.4 & 2.2 \\
              &        &  $\pm$ 0.089        &        & 18.363  $\pm$  0.006  &  1.046  $\pm$ 0.012  & 0.85  & 0.022  & 119  &   5.2  &  2.3  \\ \pagebreak
LMC-CEP-2892 & 18.520 $\pm$ 0.035 & 0.140 &  &  18.474 $\pm$ 0.012 & 1.057 $\pm$ 0.034 & 0.80 & 0.040 & 143 & 5.3 & 1.3 \\
              &        &  $\pm$ 0.083        &        & 18.605  $\pm$  0.020  &  1.043  $\pm$ 0.018  & 0.65  & 0.050  & 140  &   8.2  &  0.8  \\
LMC-CEP-0091 & 18.550 $\pm$ 0.035 & 0.082 &  &  18.608 $\pm$ 0.010 & 1.050 $\pm$ 0.026 & 1.25 & 0.045 & 187 & 6.8 & 2.4 \\
              &        &  $\pm$ 0.056        &        & 18.626  $\pm$  0.008  &  1.043  $\pm$ 0.013  & 1.25  & 0.030  & 129  &  12.0  &  1.6  \\
LMC-CEP-0281 & 18.540 $\pm$ 0.035 & 0.120 &  &  18.510 $\pm$ 0.018 & 1.048 $\pm$ 0.028 & 0.95 & 0.044 & 178 & 6.0 & 1.0 \\
              &        &  $\pm$ 0.077        &        & 18.594  $\pm$  0.020  &  1.045  $\pm$ 0.016  & 0.75  & 0.050  & 128  &   6.0  &  2.0  \\
LMC-CEP-0329 & 18.540 $\pm$ 0.035 & 0.060 &  &  18.631 $\pm$ 0.016 & 1.056 $\pm$ 0.044 & 0.90 & 0.048 & 101 & 5.5 & 1.6 \\
              &        &  $\pm$ 0.049        &        & 18.599  $\pm$  0.009  &  1.064  $\pm$ 0.011  & 1.50  & 0.017  & 112  &   5.2  &  0.7  \\
LMC-CEP-0445 & 18.550 $\pm$ 0.036 & 0.091 &  &  18.590 $\pm$ 0.030 & 1.050 $\pm$ 0.038 & 0.60 & 0.080 & 168 & 7.3 & 0.8 \\
              &        &  $\pm$ 0.046        &        & 18.616  $\pm$  0.023  &  1.044  $\pm$ 0.012  & 0.95  & 0.026  & 142  &   5.5  &  1.1  \\
LMC-CEP-0588 & 18.530 $\pm$ 0.036 & 0.058 &  &  18.600 $\pm$ 0.029 & 1.052 $\pm$ 0.026 & 1.00 & 0.037 & 195 & 8.9 & 1.3 \\
              &        &  $\pm$ 0.036        &        & 18.693  $\pm$  0.018  &  1.037  $\pm$ 0.015  & 0.75  & 0.045  & 146  &   7.7  &  0.6  \\
LMC-CEP-0794 & 18.550 $\pm$ 0.034 & 0.069 &  &  18.523 $\pm$ 0.015 & 1.069 $\pm$ 0.040 & 0.65 & 0.046 & 91 & 7.7 & 1.8 \\
              &        &  $\pm$ 0.044        &        & 18.562  $\pm$  0.013  &  1.065  $\pm$ 0.013  & 0.90  & 0.025  & 101  &   5.0  &  0.6  \\
LMC-CEP-0796 & 18.480 $\pm$ 0.030 & 0.078 &  &  18.589 $\pm$ 0.010 & 1.030 $\pm$ 0.035 & 0.80 & 0.060 & 155 & 5.1 & 2.6 \\
              &        &  $\pm$ 0.064        &        & 18.593  $\pm$  0.018  &  1.029  $\pm$ 0.018  & 0.75  & 0.035  &  81  &   5.5  &  0.9  \\
LMC-CEP-1268 & 18.460 $\pm$ 0.035 & 0.130 &  &  18.401 $\pm$ 0.010 & 1.057 $\pm$ 0.029 & 0.50 & 0.038 & 125 & 5.4 & 1.5 \\
              &        &  $\pm$ 0.074        &        & 18.418  $\pm$  0.009  &  1.056  $\pm$ 0.013  & 0.50  & 0.024  &  80  &   5.3  &  1.4  \\
LMC-CEP-1321 & 18.500 $\pm$ 0.035 & 0.063 &  &  18.540 $\pm$ 0.015 & 1.072 $\pm$ 0.027 & 0.80 & 0.044 & 194 & 5.3 & 2.8 \\
              &        &  $\pm$ 0.050        &        & 18.560  $\pm$  0.013  &  1.070  $\pm$ 0.011  & 0.85  & 0.026  & 136  &   5.5  &  1.1  \\
LMC-CEP-1640 & 18.490 $\pm$ 0.034 & 0.066 &  &  18.537 $\pm$ 0.010 & 1.068 $\pm$ 0.029 & 0.95 & 0.039 & 129 & 7.1 & 1.8 \\
              &        &  $\pm$ 0.050        &        & 18.572  $\pm$  0.008  &  1.065  $\pm$ 0.010  & 1.25  & 0.016  & 100  &   5.9  &  0.6  \\
LMC-CEP-1841 & 18.400 $\pm$ 0.035 & 0.120 &  &  18.408 $\pm$ 0.008 & 1.047 $\pm$ 0.028 & 0.50 & 0.043 & 198 & 6.1 & 1.6 \\
              &        &  $\pm$ 0.080        &        & 18.427  $\pm$  0.006  &  1.050  $\pm$ 0.011  & 0.55  & 0.023  & 126  &   5.4  &  1.8  \\
LMC-CEP-1892 & 18.450 $\pm$ 0.035 & 0.120 &  &  18.443 $\pm$ 0.014 & 1.053 $\pm$ 0.031 & 0.60 & 0.043 & 165 & 5.3 & 1.5 \\
              &        &  $\pm$ 0.064        &        & 18.499  $\pm$  0.013  &  1.050  $\pm$ 0.012  & 0.65  & 0.030  & 148  &   5.2  &  1.1  \\
LMC-CEP-1893 & 18.490 $\pm$ 0.035 & 0.083 &  &  18.523 $\pm$ 0.007 & 1.048 $\pm$ 0.029 & 0.45 & 0.048 & 183 & 6.9 & 4.5 \\
              &        &  $\pm$ 0.055        &        & 18.548  $\pm$  0.010  &  1.048  $\pm$ 0.012  & 0.50  & 0.028  & 132  &   7.9  &  3.5  \\
LMC-CEP-2171 & 18.490 $\pm$ 0.036 & 0.066 &  &  18.553 $\pm$ 0.120 & 1.057 $\pm$ 0.034 & 0.50 & 0.037 & 136 & 6.1 & 1.1 \\
              &        &  $\pm$ 0.039        &        & 18.540  $\pm$  0.012  &  1.059  $\pm$ 0.011  & 0.65  & 0.023  & 137  &   5.2  &  0.7  \\
LMC-CEP-2270 & 18.540 $\pm$ 0.036 & 0.110 &  &  18.476 $\pm$ 0.009 & 1.055 $\pm$ 0.027 & 0.75 & 0.050 & 191 & 6.7 & 1.9 \\
              &        &  $\pm$ 0.074        &        & 18.494  $\pm$  0.011  &  1.052  $\pm$ 0.009  & 0.95  & 0.022  & 143  &   5.8  &  0.4  \\
LMC-CEP-2936 & 18.510 $\pm$ 0.035 & 0.130 &  &  18.475 $\pm$ 0.016 & 1.050 $\pm$ 0.026 & 0.90 & 0.046 & 150 & 5.6 & 2.2 \\
              &        &  $\pm$ 0.044        &        & 18.504  $\pm$  0.012  &  1.047  $\pm$ 0.011  & 1.00  & 0.028  & 115  &   6.2  &  1.1  \\
LMC-CEP-2964 & 18.450 $\pm$ 0.013 & 0.054 &  &  18.537 $\pm$ 0.018 & 1.077 $\pm$ 0.045 & 0.90 & 0.045 & 93 & 5.4 & 1.6 \\
              &        &  $\pm$ 0.064        &        & 18.592  $\pm$  0.017  &  1.068  $\pm$ 0.020  & 0.90  & 0.029  &  66  &   5.4  &  1.3  \\
LMC-CEP-3207 & 18.490 $\pm$ 0.036 & 0.130 &  &  18.439 $\pm$ 0.015 & 1.046 $\pm$ 0.030 & 1.00 & 0.041 & 186 & 5.0 & 1.8 \\
              &        &  $\pm$ 0.061        &        & 18.512  $\pm$  0.006  &  1.044  $\pm$ 0.012  & 0.95  & 0.025  & 116  &   5.5  &  5.6  \\
LMC-CEP-3572 & 18.470 $\pm$ 0.048 & 0.071 &  &  18.505 $\pm$ 0.012 & 1.045 $\pm$ 0.028 & 1.75 & 0.047 & 172 & 5.9 & 1.6 \\
              &        &  $\pm$ 0.043        &        & 18.549  $\pm$  0.011  &  1.037  $\pm$ 0.011  & 2.00  & 0.029  & 147  &   7.3  &  1.7  \\
LMC-CEP-3650 & 18.440 $\pm$ 0.050 & 0.059 &  &  18.520 $\pm$ 0.013 & 1.046 $\pm$ 0.033 & 1.75 & 0.046 & 124 & 5.2 & 1.5 \\
              &        &  $\pm$ 0.071        &        & 18.569  $\pm$  0.015  &  1.033  $\pm$ 0.015  & 1.75  & 0.035  & 102  &   7.0  &  1.4  \\
LMC-CEP-3659 & 18.420 $\pm$ 0.014 & 0.037 &  &  18.645 $\pm$ 0.018 & 1.041 $\pm$ 0.042 & 0.95 & 0.080 & 164 & 6.4 & 1.4 \\
              &        &  $\pm$ 0.073        &        & 18.653  $\pm$  0.017  &  1.042  $\pm$ 0.016  & 1.25  & 0.040  & 138  &   6.5  &  1.9  \\
LMC-CEP-3833 & 18.410 $\pm$ 0.041 & 0.084 &  &  18.480 $\pm$ 0.009 & 1.050 $\pm$ 0.026 & 1.50 & 0.045 & 190 & 7.0 & 1.7 \\
              &        &  $\pm$ 0.068        &        & 18.497  $\pm$  0.011  &  1.046  $\pm$ 0.012  & 1.50  & 0.028  & 122  &   6.0  &  1.1  \\
LMC-CEP-3888 & 18.440 $\pm$ 0.014 & 0.064 &  &  18.514 $\pm$ 0.008 & 1.051 $\pm$ 0.026 & 1.75 & 0.045 & 179 & 5.8 & 2.0 \\
              &        &  $\pm$ 0.061        &        & 18.524  $\pm$  0.004  &  1.047  $\pm$ 0.009  & 2.50  & 0.016  & 124  &   5.5  &  1.5  \\
LMC-CEP-4142 & 18.440 $\pm$ 0.045 & 0.041 &  &  18.547 $\pm$ 0.005 & 1.063 $\pm$ 0.026 & 1.75 & 0.037 & 132 & 5.7 & 3.6 \\
              &        &  $\pm$ 0.062        &        & 18.564  $\pm$  0.006  &  1.050  $\pm$ 0.009  & 2.50  & 0.019  & 146  &   5.6  &  0.9  \\
LMC-CEP-0478 & 18.510 $\pm$ 0.035 & 0.100 &  &  18.514 $\pm$ 0.015 & 1.066 $\pm$ 0.033 & 0.65 & 0.060 & 143 & 5.4 & 1.2 \\
              &        &  $\pm$ 0.059        &        & 18.543  $\pm$  0.010  &  1.054  $\pm$ 0.011  & 0.90  & 0.029  & 135  &   5.2  &  0.9  \\
LMC-CEP-0772 & 18.470 $\pm$ 0.032 & 0.120 &  &  18.498 $\pm$ 0.015 & 1.042 $\pm$ 0.027 & 0.65 & 0.048 & 190 & 5.2 & 1.8 \\
              &        &  $\pm$ 0.074        &        & 18.497  $\pm$  0.009  &  1.044  $\pm$ 0.013  & 0.60  & 0.027  &  92  &   7.2  &  1.7  \\

\end{longtable}

\tablefoot{Column~1 gives the name of the system (For the none-Harvard variables prefix by OGLE-), 
with the DM (Col.~2) and reddening (Col.~3) based on \citet{Rubele18}.
Columns~4--10 contain the parameters derived in the present paper, see the footnote to Tab.~\ref{Tab-EB}.
}
}

\subsection{TRGB distances towards SMC RR Lyrae stars}
\label{SS-RRL}

No multi-wavelength study similar to \citet{Inno16} currently exists for Cepheids in the SMC that simultaneously derives reddening and distance  
(although the VMC team has studied SMC Cepheids, e.g. \citealt{Ripepi17}).
Towards the SMC we therefore used a similar approach, but using RRL from \citet{Muraveva2018} who studied 2997 fundamental mode RRL from 
the OGLE-IV survey. They derived the mean $K_{\rm s}$-mag from multi-epoch VMC data, and the reddening, $E(V-I)$, from the observed 
OGLE $V,I$ mean magnitude and the intrinsic $(V-I)_0$ colour, which they took to be a function of $V$-band pulsation amplitude 
and pulsation period following \citet{Piersimoni02}.
They then adopted (photometric) metallicities available from \citet{Skowron2016} and the period - $K$ band - magnitude - metallicity relation 
from \citet{Muraveva2015} based on 70 RRL in the LMC and calibrated using the dEB-based LMC distance \citep{Pietrzynski13} to derive 
distances to individual RRL.
%

The approach described above was used to assign 2686 RRL towards 43 los 
(21 stars with $E(V-I)$ values of less than $-0.1$ mag were excluded; 
the minimum and maximum radii of the circle that defined a los were 0.5 and 1.5\degr\ respectively, and a minimum of 50 RRL within a los was imposed).
These numbers reflect the higher surface number density of SMC RRL compared to the LMC CCs.
For each los the median and standard deviation of the DM and reddening (adopting $E(B-V)= E(V-I)/1.22$ mag) were calculated.

\begin{figure}
\centering

\includegraphics[width=0.74\hsize]{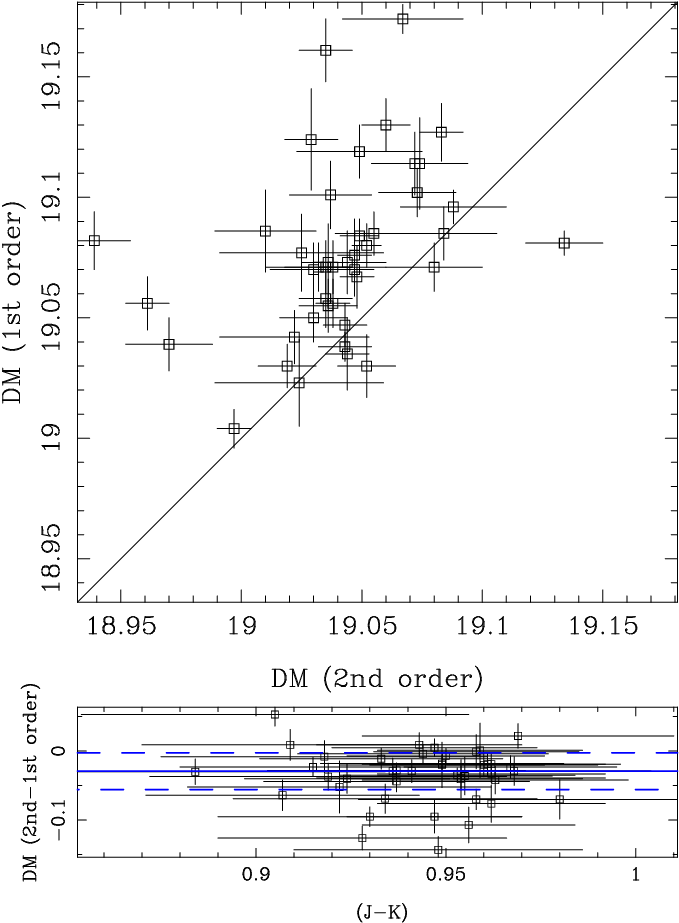}
\bigskip

\includegraphics[width=0.74\hsize]{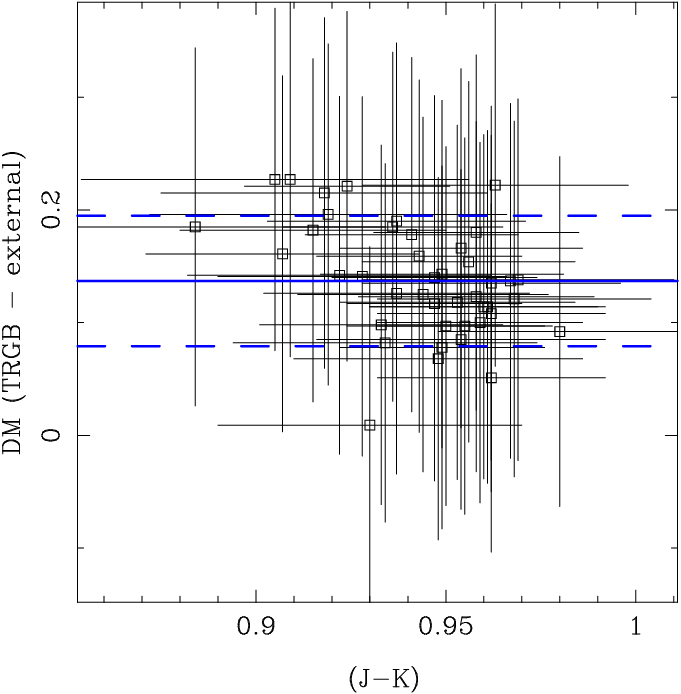}

\includegraphics[width=0.74\hsize]{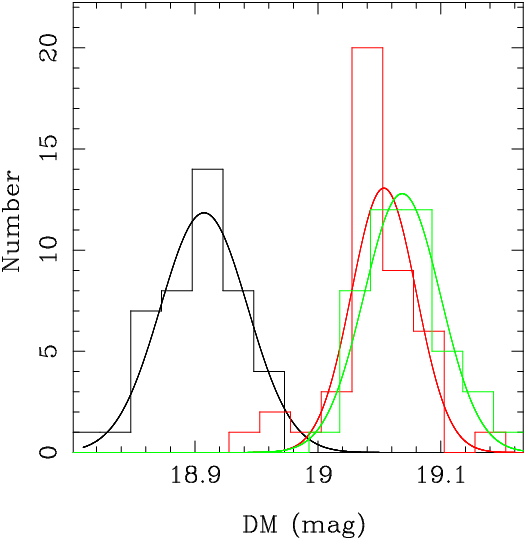}

\caption[]{ 
Same as Fig.~\ref{Fig-TRGB-CEP} for 43 los towards RRL in the SMC.
} 
\label{Fig-TRGB-RRL}
\end{figure}

The results of the calculations are listed in Table~\ref{Tab-RRL}.
Figure~\ref{Fig-TRGB-RRL} illustrates the results. In this case the difference between the second- and 
first-order-derivative-based DM is $-0.029 \pm 0.027$ mag. There is a discrepancy between the TRGB and the RRL distances of 
approximately $0.14 \pm 0.06$ mag, as illustrated in the lower two panels of Fig.~\ref{Fig-TRGB-RRL}.
We have carried out Monte Carlo simulations to find that the RRL distance distribution is described by a mean of 18.905 mag with an error 
in the mean of 0.004 mag, and a width of $\sigma = 0.042$ mag. For the second-order-derivative-based TRGB distance this 
is $19.044 \pm 0.003$, $\sigma = 0.028$ mag. The DM for the RRL is, as expected, in very good agreement with the weighted average 
of all RRL in \citet{Muraveva2018}, namely $18.88$ mag with a standard deviation of 0.20 mag.
We discuss this difference between the RRL and TRGB distances in Sect.~\ref{S-Disc}.

\longtab{
\setlength{\tabcolsep}{1.4mm}
\begin{longtable}{lcccccccrrr}
  \caption{\label{Tab-RRL} TRGB distances to SMC fields surrounding RRLs.} \\
  \hline  \hline
   System ID  &         DM        & $E(B-V)$ &    &        DM          &   $(J-K_{\rm s})$ @ TRGB   & Rlim    & bin width & $N$/bin & SNpk & $\chi^2_{\rm r}$ \\
OGLE-         &        (mag)      & (mag)  &  &            (mag)        &       (mag)      & (\degr) &   (mag)   &       &      &                 \\
\hline
\endfirsthead
\caption{continued.}\\
\hline\hline
   System ID  &         DM        & $E(B-V)$ &    &         DM          &   $(J-K_{\rm s})$ @ TRGB   & Rlim    & bin width & $N$/bin & SNpk & $\chi^2_{\rm r}$ \\
OGLE-         &        (mag)      & (mag)  &  &          (mag)        &       (mag)      & (\degr) &   (mag)   &       &      &                 \\
\hline
\endhead
\hline
\endfoot
SMC-RRLYR-1768 & 18.931 $\pm$ 0.157 & 0.041 & & 19.049 $\pm$ 0.008 & 0.953 $\pm$ 0.031 & 1.25 & 0.047 & 180 & 6.1 & 2.2 \\
              &        &  $\pm$ 0.024        &      & 19.084  $\pm$  0.007  &  0.947  $\pm$ 0.012  & 1.50  & 0.021  & 117  &   6.2  &  1.3  \\
SMC-RRLYR-5285 & 18.869 $\pm$ 0.157 & 0.057 & & 19.047 $\pm$ 0.007 & 0.941 $\pm$ 0.028 & 1.75 & 0.042 & 195 & 5.8 & 3.1 \\
              &        &  $\pm$ 0.024        &      & 19.076  $\pm$  0.015  &  0.924  $\pm$ 0.018  & 1.50  & 0.028  &  92  &   5.3  &  1.9  \\
SMC-RRLYR-1218 & 18.924 $\pm$ 0.156 & 0.066 & & 19.024 $\pm$ 0.035 & 0.959 $\pm$ 0.027 & 0.65 & 0.042 & 172 & 9.1 & 0.8 \\
              &        &  $\pm$ 0.061        &      & 19.023  $\pm$  0.018  &  0.959  $\pm$ 0.010  & 0.85  & 0.022  & 141  &   5.6  &  0.6  \\
SMC-RRLYR-0492 & 18.937 $\pm$ 0.155 & 0.041 & & 19.060 $\pm$ 0.010 & 0.958 $\pm$ 0.031 & 0.75 & 0.037 & 119 & 6.7 & 1.7 \\
              &        &  $\pm$ 0.036        &      & 19.130  $\pm$  0.011  &  0.938  $\pm$ 0.012  & 0.95  & 0.027  & 148  &   9.3  &  1.0  \\
SMC-RRLYR-1543 & 18.914 $\pm$ 0.160 & 0.049 & & 19.052 $\pm$ 0.012 & 0.969 $\pm$ 0.041 & 0.50 & 0.060 & 98 & 6.2 & 9.2 \\
              &        &  $\pm$ 0.036        &      & 19.030  $\pm$  0.013  &  0.961  $\pm$ 0.011  & 0.95  & 0.022  & 130  &   6.4  &  1.4  \\
SMC-RRLYR-1581 & 18.916 $\pm$ 0.152 & 0.057 & & 19.030 $\pm$ 0.014 & 0.960 $\pm$ 0.030 & 0.95 & 0.044 & 151 & 6.8 & 1.0 \\
              &        &  $\pm$ 0.049        &      & 19.050  $\pm$  0.010  &  0.953  $\pm$ 0.011  & 1.25  & 0.024  & 146  &   6.0  &  0.9  \\
SMC-RRLYR-1697 & 18.908 $\pm$ 0.158 & 0.049 & & 19.074 $\pm$ 0.020 & 0.954 $\pm$ 0.032 & 0.95 & 0.060 & 184 & 6.5 & 6.7 \\
              &        &  $\pm$ 0.036        &      & 19.114  $\pm$  0.019  &  0.934  $\pm$ 0.016  & 1.00  & 0.035  & 124  &   5.2  &  1.0  \\
SMC-RRLYR-0862 & 18.959 $\pm$ 0.153 & 0.057 & & 19.010 $\pm$ 0.021 & 0.962 $\pm$ 0.030 & 0.60 & 0.050 & 182 & 5.4 & 1.6 \\
              &        &  $\pm$ 0.049        &      & 19.086  $\pm$  0.017  &  0.947  $\pm$ 0.015  & 0.55  & 0.040  & 140  &   6.6  &  0.7  \\
SMC-RRLYR-5749 & 18.876 $\pm$ 0.157 & 0.066 &  & 19.037 $\pm$ 0.017 & 0.907 $\pm$ 0.036 & 2.50 & 0.070 & 242 & 5.3 & 5.8 \\
              &        &  $\pm$ 0.024        &      & 19.101  $\pm$  0.014  &  0.889  $\pm$ 0.020  & 2.50  & 0.045  & 172  &   5.3  &  1.7  \\
SMC-RRLYR-1677 & 18.957 $\pm$ 0.153 & 0.049 & & 19.049 $\pm$ 0.026 & 0.980 $\pm$ 0.032 & 0.60 & 0.046 & 142 & 27.6 & 1.9 \\
              &        &  $\pm$ 0.036        &      & 19.119  $\pm$  0.011  &  0.960  $\pm$ 0.015  & 0.60  & 0.040  & 139  &  10.5  &  1.4  \\
SMC-RRLYR-1117 & 18.922 $\pm$ 0.156 & 0.049 & & 19.036 $\pm$ 0.012 & 0.961 $\pm$ 0.031 & 0.85 & 0.070 & 198 & 7.8 & 4.6 \\
              &        &  $\pm$ 0.036        &      & 19.055  $\pm$  0.011  &  0.955  $\pm$ 0.012  & 1.00  & 0.019  &  83  &   5.2  &  0.8  \\
SMC-RRLYR-0383 & 18.904 $\pm$ 0.156 & 0.041 & & 19.084 $\pm$ 0.022 & 0.958 $\pm$ 0.027 & 0.85 & 0.050 & 191 & 8.2 & 1.3 \\
              &        &  $\pm$ 0.036        &      & 19.085  $\pm$  0.011  &  0.950  $\pm$ 0.012  & 0.95  & 0.029  & 137  &   7.8  &  0.8  \\
SMC-RRLYR-1975 & 18.850 $\pm$ 0.160 & 0.049 & & 19.072 $\pm$ 0.015 & 0.963 $\pm$ 0.035 & 0.95 & 0.060 & 174 & 5.8 & 6.8 \\
              &        &  $\pm$ 0.024        &      & 19.114  $\pm$  0.013  &  0.946  $\pm$ 0.017  & 0.95  & 0.025  &  77  &   5.2  &  1.5  \\
SMC-RRLYR-4745 & 18.871 $\pm$ 0.160 & 0.066 &  & 18.939 $\pm$ 0.015 & 0.948 $\pm$ 0.038 & 1.25 & 0.039 & 95 & 7.3 & 5.1 \\
              &        &  $\pm$ 0.012        &      & 19.082  $\pm$  0.012  &  0.924  $\pm$ 0.014  & 1.50  & 0.030  & 143  &   6.0  &  3.6  \\
SMC-RRLYR-4342 & 18.895 $\pm$ 0.154 & 0.041 &  & 19.038 $\pm$ 0.007 & 0.949 $\pm$ 0.032 & 1.25 & 0.046 & 143 & 5.1 & 3.7 \\
              &        &  $\pm$ 0.024        &      & 19.056  $\pm$  0.010  &  0.942  $\pm$ 0.016  & 1.25  & 0.027  &  86  &   7.2  &  1.6  \\
SMC-RRLYR-1867 & 18.898 $\pm$ 0.157 & 0.057 & & 19.035 $\pm$ 0.011 & 0.967 $\pm$ 0.028 & 0.95 & 0.047 & 188 & 5.1 & 1.5 \\
              &        &  $\pm$ 0.036        &      & 19.058  $\pm$  0.012  &  0.959  $\pm$ 0.013  & 0.95  & 0.025  & 103  &   5.9  &  1.1  \\
SMC-RRLYR-0216 & 18.947 $\pm$ 0.150 & 0.041 &  & 19.032 $\pm$ 0.014 & 0.954 $\pm$ 0.038 & 0.90 & 0.041 & 90 & 7.1 & 3.8 \\
              &        &  $\pm$ 0.024        &      & 19.071  $\pm$  0.010  &  0.940  $\pm$ 0.016  & 1.00  & 0.020  &  60  &   5.9  &  2.2  \\
SMC-RRLYR-3606 & 18.903 $\pm$ 0.161 & 0.041 & & 19.043 $\pm$ 0.011 & 0.947 $\pm$ 0.027 & 1.50 & 0.047 & 189 & 11.5 & 1.3 \\
              &        &  $\pm$ 0.024        &      & 19.038  $\pm$  0.006  &  0.947  $\pm$ 0.011  & 1.75  & 0.017  & 107  &   7.4  &  1.0  \\
SMC-RRLYR-5163 & 18.840 $\pm$ 0.151 & 0.074 &  & 19.036 $\pm$ 0.011 & 0.919 $\pm$ 0.047 & 1.25 & 0.070 & 114 & 5.0 & 3.2 \\
              &        &  $\pm$ 0.036        &      & 19.073  $\pm$  0.016  &  0.921  $\pm$ 0.015  & 1.75  & 0.030  & 136  &   5.5  &  1.1  \\
SMC-RRLYR-1108 & 18.913 $\pm$ 0.158 & 0.041 & & 19.067 $\pm$ 0.025 & 0.956 $\pm$ 0.028 & 0.95 & 0.044 & 197 & 5.0 & 1.1 \\
              &        &  $\pm$ 0.024        &      & 19.174  $\pm$  0.006  &  0.928  $\pm$ 0.017  & 0.80  & 0.035  & 125  &   5.5  &  4.6  \\
SMC-RRLYR-0063 & 18.944 $\pm$ 0.158 & 0.049 &  & 19.022 $\pm$ 0.031 & 0.949 $\pm$ 0.027 & 1.50 & 0.042 & 193 & 5.1 & 5.7 \\
              &        &  $\pm$ 0.024        &      & 19.042  $\pm$  0.011  &  0.943  $\pm$ 0.011  & 1.75  & 0.019  & 132  &   5.2  &  0.8  \\
SMC-RRLYR-4691 & 18.862 $\pm$ 0.158 & 0.041 &  & 19.052 $\pm$ 0.006 & 0.937 $\pm$ 0.034 & 1.50 & 0.047 & 148 & 5.8 & 12.6 \\
              &        &  $\pm$ 0.024        &      & 19.080  $\pm$  0.009  &  0.934  $\pm$ 0.014  & 1.75  & 0.024  & 112  &   5.4  &  1.4  \\
SMC-RRLYR-4924 & 18.883 $\pm$ 0.155 & 0.074 &  & 19.025 $\pm$ 0.034 & 0.922 $\pm$ 0.040 & 1.50 & 0.050 & 105 & 5.2 & 3.9 \\
              &        &  $\pm$ 0.036        &      & 19.077  $\pm$  0.016  &  0.905  $\pm$ 0.021  & 1.50  & 0.050  & 112  &   8.7  &  2.4  \\
SMC-RRLYR-0767 & 18.957 $\pm$ 0.160 & 0.049 &  & 19.083 $\pm$ 0.009 & 0.937 $\pm$ 0.035 & 0.75 & 0.041 & 88 & 13.5 & 4.7 \\
              &        &  $\pm$ 0.024        &      & 19.127  $\pm$  0.012  &  0.920  $\pm$ 0.016  & 0.90  & 0.023  &  76  &   5.2  &  0.9  \\
SMC-RRLYR-2148 & 18.913 $\pm$ 0.157 & 0.049 &  & 19.048 $\pm$ 0.007 & 0.962 $\pm$ 0.034 & 1.00 & 0.039 & 114 & 5.5 & 2.7 \\
              &        &  $\pm$ 0.024        &      & 19.067  $\pm$  0.013  &  0.953  $\pm$ 0.018  & 0.90  & 0.026  &  63  &   5.2  &  1.4  \\
SMC-RRLYR-4332 & 18.888 $\pm$ 0.154 & 0.049 &  & 19.073 $\pm$ 0.016 & 0.936 $\pm$ 0.029 & 1.50 & 0.048 & 177 & 5.3 & 7.2 \\
              &        &  $\pm$ 0.024        &      & 19.102  $\pm$  0.010  &  0.924  $\pm$ 0.013  & 1.75  & 0.025  & 133  &   7.2  &  2.2  \\
SMC-RRLYR-0165 & 18.921 $\pm$ 0.159 & 0.049 &  & 19.019 $\pm$ 0.012 & 0.933 $\pm$ 0.032 & 2.00 & 0.037 & 116 & 5.8 & 2.8 \\
              &        &  $\pm$ 0.024        &      & 19.030  $\pm$  0.009  &  0.925  $\pm$ 0.016  & 2.00  & 0.035  & 112  &  10.2  &  2.3  \\
SMC-RRLYR-2293 & 18.885 $\pm$ 0.156 & 0.090 & & 19.044 $\pm$ 0.009 & 0.943 $\pm$ 0.027 & 1.25 & 0.047 & 178 & 5.2 & 4.0 \\
              &        &  $\pm$ 0.049        &      & 19.035  $\pm$  0.015  &  0.937  $\pm$ 0.019  & 1.00  & 0.027  &  57  &   5.0  &  1.6  \\
SMC-RRLYR-5451 & 18.853 $\pm$ 0.156 & 0.074 &  & 19.080 $\pm$ 0.020 & 0.909 $\pm$ 0.039 & 1.50 & 0.050 & 129 & 4.1 & 2.5 \\
              &        &  $\pm$ 0.036        &      & 19.071  $\pm$  0.010  &  0.916  $\pm$ 0.013  & 2.00  & 0.021  & 120  &   5.6  &  3.3  \\ \pagebreak
SMC-RRLYR-3860 & 18.930 $\pm$ 0.158 & 0.049 &  & 19.038 $\pm$ 0.006 & 0.962 $\pm$ 0.030 & 0.70 & 0.060 & 166 & 11.8 & 8.5 \\
              &        &  $\pm$ 0.024        &      & 19.071  $\pm$  0.011  &  0.952  $\pm$ 0.012  & 0.85  & 0.035  & 147  &   8.5  &  0.8  \\
SMC-RRLYR-3890 & 18.952 $\pm$ 0.158 & 0.041 &  & 18.961 $\pm$ 0.009 & 0.930 $\pm$ 0.040 & 2.00 & 0.037 & 92 & 8.4 & 5.9 \\
              &        &  $\pm$ 0.024        &      & 19.056  $\pm$  0.011  &  0.933  $\pm$ 0.013  & 2.50  & 0.025  & 138  &   5.1  &  0.7  \\
SMC-RRLYR-3551 & 18.938 $\pm$ 0.165 & 0.049 &  & 19.035 $\pm$ 0.025 & 0.955 $\pm$ 0.023 & 1.25 & 0.035 & 193 & 10.8 & 0.6 \\
              &        &  $\pm$ 0.024        &      & 19.071  $\pm$  0.011  &  0.944  $\pm$ 0.011  & 1.25  & 0.025  & 144  &   7.0  &  0.8  \\
SMC-RRLYR-2066 & 18.923 $\pm$ 0.157 & 0.074 &  & 19.044 $\pm$ 0.016 & 0.968 $\pm$ 0.036 & 0.75 & 0.050 & 106 & 5.5 & 2.8 \\
              &        &  $\pm$ 0.036        &      & 19.073  $\pm$  0.013  &  0.954  $\pm$ 0.016  & 0.80  & 0.026  &  67  &   5.0  &  1.3  \\
SMC-RRLYR-3026 & 18.894 $\pm$ 0.159 & 0.049 &  & 19.035 $\pm$ 0.011 & 0.928 $\pm$ 0.038 & 2.00 & 0.039 & 108 & 5.7 & 4.5 \\
              &        &  $\pm$ 0.024        &      & 19.161  $\pm$  0.013  &  0.908  $\pm$ 0.013  & 2.50  & 0.035  & 233  &   9.0  &  1.5  \\
SMC-RRLYR-4299 & 18.888 $\pm$ 0.158 & 0.033 &  & 18.970 $\pm$ 0.018 & 0.934 $\pm$ 0.040 & 2.50 & 0.037 & 110 & 5.7 & 4.0 \\
              &        &  $\pm$ 0.036        &      & 19.039  $\pm$  0.011  &  0.927  $\pm$ 0.018  & 2.50  & 0.029  &  96  &   7.1  &  2.8  \\
SMC-RRLYR-2766 & 18.912 $\pm$ 0.157 & 0.041 &  & 19.029 $\pm$ 0.011 & 0.947 $\pm$ 0.023 & 3.00 & 0.039 & 278 & 6.2 & 4.0 \\
              &        &  $\pm$ 0.036        &      & 19.124  $\pm$  0.021  &  0.925  $\pm$ 0.011  & 3.00  & 0.026  & 213  &   4.7  &  1.1  \\
SMC-RRLYR-5000 & 18.873 $\pm$ 0.154 & 0.041 &  & 19.088 $\pm$ 0.022 & 0.918 $\pm$ 0.043 & 1.50 & 0.080 & 170 & 9.5 & 9.0 \\
              &        &  $\pm$ 0.024        &      & 19.096  $\pm$  0.007  &  0.927  $\pm$ 0.015  & 2.00  & 0.023  &  99  &   5.2  &  2.8  \\
SMC-RRLYR-5354 & 18.865 $\pm$ 0.152 & 0.057 &  & 19.047 $\pm$ 0.008 & 0.915 $\pm$ 0.035 & 2.00 & 0.060 & 183 & 7.8 & 3.4 \\
              &        &  $\pm$ 0.036        &      & 19.070  $\pm$  0.011  &  0.924  $\pm$ 0.013  & 2.50  & 0.025  & 149  &   6.1  &  1.8  \\
SMC-RRLYR-5723 & 18.809 $\pm$ 0.154 & 0.066 &  & 19.030 $\pm$ 0.018 & 0.924 $\pm$ 0.027 & 2.50 & 0.060 & 324 & 6.9 & 4.9 \\
              &        &  $\pm$ 0.024        &      & 19.070  $\pm$  0.011  &  0.910  $\pm$ 0.015  & 2.50  & 0.021  & 120  &   5.4  &  1.1  \\
SMC-RRLYR-3304 & 18.907 $\pm$ 0.151 & 0.057 &  & 19.134 $\pm$ 0.016 & 0.905 $\pm$ 0.051 & 1.50 & 0.080 & 125 & 6.8 & 20.8 \\
              &        &  $\pm$ 0.024        &      & 19.081  $\pm$  0.005  &  0.909  $\pm$ 0.021  & 1.75  & 0.019  &  46  &   5.3  &  5.4  \\
SMC-RRLYR-0679 & 18.918 $\pm$ 0.157 & 0.033 &  & 19.043 $\pm$ 0.009 & 0.944 $\pm$ 0.033 & 1.50 & 0.042 & 139 & 11.5 & 2.6 \\
              &        &  $\pm$ 0.024        &      & 19.047  $\pm$  0.009  &  0.948  $\pm$ 0.011  & 2.00  & 0.021  & 145  &   7.5  &  1.0  \\
SMC-RRLYR-0045 & 18.900 $\pm$ 0.159 & 0.049 &  & 18.997 $\pm$ 0.007 & 0.950 $\pm$ 0.026 & 1.75 & 0.035 & 150 & 6.2 & 2.8 \\
              &        &  $\pm$ 0.024        &      & 19.004  $\pm$  0.008  &  0.935  $\pm$ 0.017  & 1.50  & 0.026  &  71  &   5.3  &  3.1  \\
SMC-RRLYR-5929 & 18.870 $\pm$ 0.158 & 0.057 &  & 19.055 $\pm$ 0.016 & 0.884 $\pm$ 0.050 & 2.00 & 0.060 & 109 & 7.1 & 7.1 \\
              &        &  $\pm$ 0.024        &      & 19.085  $\pm$  0.009  &  0.906  $\pm$ 0.015  & 3.00  & 0.018  & 104  &   5.9  &  4.4  \\

\end{longtable}
\tablefoot{Column~1 gives the name of the system, 
with the DM (Col.~2) and reddening (Col.~3) based on \citet{Muraveva2018}.
Columns~4--10 contain the parameters derived in the present paper, see the note to Tab.~\ref{Tab-EB}.
}
}

\subsection{TRGB distances towards other SMC fields}
\label{SS-SFH}

\citet{Rubele18} used VMC data to derive the SFH 
in the main body and the wing of the SMC. In total they analysed 168 sub-regions covering about 24 square degrees. 
As part of their method the DM and visual extinction are derived simultaneously with the SFH. Here we use the values based on the analysis
of the $K_{\rm s}, (J-K_{\rm s})$ CMD, as they consider these to give the most reliable values for the 
reddening (we use $E(B-V) = A_{V}/3.1$ mag).

\begin{figure}[ht]
\centering

\includegraphics[width=0.74\hsize]{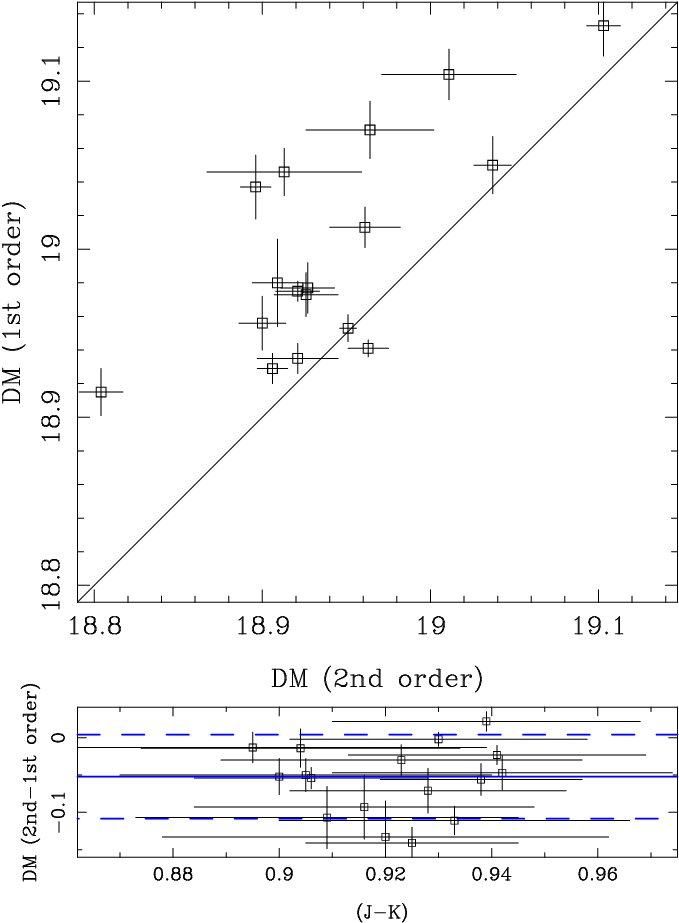}
\bigskip

\includegraphics[width=0.74\hsize]{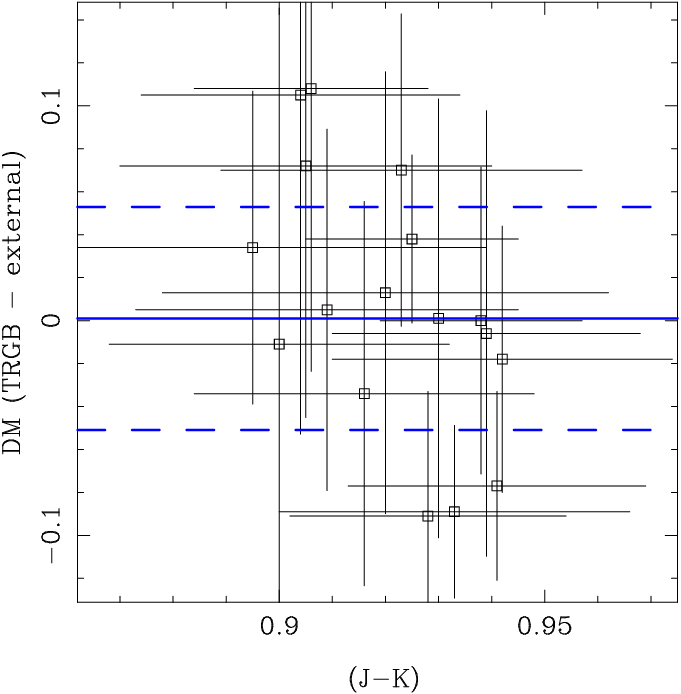}

\includegraphics[width=0.74\hsize]{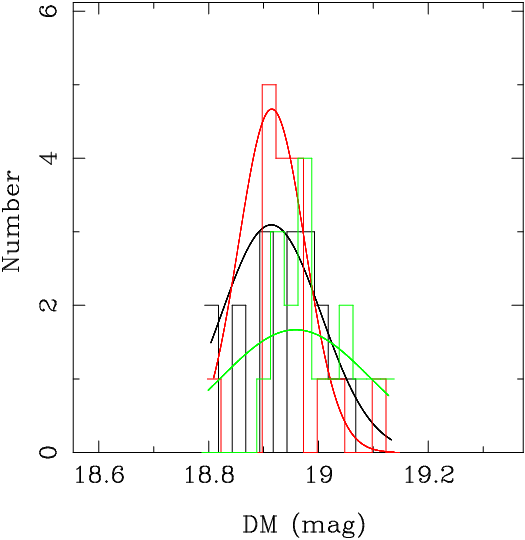}

\caption[]{ 
As in Fig.~\ref{Fig-TRGB-EB} but for 17 los towards fields in the SMC.
} 
\label{Fig-TRGB-SFH} 
\end{figure}

As before we constructed 17 los towards the SMC using the coordinates of the sub-regions as input and averaging over a 
number of them (between 5 and 19) to have sufficient statistics to carry out the TRGB analysis.
The results are displayed in Fig.~\ref{Fig-TRGB-SFH} and Table~\ref{Tab-SFH}.

In this case the difference between the second- and first-order-derivative-based DM is $-0.052 \pm 0.056$ mag. 
The TRGB and the distance derived from the SFH analysis are in excellent agreement, the weighted mean difference being $0.001 \pm 0.052$ mag.
Again we carried out Monte Carlo simulations to find that the distance distribution based on the SFH analysis is described 
by a mean of 18.95 with an error in the mean of 0.04, and a width of $\sigma = 0.14$ mag. For the second-order-derivative-based 
TRGB distance this is $18.93 \pm 0.02$, $\sigma = 0.09$ mag. The DM from the SFH analysis is, as expected, in very good agreement 
with the $18.910 \pm 0.064$ mag given by Rubele et al. (2018) as the DM to the mass-weighted centre of the SMC.

\begin{table*}
\setlength{\tabcolsep}{1.5mm}

\caption{TRGB distances to SMC fields.}
\label{Tab-SFH}
\centering
  \begin{tabular}{lccccccrrrr}
  \hline\hline
 RA \hspace{10mm} Dec  &      DM        & $E(B-V)$ &         DM          &   $(J-K_{\rm s})_0$@TRGB   & Rlim    & bin width & $N$/bin & SNpk & $\chi^2_{\rm r}$ \\ 
                       &    (mag)       & (mag)    &        (mag)        &       (mag)              & (\degr) &   (mag)   &         &      &                 \\
\hline
013.2281 $-$73.1258 & 18.893 $\pm$ 0.038 & 0.161 &  18.804 $\pm$ 0.013 & 0.933 $\pm$ 0.033 & 0.53 & 0.040 & 94 & 5.2 & 2.7 \\
              &        &  $\pm$ 0.017        &      18.915  $\pm$  0.014  &  0.923  $\pm$ 0.012  & 0.53  & 0.040  & 114  &   7.2  &  1.6  \\
010.7685 $-$72.7243 & 18.944 $\pm$ 0.059 & 0.117 &  18.926 $\pm$ 0.019 & 0.942 $\pm$ 0.032 & 0.51 & 0.044 & 87 & 5.5 & 1.9 \\
              &        &  $\pm$ 0.045        &      18.973  $\pm$  0.013  &  0.935  $\pm$ 0.014  & 0.51  & 0.070  & 146  &   9.4  &  2.6  \\
015.7545 $-$73.1000 & 18.858 $\pm$ 0.038 & 0.188 &  18.896 $\pm$ 0.009 & 0.925 $\pm$ 0.020 & 0.69 & 0.060 & 195 & 5.4 & 8.9 \\
              &        &  $\pm$ 0.014        &      19.037  $\pm$  0.019  &  0.894  $\pm$ 0.010  & 0.69  & 0.070  & 271  &   9.3  &  2.7  \\
013.0249 $-$72.4332 & 18.983 $\pm$ 0.043 & 0.129 &  18.906 $\pm$ 0.009 & 0.941 $\pm$ 0.028 & 0.58 & 0.060 & 142 & 11.3 & 6.2 \\
              &        &  $\pm$ 0.088        &      18.929  $\pm$  0.009  &  0.937  $\pm$ 0.013  & 0.58  & 0.050  & 124  &   7.8  &  3.7  \\
008.0189 $-$73.7714 & 19.033 $\pm$ 0.072 & 0.089 &  19.103 $\pm$ 0.010 & 0.923 $\pm$ 0.034 & 0.74 & 0.080 & 165 & 7.9 & 11.9 \\
              &        &  $\pm$ 0.022        &      19.133  $\pm$  0.018  &  0.903  $\pm$ 0.020  & 0.74  & 0.070  & 149  &  10.0  &  2.2  \\
013.3053 $-$74.6134 & 18.959 $\pm$ 0.075 & 0.131 &  18.964 $\pm$ 0.038 & 0.909 $\pm$ 0.036 & 0.75 & 0.075 & 107 & 7.0 & 14.9 \\
              &        &  $\pm$ 0.009        &      19.071  $\pm$  0.017  &  0.880  $\pm$ 0.020  & 0.75  & 0.070  & 116  &  11.8  &  2.8  \\
013.2618 $-$73.8223 & 19.000 $\pm$ 0.056 & 0.139 &  18.909 $\pm$ 0.015 & 0.928 $\pm$ 0.026 & 0.73 & 0.060 & 183 & 8.8 & 6.0 \\
              &        &  $\pm$ 0.031        &      18.980  $\pm$  0.026  &  0.914  $\pm$ 0.012  & 0.73  & 0.040  & 136  &   4.1  &  1.9  \\
017.9103 $-$71.9610 & 18.900 $\pm$ 0.092 & 0.143 &  18.913 $\pm$ 0.046 & 0.920 $\pm$ 0.042 & 0.71 & 0.070 & 86 & 6.7 & 9.6 \\
              &        &  $\pm$ 0.015        &      19.046  $\pm$  0.014  &  0.895  $\pm$ 0.020  & 0.71  & 0.070  & 102  &  15.0  &  3.7  \\
007.7842 $-$74.5600 & 19.003 $\pm$ 0.072 & 0.144 &  19.037 $\pm$ 0.011 & 0.895 $\pm$ 0.044 & 0.83 & 0.080 & 94 & 8.6 & 18.5 \\
              &        &  $\pm$ 0.007        &      19.050  $\pm$  0.017  &  0.883  $\pm$ 0.024  & 0.83  & 0.060  &  71  &   7.3  &  4.6  \\
006.7276 $-$73.7388 & 19.045 $\pm$ 0.080 & 0.112 &  19.011 $\pm$ 0.040 & 0.916 $\pm$ 0.032 & 0.99 & 0.050 & 109 & 5.7 & 7.2 \\
              &        &  $\pm$ 0.049        &      19.104  $\pm$  0.015  &  0.889  $\pm$ 0.018  & 0.99  & 0.050  & 123  &   8.6  &  1.5  \\
016.9988 $-$73.0766 & 18.900 $\pm$ 0.070 & 0.160 &  18.900 $\pm$ 0.014 & 0.938 $\pm$ 0.019 & 0.99 & 0.060 & 256 & 7.1 & 7.5 \\
              &        &  $\pm$ 0.023        &      18.956  $\pm$  0.016  &  0.925  $\pm$ 0.009  & 0.99  & 0.035  & 166  &   5.2  &  5.9  \\
010.6618 $-$72.0298 & 18.969 $\pm$ 0.103 & 0.094 &  18.963 $\pm$ 0.012 & 0.939 $\pm$ 0.029 & 0.81 & 0.042 & 92 & 5.4 & 2.3 \\
              &        &  $\pm$ 0.004        &      18.941  $\pm$  0.005  &  0.938  $\pm$ 0.014  & 0.81  & 0.022  &  46  &   9.6  & 15.9  \\
013.0024 $-$71.6420 & 18.950 $\pm$ 0.102 & 0.119 &  18.951 $\pm$ 0.005 & 0.930 $\pm$ 0.028 & 0.86 & 0.042 & 90 & 7.5 & 9.7 \\
              &        &  $\pm$ 0.029        &      18.953  $\pm$  0.008  &  0.924  $\pm$ 0.014  & 0.86  & 0.070  & 151  &  12.5  &  6.9  \\
018.7862 $-$74.5282 & 18.855 $\pm$ 0.116 & 0.156 &  18.927 $\pm$ 0.016 & 0.905 $\pm$ 0.035 & 1.07 & 0.080 & 87 & 11.5 & 26.9 \\
              &        &  $\pm$ 0.025        &      18.977  $\pm$  0.015  &  0.883  $\pm$ 0.021  & 1.07  & 0.070  &  83  &   9.8  & 11.9  \\
013.1518 $-$70.5491 & 18.972 $\pm$ 0.155 & 0.113 &  18.961 $\pm$ 0.021 & 0.900 $\pm$ 0.032 & 1.37 & 0.070 & 115 & 8.3 & 8.0 \\
              &        &  $\pm$ 0.045        &      19.013  $\pm$  0.012  &  0.881  $\pm$ 0.018  & 1.37  & 0.050  &  88  &  11.6  &  3.6  \\
017.6619 $-$70.8613 & 18.816 $\pm$ 0.156 & 0.138 &  18.921 $\pm$ 0.024 & 0.904 $\pm$ 0.030 & 1.44 & 0.080 & 168 & 8.2 & 4.7 \\
              &        &  $\pm$ 0.013        &      18.935  $\pm$  0.009  &  0.896  $\pm$ 0.015  & 1.44  & 0.027  &  58  &   6.2  &  9.3  \\
024.1335 $-$74.3093 & 18.813 $\pm$ 0.131 & 0.157 &  18.921 $\pm$ 0.013 & 0.906 $\pm$ 0.022 & 2.50 & 0.065 & 216 & 4.3 & 8.6 \\
              &        &  $\pm$ 0.023        &      18.975  $\pm$  0.006  &  0.882  $\pm$ 0.013  & 2.50  & 0.027  & 100  &   6.4  &  5.2  \\

\hline
\end{tabular}

\tablefoot{Columns~1 and 2 gives the RA and Dec of the los, with the DM (Col.~3) and reddening (Col.~4) based on \citet{Rubele18}.
Columns~5--11 contain the parameters derived in the present paper, see the note to Tab.~\ref{Tab-EB}.
}

\end{table*}

\subsection{TRGB distances towards VMC fields}
\label{SS-VMC}

In a final application we used the VMC data themselves to generate los towards SMC, LMC and the MB.
The minimum and maximum radii of the circle that defined a los were 0.45 and 2.0\degr, respectively.
A total of 17 los towards the SMC, and 55 towards the LMC were defined. 
In the direction of the MB three los were placed, spaced at 10\degr\ intervals in RA with larger radii of 5--9\degr.   

The reddening was calculated from the procedure used in Sects.~\ref{SS-CEP} and \ref{SS-RRL} for LMC and SMC, respectively.
The field in the MB closest to the SMC had a $E(B-V)$ value of 0.049 mag determined in this way,
while the field in the LMC closest to the MB had a value of 0.043. For the two fields in the MB in between these two pointings
a value of 0.045 mag was adopted.

The code was run and the results are listed in Table~\ref{Tab-VMC}.
Contrary to the previous applications the radius of the area was fixed and the code only considered different bin widths to determine the best fit.

As before Monte Carlo simulations were carried out to find the mean DM of $18.518 \pm 0.008$ (LMC) and $19.057 \pm 0.014$ mag (SMC).
The simple weighted average of the three fields in the MS is  $18.97 \pm 0.01$ mag; also see Fig.~\ref{Fig-VMC-DIST} and Sect.~\ref{SS-M}.

For the SMC we also ran models taking the reddening of the closest SMC subfield from \citet{Rubele15} (median value over the los of $E(B-V)= 0.118$) 
instead of that found from the RRL (median value of 0.049) reducing the DM to $18.97 \pm 0.07$ mag.

\longtab{
\setlength{\tabcolsep}{1.4mm}
\begin{longtable}{lccccccrrr}
  \caption{\label{Tab-VMC} TRGB distances to MC fields.} \\
  \hline  \hline
\hspace{4mm} RA \hspace{10mm} Dec  &          $E(B-V)$ &  &       DM          &   $(J-K_{\rm s})$ @ TRGB   & Rlim    & bin width & $N$/bin & SNpk & $\chi^2_{\rm r}$ \\
                                   &           (mag)   &  &      (mag)        &       (mag)      & (\degr) &   (mag)   &       &      &                 \\
\hline
\endfirsthead
\caption{continued.}\\
\hline\hline
\hspace{4mm} RA \hspace{10mm} Dec  &          $E(B-V)$ &  &        DM          &   $(J-K_{\rm s})$ @ TRGB   & Rlim    & bin width & $N$/bin & SNpk & $\chi^2_{\rm r}$ \\
                                   &           (mag)   &  &       (mag)        &       (mag)      & (\degr) &   (mag)   &       &      &                 \\
\hline
\endhead
\hline
\endfoot
01.008167 -73.494019 &  0.041 &  & 19.227 $\pm$ 0.010 & 0.900 $\pm$ 0.048 & 2.00 & 0.080 & 178 & 6.8 & 4.8 \\
              & &        & 19.235  $\pm$  0.013  &  0.880  $\pm$ 0.033  & 2.00  & 0.070  & 157  &  10.3  &  3.4  \\
04.453103 -71.679398 &  0.041 &  & 19.104 $\pm$ 0.010 & 0.918 $\pm$ 0.038 & 1.93 & 0.065 & 164 & 17.0 & 16.1 \\
              & &        & 19.137  $\pm$  0.011  &  0.901  $\pm$ 0.022  & 1.93  & 0.060  & 158  &  16.1  &  2.5  \\
05.012290 -75.248932 &  0.049 &  & 19.166 $\pm$ 0.007 & 0.910 $\pm$ 0.040 & 1.88 & 0.080 & 205 & 8.7 & 21.8 \\
              & &        & 19.175  $\pm$  0.017  &  0.894  $\pm$ 0.024  & 1.88  & 0.060  & 156  &  11.0  &  1.2  \\
08.007868 -73.256058  &  0.041 &  & 19.108 $\pm$ 0.008 & 0.949 $\pm$ 0.033 & 0.80 & 0.075 & 192 & 16.7 & 6.3 \\
              & &        & 19.188  $\pm$  0.017  &  0.926  $\pm$ 0.018  & 0.80  & 0.060  & 173  &  10.2  &  1.3  \\
09.290643 -70.451866  &  0.041 &  & 18.994 $\pm$ 0.007 & 0.925 $\pm$ 0.040 & 1.89 & 0.039 & 91 & 5.2 & 4.6 \\
              & &        & 18.999  $\pm$  0.010  &  0.920  $\pm$ 0.020  & 1.89  & 0.021  &  49  &   6.0  &  3.8  \\
09.889688 -72.642288  &  0.049 &  & 19.047 $\pm$ 0.010 & 0.962 $\pm$ 0.029 & 0.67 & 0.042 & 111 & 5.8 & 2.7 \\
              & &        & 19.079  $\pm$  0.012  &  0.953  $\pm$ 0.014  & 0.67  & 0.070  & 191  &  10.4  &  2.9  \\
10.010288 -73.857109 &  0.041 &  & 19.080 $\pm$ 0.017 & 0.952 $\pm$ 0.032 & 0.71 & 0.042 & 121 & 6.8 & 5.7 \\
              & &        & 19.105  $\pm$  0.015  &  0.941  $\pm$ 0.016  & 0.71  & 0.027  &  81  &   5.7  &  0.8  \\
11.165653 -73.218826 &  0.057 &  & 19.037 $\pm$ 0.019 & 0.961 $\pm$ 0.030 & 0.58 & 0.060 & 212 & 7.7 & 3.5 \\
              & &        & 19.107  $\pm$  0.019  &  0.943  $\pm$ 0.014  & 0.58  & 0.040  & 154  &   6.0  &  0.9  \\
11.551089 -72.196823 &  0.049 &  & 19.049 $\pm$ 0.008 & 0.961 $\pm$ 0.025 & 0.86 & 0.055 & 216 & 6.7 & 4.5 \\
              & &        & 19.060  $\pm$  0.010  &  0.953  $\pm$ 0.012  & 0.86  & 0.035  & 139  &   9.1  &  1.3  \\
11.896346 -74.379608 &  0.049 &  & 19.070 $\pm$ 0.032 & 0.945 $\pm$ 0.029 & 0.98 & 0.055 & 202 & 6.5 & 5.7 \\
              & &        & 19.162  $\pm$  0.014  &  0.916  $\pm$ 0.015  & 0.98  & 0.050  & 207  &   5.3  &  1.4  \\
12.995682 -72.964790 &  0.057 &  & 19.005 $\pm$ 0.025 & 0.961 $\pm$ 0.033 & 0.51 & 0.070 & 202 & 8.1 & 4.4 \\
              & &        & 19.101  $\pm$  0.017  &  0.940  $\pm$ 0.015  & 0.51  & 0.040  & 132  &   5.1  &  1.7  \\
13.567457 -75.465141  &  0.049 &  & 19.102 $\pm$ 0.028 & 0.934 $\pm$ 0.025 & 1.98 & 0.055 & 293 & 7.2 & 4.2 \\
              & &        & 19.105  $\pm$  0.010  &  0.921  $\pm$ 0.013  & 1.98  & 0.035  & 187  &   7.5  &  5.6  \\
13.894644 -73.422371 &  0.049 &  & 19.051 $\pm$ 0.014 & 0.964 $\pm$ 0.024 & 0.81 & 0.065 & 362 & 3.9 & 7.7 \\
              & &        & 19.040  $\pm$  0.023  &  0.959  $\pm$ 0.012  & 0.81  & 0.024  & 130  &   4.0  &  1.1  \\
13.898738 -71.724915  &  0.049 &  & 19.035 $\pm$ 0.008 & 0.961 $\pm$ 0.035 & 0.81 & 0.065 & 156 & 12.3 & 2.6 \\
              & &        & 19.053  $\pm$  0.007  &  0.953  $\pm$ 0.017  & 0.81  & 0.050  & 124  &  14.3  &  4.2  \\
14.439657 -72.623688  &  0.049 &  & 19.038 $\pm$ 0.018 & 0.973 $\pm$ 0.022 & 0.89 & 0.050 & 318 & 6.0 & 0.9 \\
              & &        & 19.084  $\pm$  0.016  &  0.959  $\pm$ 0.010  & 0.89  & 0.040  & 273  &   7.1  &  1.4  \\
15.134129 -70.933929 &  0.041 &  & 19.063 $\pm$ 0.017 & 0.951 $\pm$ 0.022 & 1.95 & 0.050 & 351 & 7.7 & 1.5 \\
              & &        & 19.082  $\pm$  0.009  &  0.942  $\pm$ 0.011  & 1.95  & 0.030  & 216  &   9.1  &  3.0  \\
16.689327 -73.251724 &  0.057 &  & 19.026 $\pm$ 0.020 & 0.963 $\pm$ 0.017 & 1.63 & 0.050 & 567 & 4.8 & 1.3 \\
              & &        & 19.105  $\pm$  0.015  &  0.942  $\pm$ 0.008  & 1.63  & 0.035  & 450  &   5.6  &  1.1  \\
28.000834 -73.349922 &  0.066 &  & 19.050 $\pm$ 0.030 & 0.890 $\pm$ 0.036 & 3.00 & 0.065 & 247 & 4.4 & 2.0 \\
              & &        & 19.217  $\pm$  0.025  &  0.853  $\pm$ 0.026  & 3.00  & 0.060  & 285  &   3.4  &  0.6  \\
26.000000 -73.000000  &  0.049 &  & 19.054 $\pm$ 0.008 & 0.944 $\pm$ 0.013 & 5.00 & 0.044 & 952 & 4.8 & 1.9 \\
              & &        & 19.091  $\pm$  0.012  &  0.929  $\pm$ 0.007  & 5.00  & 0.035  & 800  &   7.9  &  2.2  \\
36.000000 -73.000000  &  0.045 &  & 19.046 $\pm$ 0.013 & 0.940 $\pm$ 0.012 & 9.00 & 0.060 & 1523 & 9.5 & 4.2 \\
              & &        & 19.092  $\pm$  0.017  &  0.925  $\pm$ 0.007  & 9.00  & 0.035  & 949  &   5.6  &  2.0  \\
46.000000 -73.000000 &  0.045 &  & 18.892 $\pm$ 0.007 & 0.864 $\pm$ 0.098 & 5.00 & 0.065 & 87 & 4.6 & 45.5 \\
              & &        & 18.916  $\pm$  0.004  &  0.859  $\pm$ 0.059  & 5.00  & 0.050  &  70  &  10.4  & 55.1  \\
61.387905 -72.609306 &  0.043 &  & 18.883 $\pm$ 0.013 & 0.957 $\pm$ 0.071 & 2.96 & 0.075 & 96 & 11.6 & 12.5 \\
              & &        & 18.886  $\pm$  0.010  &  0.942  $\pm$ 0.042  & 2.96  & 0.050  &  64  &  10.1  &  5.6  \\
66.588913 -69.946556 &  0.069 &  & 18.646 $\pm$ 0.020 & 1.062 $\pm$ 0.044 & 1.88 & 0.080 & 158 & 8.0 & 3.0 \\
              & &        & 18.781  $\pm$  0.020  &  1.033  $\pm$ 0.019  & 1.88  & 0.070  & 162  &  10.3  &  2.2  \\
68.576553 -68.179131 &  0.090 &  & 18.566 $\pm$ 0.020 & 1.052 $\pm$ 0.033 & 1.74 & 0.075 & 186 & 8.8 & 2.4 \\
              & &        & 18.602  $\pm$  0.009  &  1.038  $\pm$ 0.015  & 1.74  & 0.024  &  63  &   8.2  &  3.2  \\
70.646675 -74.148064  &  0.041 &  & 18.665 $\pm$ 0.010 & 1.064 $\pm$ 0.046 & 1.88 & 0.055 & 95 & 6.4 & 5.3 \\
              & &        & 18.829  $\pm$  0.017  &  1.027  $\pm$ 0.020  & 1.88  & 0.060  & 134  &  12.4  &  2.1  \\
70.651901 -71.526146  &  0.074 &  & 18.615 $\pm$ 0.016 & 1.061 $\pm$ 0.039 & 1.46 & 0.040 & 109 & 5.2 & 1.9 \\
              & &        & 18.705  $\pm$  0.016  &  1.046  $\pm$ 0.016  & 1.46  & 0.035  & 110  &   7.0  &  0.5  \\
70.916809 -66.650620  &  0.057 &  & 18.570 $\pm$ 0.016 & 1.052 $\pm$ 0.031 & 1.84 & 0.075 & 219 & 10.1 & 2.0 \\
              & &        & 18.619  $\pm$  0.010  &  1.036  $\pm$ 0.015  & 1.84  & 0.027  &  84  &   6.7  &  2.4  \\
71.947662 -69.389572  &  0.120 &  & 18.541 $\pm$ 0.020 & 1.054 $\pm$ 0.032 & 1.02 & 0.050 & 139 & 11.0 & 2.9 \\
              & &        & 18.596  $\pm$  0.013  &  1.042  $\pm$ 0.014  & 1.02  & 0.030  &  91  &   7.3  &  1.6  \\
73.229797 -68.374863 &  0.096 &  & 18.582 $\pm$ 0.019 & 1.047 $\pm$ 0.028 & 0.97 & 0.065 & 235 & 6.2 & 1.0 \\
              & &        & 18.619  $\pm$  0.019  &  1.039  $\pm$ 0.013  & 0.97  & 0.060  & 229  &   5.6  &  1.2  \\  \pagebreak
73.492004 -70.261658  &  0.074 &  & 18.627 $\pm$ 0.024 & 1.061 $\pm$ 0.031 & 0.96 & 0.044 & 140 & 5.3 & 2.0 \\
              & &        & 18.658  $\pm$  0.013  &  1.056  $\pm$ 0.014  & 0.96  & 0.040  & 131  &   7.3  &  1.2  \\
74.165871 -72.483688 &  0.060 &  & 18.609 $\pm$ 0.016 & 1.064 $\pm$ 0.030 & 1.56 & 0.044 & 158 & 5.3 & 0.6 \\
              & &        & 18.651  $\pm$  0.013  &  1.054  $\pm$ 0.014  & 1.56  & 0.027  & 103  &   5.5  &  1.3  \\
74.812073 -69.212502 &  0.100 &  & 18.517 $\pm$ 0.012 & 1.058 $\pm$ 0.031 & 0.62 & 0.070 & 199 & 8.2 & 2.9 \\
              & &        & 18.655  $\pm$  0.029  &  1.039  $\pm$ 0.014  & 0.62  & 0.070  & 231  &   7.2  &  1.4  \\
74.935501 -67.595383 &  0.053 &  & 18.637 $\pm$ 0.009 & 1.041 $\pm$ 0.029 & 0.98 & 0.065 & 249 & 10.8 & 3.3 \\
              & &        & 18.694  $\pm$  0.017  &  1.029  $\pm$ 0.014  & 0.98  & 0.060  & 246  &  10.2  &  1.0  \\
74.940865 -65.570137 &  0.053 &  & 18.478 $\pm$ 0.013 & 1.054 $\pm$ 0.037 & 1.61 & 0.036 & 86 & 5.5 & 2.6 \\
              & &        & 18.659  $\pm$  0.016  &  1.018  $\pm$ 0.015  & 1.61  & 0.040  & 127  &   9.6  &  1.6  \\
75.849228 -68.684906 &  0.093 &  & 18.543 $\pm$ 0.013 & 1.049 $\pm$ 0.027 & 0.70 & 0.065 & 275 & 5.7 & 3.9 \\
              & &        & 18.629  $\pm$  0.022  &  1.037  $\pm$ 0.012  & 0.70  & 0.060  & 278  &   7.8  &  1.0  \\
75.946381 -69.692665 &  0.086 &  & 18.517 $\pm$ 0.015 & 1.067 $\pm$ 0.024 & 0.80 & 0.050 & 257 & 6.0 & 3.7 \\
              & &        & 18.675  $\pm$  0.022  &  1.043  $\pm$ 0.010  & 0.80  & 0.060  & 379  &   7.7  &  1.0  \\
76.095901 -70.811180 &  0.052 &  & 18.573 $\pm$ 0.017 & 1.070 $\pm$ 0.034 & 0.84 & 0.060 & 179 & 5.4 & 2.9 \\
              & &        & 18.698  $\pm$  0.015  &  1.052  $\pm$ 0.014  & 0.84  & 0.050  & 179  &   7.2  &  2.6  \\
76.949211 -66.970993 &  0.052 &  & 18.550 $\pm$ 0.017 & 1.057 $\pm$ 0.030 & 1.15 & 0.045 & 174 & 5.2 & 0.9 \\
              & &        & 18.633  $\pm$  0.014  &  1.042  $\pm$ 0.014  & 1.15  & 0.035  & 153  &   5.6  &  1.3  \\
77.280975 -69.180901 &  0.110 &  & 18.558 $\pm$ 0.024 & 1.051 $\pm$ 0.029 & 0.48 & 0.070 & 234 & 6.5 & 8.4 \\
              & &        & 18.603  $\pm$  0.020  &  1.046  $\pm$ 0.013  & 0.48  & 0.070  & 245  &   9.7  &  1.6  \\
77.418922 -68.119576 &  0.081 &  & 18.527 $\pm$ 0.008 & 1.053 $\pm$ 0.028 & 0.81 & 0.037 & 162 & 7.1 & 1.4 \\
              & &        & 18.643  $\pm$  0.014  &  1.036  $\pm$ 0.012  & 0.81  & 0.060  & 299  &  11.9  &  1.2  \\
77.509560 -74.509254 &  0.044 &  & 18.643 $\pm$ 0.013 & 1.067 $\pm$ 0.037 & 1.97 & 0.037 & 92 & 5.0 & 2.6 \\
              & &        & 18.802  $\pm$  0.015  &  1.031  $\pm$ 0.017  & 1.97  & 0.060  & 188  &  13.9  &  2.4  \\
77.561249 -70.085625 &  0.070 &  & 18.514 $\pm$ 0.011 & 1.077 $\pm$ 0.033 & 0.54 & 0.050 & 130 & 5.9 & 4.1 \\
              & &        & 18.689  $\pm$  0.022  &  1.049  $\pm$ 0.014  & 0.54  & 0.070  & 225  &   9.2  &  1.8  \\
78.091522 -69.566551 &  0.120 &  & 18.413 $\pm$ 0.014 & 1.065 $\pm$ 0.031 & 0.46 & 0.035 & 111 & 6.8 & 2.0 \\
              & &        & 18.578  $\pm$  0.022  &  1.045  $\pm$ 0.012  & 0.46  & 0.070  & 277  &   8.4  &  0.8  \\
78.132858 -71.399323 &  0.055 &  & 18.582 $\pm$ 0.007 & 1.070 $\pm$ 0.030 & 0.91 & 0.080 & 260 & 16.3 & 4.8 \\
              & &        & 18.600  $\pm$  0.009  &  1.066  $\pm$ 0.014  & 0.91  & 0.060  & 201  &  11.1  &  2.5  \\
78.219818 -64.540184 &  0.058 &  & 18.509 $\pm$ 0.015 & 1.049 $\pm$ 0.032 & 1.82 & 0.045 & 118 & 5.7 & 3.2 \\
              & &        & 18.555  $\pm$  0.013  &  1.037  $\pm$ 0.015  & 1.82  & 0.035  &  99  &   8.1  &  2.5  \\
78.354759 -68.878601 &  0.110 &  & 18.513 $\pm$ 0.009 & 1.049 $\pm$ 0.028 & 0.52 & 0.041 & 169 & 5.7 & 13.9 \\
              & &        & 18.537  $\pm$  0.006  &  1.048  $\pm$ 0.012  & 0.52  & 0.030  & 128  &   6.8  & 11.6  \\
78.653793 -70.491211  &  0.070 &  & 18.514 $\pm$ 0.015 & 1.072 $\pm$ 0.027 & 0.71 & 0.037 & 159 & 5.1 & 1.5 \\
              & &        & 18.542  $\pm$  0.012  &  1.068  $\pm$ 0.012  & 0.71  & 0.025  & 113  &   5.5  &  1.2  \\
79.205589 -69.304588  &  0.120 &  & 18.444 $\pm$ 0.011 & 1.052 $\pm$ 0.031 & 0.45 & 0.055 & 210 & 6.1 & 1.8 \\
              & &        & 18.507  $\pm$  0.018  &  1.052  $\pm$ 0.013  & 0.45  & 0.060  & 248  &   6.8  &  1.3  \\
79.240952 -69.799301  &  0.100 &  & 18.457 $\pm$ 0.024 & 1.064 $\pm$ 0.023 & 0.64 & 0.042 & 267 & 8.3 & 2.0 \\
              & &        & 18.519  $\pm$  0.012  &  1.058  $\pm$ 0.010  & 0.64  & 0.030  & 206  &   6.9  &  0.8  \\
79.351082 -67.736267  &  0.089 &  & 18.478 $\pm$ 0.015 & 1.057 $\pm$ 0.029 & 1.08 & 0.044 & 196 & 5.0 & 2.1 \\
              & &        & 18.481  $\pm$  0.007  &  1.059  $\pm$ 0.013  & 1.08  & 0.025  & 112  &   5.0  &  1.8  \\
79.401405 -72.342812  &  0.057 &  & 18.574 $\pm$ 0.010 & 1.070 $\pm$ 0.027 & 1.17 & 0.037 & 135 & 7.3 & 1.3 \\
              & &        & 18.603  $\pm$  0.009  &  1.061  $\pm$ 0.013  & 1.17  & 0.025  &  95  &   7.4  &  1.7  \\
79.489983 -66.370186  &  0.059 &  & 18.506 $\pm$ 0.013 & 1.056 $\pm$ 0.024 & 1.76 & 0.041 & 263 & 5.2 & 1.5 \\
              & &        & 18.557  $\pm$  0.012  &  1.048  $\pm$ 0.010  & 1.76  & 0.027  & 189  &   5.1  &  1.2  \\
79.966331 -68.855019  &  0.120 &  & 18.450 $\pm$ 0.009 & 1.051 $\pm$ 0.029 & 0.63 & 0.055 & 246 & 8.1 & 3.7 \\
              & &        & 18.414  $\pm$  0.005  &  1.061  $\pm$ 0.013  & 0.63  & 0.022  &  93  &   5.0  &  5.5  \\
80.498726 -70.870071  &  0.076 &  & 18.546 $\pm$ 0.011 & 1.059 $\pm$ 0.033 & 0.58 & 0.070 & 197 & 5.3 & 2.4 \\
              & &        & 18.589  $\pm$  0.009  &  1.057  $\pm$ 0.014  & 0.58  & 0.025  &  74  &   7.7  &  2.4  \\
80.721550 -70.254166  &  0.076 &  & 18.531 $\pm$ 0.011 & 1.051 $\pm$ 0.032 & 0.45 & 0.065 & 216 & 4.8 & 3.0 \\
              & &        & 18.639  $\pm$  0.025  &  1.039  $\pm$ 0.014  & 0.45  & 0.070  & 266  &   8.0  &  1.6  \\
80.750961 -69.417130  &  0.120 &  & 18.428 $\pm$ 0.030 & 1.046 $\pm$ 0.032 & 0.45 & 0.060 & 230 & 5.1 & 3.4 \\
              & &        & 18.439  $\pm$  0.008  &  1.050  $\pm$ 0.014  & 0.45  & 0.030  & 117  &   5.2  &  1.8  \\
81.319168 -69.833809 &  0.093 &  & 18.436 $\pm$ 0.029 & 1.050 $\pm$ 0.027 & 0.45 & 0.060 & 268 & 5.2 & 3.0 \\
              & &        & 18.505  $\pm$  0.012  &  1.043  $\pm$ 0.012  & 0.45  & 0.040  & 198  &   7.6  &  4.3  \\
81.493805 -68.543480 &  0.140 &  & 18.518 $\pm$ 0.010 & 1.036 $\pm$ 0.029 & 0.80 & 0.060 & 253 & 5.2 & 4.8 \\
              & &        & 18.555  $\pm$  0.018  &  1.032  $\pm$ 0.013  & 0.80  & 0.060  & 265  &   7.2  &  2.9  \\
81.612061 -71.347397 &  0.068 &  & 18.542 $\pm$ 0.012 & 1.068 $\pm$ 0.028 & 0.78 & 0.055 & 211 & 5.3 & 1.5 \\
              & &        & 18.555  $\pm$  0.012  &  1.066  $\pm$ 0.013  & 0.78  & 0.027  & 105  &   5.2  &  0.8  \\
81.929054 -70.511787 &  0.064 &  & 18.532 $\pm$ 0.013 & 1.056 $\pm$ 0.027 & 0.51 & 0.070 & 299 & 7.6 & 1.3 \\
              & &        & 18.552  $\pm$  0.015  &  1.056  $\pm$ 0.012  & 0.51  & 0.030  & 131  &   5.7  &  1.0  \\
82.057449 -69.381203  &  0.120 &  & 18.408 $\pm$ 0.015 & 1.045 $\pm$ 0.023 & 0.69 & 0.043 & 297 & 9.8 & 1.9 \\
              & &        & 18.565  $\pm$  0.024  &  1.029  $\pm$ 0.010  & 0.69  & 0.060  & 504  &   7.3  &  0.4  \\
82.335655 -67.794502  &  0.100 &  & 18.536 $\pm$ 0.007 & 1.049 $\pm$ 0.028 & 1.17 & 0.065 & 295 & 9.1 & 7.8 \\
              & &        & 18.613  $\pm$  0.009  &  1.034  $\pm$ 0.012  & 1.17  & 0.027  & 135  &   5.5  &  3.9  \\
82.569061 -70.049217  &  0.095 &  & 18.453 $\pm$ 0.024 & 1.048 $\pm$ 0.021 & 0.66 & 0.055 & 412 & 7.7 & 3.4 \\
              & &        & 18.495  $\pm$  0.013  &  1.045  $\pm$ 0.009  & 0.66  & 0.030  & 240  &   6.5  &  0.7  \\
82.726616 -65.006783 &  0.064 &  & 18.492 $\pm$ 0.008 & 1.054 $\pm$ 0.025 & 2.00 & 0.065 & 253 & 8.9 & 4.5 \\
              & &        & 18.542  $\pm$  0.008  &  1.043  $\pm$ 0.011  & 2.00  & 0.018  &  76  &   5.8  &  2.0  \\
83.191772 -72.059097 &  0.069 &  & 18.512 $\pm$ 0.027 & 1.072 $\pm$ 0.030 & 0.91 & 0.039 & 129 & 5.2 & 2.2 \\
              & &        & 18.729  $\pm$  0.018  &  1.038  $\pm$ 0.012  & 0.91  & 0.060  & 277  &  10.9  &  1.1  \\
83.276558 -70.768044  &  0.110 &  & 18.512 $\pm$ 0.056 & 1.039 $\pm$ 0.024 & 0.70 & 0.037 & 213 & 5.1 & 1.3 \\
              & &        & 18.502  $\pm$  0.014  &  1.044  $\pm$ 0.011  & 0.70  & 0.025  & 142  &   5.1  &  0.8  \\
83.462379 -68.890724 &  0.130 &  & 18.464 $\pm$ 0.016 & 1.050 $\pm$ 0.023 & 1.02 & 0.060 & 411 & 9.4 & 6.2 \\
              & &        & 18.542  $\pm$  0.015  &  1.041  $\pm$ 0.010  & 1.02  & 0.050  & 382  &   7.7  & 11.1  \\
83.880829 -73.012154  &  0.076 &  & 18.530 $\pm$ 0.010 & 1.069 $\pm$ 0.021 & 1.74 & 0.034 & 199 & 5.0 & 2.0 \\
              & &        & 18.551  $\pm$  0.012  &  1.064  $\pm$ 0.010  & 1.74  & 0.018  & 110  &   5.0  &  0.7  \\
84.427254 -69.864265  &  0.130 &  & 18.419 $\pm$ 0.030 & 1.044 $\pm$ 0.024 & 0.90 & 0.034 & 242 & 9.0 & 1.4 \\
              & &        & 18.495  $\pm$  0.014  &  1.039  $\pm$ 0.010  & 0.90  & 0.027  & 214  &   5.3  &  1.1  \\
84.445259 -66.893578  &  0.052 &  & 18.548 $\pm$ 0.018 & 1.057 $\pm$ 0.025 & 1.57 & 0.045 & 220 & 6.5 & 1.6 \\
              & &        & 18.581  $\pm$  0.011  &  1.048  $\pm$ 0.012  & 1.57  & 0.026  & 134  &   6.9  &  0.9  \\
84.803436 -71.260735  &  0.120 &  & 18.439 $\pm$ 0.047 & 1.051 $\pm$ 0.024 & 0.99 & 0.036 & 223 & 4.7 & 2.3 \\
              & &        & 18.556  $\pm$  0.011  &  1.042  $\pm$ 0.009  & 0.99  & 0.035  & 258  &   5.7  &  0.8  \\
86.247414 -69.125412 &  0.140 &  & 18.402 $\pm$ 0.044 & 1.062 $\pm$ 0.030 & 1.13 & 0.041 & 185 & 5.2 & 3.4 \\
              & &        & 18.465  $\pm$  0.013  &  1.057  $\pm$ 0.012  & 1.13  & 0.030  & 153  &   6.5  & 13.1  \\
86.570396 -70.388313 &  0.130 &  & 18.421 $\pm$ 0.022 & 1.048 $\pm$ 0.020 & 1.36 & 0.033 & 306 & 5.2 & 2.0 \\
              & &        & 18.510  $\pm$  0.010  &  1.043  $\pm$ 0.008  & 1.36  & 0.026  & 278  &   7.0  &  2.5  \\
87.361679 -67.973282 &  0.066 &  & 18.543 $\pm$ 0.005 & 1.058 $\pm$ 0.024 & 1.70 & 0.050 & 296 & 8.6 & 4.7 \\
              & &        & 18.562  $\pm$  0.011  &  1.054  $\pm$ 0.011  & 1.70  & 0.026  & 159  &   6.5  &  0.8  \\
88.161560 -71.762009 &  0.120 &  & 18.421 $\pm$ 0.013 & 1.052 $\pm$ 0.018 & 2.00 & 0.033 & 312 & 7.2 & 1.4 \\
              & &        & 18.457  $\pm$  0.010  &  1.052  $\pm$ 0.007  & 2.00  & 0.018  & 182  &   5.1  &  1.1  \\
89.149704 -65.759651 &  0.042 &  & 18.661 $\pm$ 0.015 & 1.027 $\pm$ 0.024 & 2.00 & 0.065 & 272 & 3.5 & 3.5 \\
              & &        & 18.633  $\pm$  0.020  &  1.024  $\pm$ 0.012  & 2.00  & 0.070  & 284  &   7.9  &  2.3  \\
\end{longtable}
\tablefoot{Columns~1 and 2 gives the RA and Dec of the los, Col.~3 the reddening as outlined in Sect.~\ref{SS-VMC}.
Columns~4--10 contain the parameters derived in the present paper, see the note to Tab.~\ref{Tab-EB}.
}
}

\section{Discussion}
\label{S-Disc}

\subsection{The internal errors}
\label{SS-Int}

The errors quoted for the TRGB distances are formal errors as given by the minimisation routine.
The fitting routine takes into account the error bars in the luminosity function, as explained in Appendix~\ref{App-Simul}.
The fact that the reduced $\chi^2$ in Tables~\ref{Tab-EB},\ref{Tab-CEP}, and \ref{Tab-RRL} scatter around unity 
indicates that this procedure seems to give reliable estimates of the error bars.

As explained in Sect.~\ref{S-Model} the best model was assumed to be the one with the lowest reduced $\chi^2$ among all models 
that met certain criteria. As an independent check the scatter in the DM was investigated among the models with a 
reduced $\chi^2$ less than twice the minimum value. If there were five or more such  models the dispersion 
(actually 1.48 $\cdot$ MAD) around the median was determined and compared with the formal error.
This exercise revealed no systematic effects and the errors estimated in such a way are consistent with the formal errors.

\subsection{Comparing dEBs and TRGB with Cepheid and RR Lyrae distances}
\label{SS-Comp}

In Sect.~\ref{SS-EB} the TRGB distances are compared with the distances to 14 dEBs. 
One can also compare the TRGB distances with other independent distance estimates, as we did in Sects.~\ref{SS-CEP}, \ref{SS-RRL} and \ref{SS-SFH}.
We therefore took an identical approach as in Sects.~\ref{SS-CEP} and \ref{SS-RRL} and determined the median DM and reddening value
of CCs (in the LMC), and RRL (in the SMC) in the direction of the dEBs.
The results are listed in Table~\ref{Tab-EBCEPRRL} which first repeats the DM and reddening derived in the literature 
for the dEBs and the TRGB distance (based on the second-order derivative method) from Table~\ref{Tab-EB}.
Columns 5 and 6 give the DM and reddening values based on the CCs and RRL in those fields.

It is evident that the reddening estimates are smaller than adopted in the dEB analysis.
In the SMC this is the case for all five objects. Although the differences are within the respective error bars it appears to 
be a systematic effect. In the LMC this is the case for eight out of nine objects but the differences appear to be smaller 
on average than for the SMC.

To test the effect of reddening, the TRGB distance was derived using the $E(B-V)$ from column~6, and the results are listed in column~7.
It is clear that the effect on the DM is roughly inversely proportional to a change in $E(B-V)$.
Based on the definition of the sharpened magnitude, the absolute calibration equation (Eq.~\ref{Eq-AC}) and the reddening coefficients one
expects a relation $\Delta {\rm DM} / \Delta E(B-V)= -1.1$.

The overall effect is noticeable however. The weighted mean DM of the nine LMC dEBs is shifted from $18.483 \pm 0.006$ mag 
to $18.523 \pm 0.005$ mag, and that of the five SMC binaries is shifted from $19.023 \pm 0.007$ mag to $19.051 \pm 0.009$ mag.

In a similar way we used the data of Rubele et al. and took the sub-region closest to the dEBs in the SMC. 
The DM and reddening they report are listed in columns.~8 and 9. The reddenings are significantly larger than those used 
for the dEBs and RRL studies.
Column~10 gives the TRGB distance based on these reddenings, and they are significantly shorter on average.
The weighted mean DM of the five SMC binaries is $18.920 \pm 0.007$ mag.

As a final test the reddening of \citet{Haschke11} was used, taking the value of the closest positional match from their tables. 
This reddening is listed in column~11. 
These reddenings are significantly smaller than those used in the other studies.
Column~12 gives the TRGB distance based on these reddenings, and they are significantly longer on average.
The weighted mean DM of the  nine LMC dEBs is $18.574 \pm 0.005$ mag, and that of the five SMC binaries is $19.071 \pm 0.008$ mag.

Regarding the SMC, \citet{Marconi17} modelled the optical and NIR light curves ($JK$ data from VMC, see \citealt{Ripepi16}, 
corrected for reddening using \citealt{Haschke11}) and radial velocity curves of nine fundamental and three first overtone CCs 
to quote a mean DM of 19.01 mag with 0.08 mag dispersion.
The weighted mean value and the error on the mean for this sample are 18.99 mag, and 0.02 mag, respectively.

\begin{sidewaystable}
\setlength{\tabcolsep}{1.1mm}

\caption{Comparison of distances for the dEBs.}
\label{Tab-EBCEPRRL}
\centering
\small
  \begin{tabular}{lccccccccrrr}
  \hline\hline
   System     &       DM (EB)      &       $E(B-V)$    &      DM (TRGB)     &    DM (CC/RRL)     &    $E(B-V)$       &    DM (TRGB)  &     DM (SFH)    & $E(B-V)$    &    DM (TRGB) &    $E(B-V)$ &    DM (TRGB) \\
              &        (mag)       &       (mag)       &        (mag)       &       (mag)        &     (mag)         &       (mag)   &        (mag)    &  (mag)      &      (mag)   &     (mag)   &       (mag)         \\
\hline
LMC-ECL-01866 & 18.496 $\pm$ 0.028 & 0.115 $\pm$ 0.020 & 18.555 $\pm$ 0.024 & 18.520 $\pm$ 0.089 & 0.090 $\pm$ 0.058 & 18.585 $\pm$ 0.022 &               &                   &                   & 0.051 & 18.646 $\pm$ 0.017 \\
LMC-ECL-03160 & 18.505 $\pm$ 0.029 & 0.123 $\pm$ 0.020 & 18.557 $\pm$ 0.025 & 18.520 $\pm$ 0.035 & 0.067 $\pm$ 0.049 & 18.618 $\pm$ 0.016 &               &                   &                   & 0.080 & 18.642 $\pm$ 0.024 \\
LMC-ECL-06575 & 18.497 $\pm$ 0.019 & 0.107 $\pm$ 0.020 & 18.468 $\pm$ 0.033 & 18.500 $\pm$ 0.035 & 0.100 $\pm$ 0.058 & 18.500 $\pm$ 0.028 &               &                   &                   & 0.036 & 18.657 $\pm$ 0.023 \\
LMC-ECL-09114 & 18.465 $\pm$ 0.021 & 0.160 $\pm$ 0.020 & 18.459 $\pm$ 0.019 & 18.470 $\pm$ 0.033 & 0.110 $\pm$ 0.074 & 18.524 $\pm$ 0.009 &               &                   &                   & 0.051 & 18.568 $\pm$ 0.010 \\
LMC-ECL-09660 & 18.489 $\pm$ 0.025 & 0.127 $\pm$ 0.020 & 18.437 $\pm$ 0.012 & 18.480 $\pm$ 0.034 & 0.058 $\pm$ 0.075 & 18.517 $\pm$ 0.012 &               &                   &                   & 0.058 & 18.511 $\pm$ 0.014 \\
LMC-ECL-10567 & 18.490 $\pm$ 0.027 & 0.102 $\pm$ 0.020 & 18.513 $\pm$ 0.010 & 18.460 $\pm$ 0.035 & 0.098 $\pm$ 0.077 & 18.514 $\pm$ 0.009 &               &                   &                   & 0.051 & 18.573 $\pm$ 0.009 \\
LMC-ECL-15260 & 18.509 $\pm$ 0.021 & 0.100 $\pm$ 0.020 & 18.439 $\pm$ 0.028 & 18.410 $\pm$ 0.035 & 0.120 $\pm$ 0.074 & 18.412 $\pm$ 0.025 &               &                   &                   & 0.029 & 18.540 $\pm$ 0.025  \\
LMC-ECL-25658 & 18.452 $\pm$ 0.051 & 0.091 $\pm$ 0.030 & 18.493 $\pm$ 0.019 & 18.400 $\pm$ 0.035 & 0.039 $\pm$ 0.060 & 18.550 $\pm$ 0.029 &               &                   &                   & 0.036 & 18.561 $\pm$ 0.017 \\
LMC-ECL-26122 & 18.469 $\pm$ 0.025 & 0.140 $\pm$ 0.020 & 18.426 $\pm$ 0.023 & 18.460 $\pm$ 0.034 & 0.120 $\pm$ 0.082 & 18.442 $\pm$ 0.022 &               &                   &                   & 0.080 & 18.518 $\pm$ 0.034 \\

SMC-ECL-0195  & 18.948 $\pm$ 0.023 & 0.079 $\pm$ 0.020 & 19.020 $\pm$ 0.020 & 18.917 $\pm$ 0.156 & 0.033 $\pm$ 0.024 & 19.122 $\pm$ 0.019 & 18.99 $\pm$ 0.07 & 0.084 $\pm$ 0.023 & 19.045 $\pm$ 0.017   & 0.029 & 19.126 $\pm$ 0.020 \\
SMC-ECL-0708  & 18.979 $\pm$ 0.025 & 0.080 $\pm$ 0.020 & 19.027 $\pm$ 0.013 & 18.948 $\pm$ 0.153 & 0.057 $\pm$ 0.049 & 19.008 $\pm$ 0.045 & 18.94 $\pm$ 0.06 & 0.107 $\pm$ 0.010 & 18.921 $\pm$ 0.017   & 0.022 & 19.041 $\pm$ 0.019 \\
SMC-ECL-1421  & 19.057 $\pm$ 0.049 & 0.067 $\pm$ 0.020 & 19.009 $\pm$ 0.026 & 18.948 $\pm$ 0.154 & 0.057 $\pm$ 0.061 & 19.026 $\pm$ 0.024 & 18.87 $\pm$ 0.07 & 0.145 $\pm$ 0.020 & 18.907 $\pm$ 0.014   & 0.036 & 19.047 $\pm$ 0.035 \\
SMC-ECL-4152  & 19.032 $\pm$ 0.019 & 0.093 $\pm$ 0.020 & 18.978 $\pm$ 0.020 & 18.919 $\pm$ 0.155 & 0.066 $\pm$ 0.049 & 19.032 $\pm$ 0.019 & 18.91 $\pm$ 0.07 & 0.167 $\pm$ 0.011 & 18.867 $\pm$ 0.011   & 0.043 & 19.065 $\pm$ 0.019 \\
SMC-ECL-5123  & 18.830 $\pm$ 0.054 & 0.060 $\pm$ 0.030 & 19.039 $\pm$ 0.012 & 18.916 $\pm$ 0.155 & 0.057 $\pm$ 0.036 & 19.035 $\pm$ 0.014 & 18.89 $\pm$ 0.05 & 0.171 $\pm$ 0.023 & 18.960 $\pm$ 0.025   & 0.036 & 19.067 $\pm$ 0.013 \\

\hline
\end{tabular}
\tablefoot{
Columns~1--4 are taken from Tab.~\ref{Tab-EB}. They indicate the name of the system, the DM and reddening based on the works 
listed in Col.~4 of Tab.~\ref{Tab-EB}, and the TRGB distance using that reddening.
Columns~5 and 6 list the DM and reddening of CCs (for the LMC objects) and RRL (for the SMC objects) in the direction of the EBs, and
Col.~7 lists the TRGB distance using the reddening in Col.~6.
Similarly, Columns~8 and 9 give the DM and reddening in \citet{Rubele18} in the direction of the EBs, and
Col.~10 lists the TRGB distance using the reddening in Col.~9.
Finally, Col.~11 lists the reddening derived from \citet{Haschke11} in the direction of the EBs, and
Col.~12 lists the TRGB distance using that reddening.
}
\end{sidewaystable}

\subsection{The absolute calibration relation}
\label{SS-Calib}

As outlined in Sect.~\ref{S-Calib} the absolute calibration of the TRGB is a linear relation
$M_{K} = \alpha + \beta \cdot (J-K_{\rm s})$, calibrated using the theoretical calculations by \citet{Serenelli17}.
The default relation is based on a linear fit in the colour range $0.75 < (J-K_{\rm s}) < 1.3$ mag and reads 
$M_{K_{\rm s}} =  -4.196 -2.013 \, (J-K_{\rm s})$ (Eq.~\ref{Eq-AC}).
An alternative fit in a more restricted colour range is 
$M_{K_{\rm s}} =  -4.331 -1.873 \, (J-K_{\rm s})$ (Sect.~\ref{S-Calib}).
At a colour typical for the SMC ($J-K_{\rm s}= 0.95$ mag) this relation gives a brighter tip by a negligible amount of 2 millimag; 
at a colour typical for the LMC ($J-K_{\rm s}= 1.05$ mag) this relation gives a fainter tip by 0.01 mag.

Although one therefore expects relatively small differences due to the calibration equation there are differences in $(J-K_{\rm s})$ colour
over the different los in both galaxies, and therefore all five applications considered in Sect.~\ref{S-Appl} were re-run 
with the alternative calibration.

These calculations largely confirm the expectations. The mean distance to the LMC is reduced by 10-15 millimag, while the distance 
to the SMC increased by 4-9 millimag using the alternative calibration. 
These differences are of the same order as or smaller than the formal error in the DM for any given los, and are also smaller 
than the dispersion in the calibrating relation itself.

\subsection{Morphology of the MC system}
\label{SS-M}

\begin{figure}
\centering
\resizebox{\hsize}{!}{\includegraphics{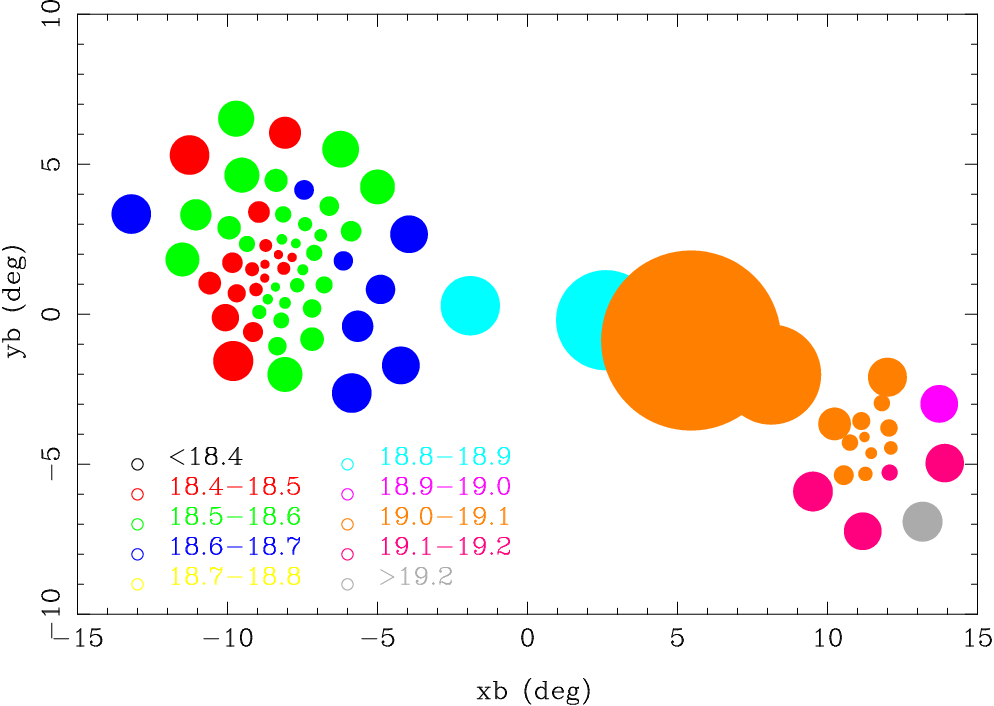}}
\caption[]{ 
Distribution of DM across the MCs
%
based on the VMC data themselves, 
with coordinates deprojected relative to RA= 55\degr, Dec= $-73$\degr.
%
The size of the circles is proportional to the area used in calculating the TRGB distance.
} 
\label{Fig-VMC-DIST}
\end{figure}

Figure~\ref{Fig-VMC-DIST} shows the distribution of the DM over the MC system
for the los chosen from the VMC data (Sect.~\ref{SS-VMC}).
It is beyond the scope of this paper to discuss the structure of the MC system in detail, but
one can notice a gradient across the western part of the LMC, the fields in the Bridge, and the SMC.
This is roughly consistent with what other recent papers found; for example \cite{Subramanian12} based on RC stars, 
\cite{Ripepi17} based on CCs, \cite{Muraveva2018} based on RRL, and \cite{Rubele18} for the SMC, 
and the work using RRL and CCs from OGLE-IV for the MC system \citep{Jacyszyn16,Jacyszyn17}.
The disadvantage of the TRGB method compared to other methods is that a relatively large area needs to be sampled to obtain
a sufficient number of TRGB stars and a high precision for the DM. The number of los that the RRL, CC or RC-based
methods can study in the direction of the MCs is an order of magnitude larger.

\section{Summary and conclusions}
\label{S-Sum}

In this paper we discuss the use of the TRGB in the NIR, and apply it to VMC data in the MCs.
The basis of our work is the theoretical work by \citet{Serenelli17} and the relation $M_{K_{\rm s}} =  -4.196 -2.013 \, (J-K_{\rm s})$ 
we derive for their standard model in the colour range $0.75 < (J-K_{\rm s}) < 1.3$. An alternative calibration in the colour range
$0.82 < (J-K_{\rm s}) < 1.2$ is $M_{K_{\rm s}} = -4.331 - 1.873 \, (J-K_{\rm s})$, which gives nearly identical DM to the LMC and SMC.
The recent empirical determination of the slope based on data in IC~1613 by \cite{Madore18} is $-1.85 \pm 0.27$, which is consistent 
with both relations.

\citet{Serenelli17} state that the colour transformations introduce larger uncertainties than the differences between the
two stellar evolution codes they consider. Their figure~9 shows how the absolute $K$-magnitude depends on the different 
adopted bolometric corrections.
In the range covered by the SMC and LMC ($(J-K_{\rm s}) \sim 0.95-1.05$ mag) these differences are small (at the same level as the scatter in the relation 
judging from their plot), but for $(J-K_{\rm s}) \more 1.2$  mag they become noticeable.
When in the future Gaia data provide reliable and accurate parallaxes, metallicity and reddening estimates for the brightest objects, 
it may well be possible to select TRGB stars with accurate parallaxes and empirically determine the colour dependence of the 
calibration relation towards redder colours (higher metallicities).

The scatter in the calibrating relation is 0.030 mag, which we consider as one source of the systematic uncertainty.
The methodology is another possible source of uncertainty. The simulations in the Appendix show that criteria related to the number 
of stars per bin and the significance of detection of the peak of the filter response curve can be defined in such a way as to give 
unbiased DM to a level of $\sim$0.005 mag.
The second-order derivative filter requires about twice as many stars per bin as the first-order derivative filter to achieve this.
The empirical results derived in this paper however show that the DM based on the second- and first-order derivative filters give
marginally different results. The weighted mean of the four estimates is $-0.033 \pm 0.017$ mag.
We do not have a ready explanation for this. Although depth effects were considered, the modelling of the number density of stars 
by a Gaussian distribution with different scale lengths is probably too simple, and the first- and second-order derivative filters 
may behave differently to this. 
For example, \citet{Subramanian17} find a bimodal magnitude distribution of RC stars in the eastern part of the SMC, interpreted 
as a population at a distance of about 12 kpc in front of the main body. To a lesser extent, \citet{Subramanian13} found extra-planar features both in front 
and behind the main disc of the LMC from an analysis of RC stars.
In addition, differential reddening along a los and reddening differences across a field-of-view may play a role.
At this point we consider this difference in results between the two filters as a measure of a potential systematic uncertainty in the method.

If the condition on the number of stars per bin and the significance of detection of the peak of the filter response curve are met the 
statistical error in the method is small. Of all the random errors in the DM listed in 
Tables~\ref{Tab-EB},\ref{Tab-CEP},\ref{Tab-RRL}, and \ref{Tab-SFH},  50\% are 0.015 mag or smaller (91\% are less than 0.03 mag).

Therefore, our preferred absolute calibration relation of the TRGB in the $K_{\rm s}$-band (in the 2MASS system) in 
the colour range $0.75 < (J-K_{\rm s})_0 < 1.3$ mag 
is $M_{K_{\rm s}} =  -4.196 -2.013 \, (J-K_{\rm s})_0$ with a systematic error of 0.045 mag, and where statistical errors of $\sim 0.015$ mag are possible 
if the criteria on the number of TRGB stars and the quality of the fit are respected.

In practice, the choice of reddening also plays an important role in determining the distance to any stellar system.
Table~\ref{Tab-EBCEPRRL} illustrates this for the dEBs. For typical (median) reddenings of
$\sim 0.04$ \citep{Haschke11}, $\sim 0.06$ (based on the RRL study), $\sim 0.08$ (based on the EB studies), and $\sim 0.15$ mag (based on the SFH study), 
the weighted mean DM of the systems in the SMC is
$19.071 \pm 0.008$,            $19.051 \pm 0.009$,                   $19.023 \pm 0.007$, and                $18.920 \pm 0.007$ mag, respectively.
Similarly, for the LMC systems, with typical reddenings of
$\sim 0.05$ \citep{Haschke11}, $\sim 0.10$ (based on the CCs study), and $\sim 0.12$ mag (based on the EB studies),
the weighted mean DM is
$18.574 \pm 0.005$,            $18.523 \pm 0.005$,                   and $18.483 \pm 0.006$ mag, respectively. 

Considering the systematic uncertainty quoted above these estimates are consistent within $2~\sigma$ 
with the recommended DM of 
18.96 $\pm$ 0.02 mag (formal error only; \citealt{deGB2015}. For typical reddening $\less 0.08$) to the SMC and
18.49 $\pm$ 0.09 mag \citep{deGB2014} to the LMC.

\begin{acknowledgements}
This paper is based on observations collected at the European Organisation 
for Astronomical Research in the Southern Hemisphere under ESO programme 179.B-2003.
We thank the CASU and the WFAU in Edinburgh for providing calibrated data products under the support
of the Science and Technology Facility Council (STFC) in the UK.
Maurizio Salaris (Liverpool John Moores University) is thanked for providing the results from \citet{Serenelli17} in electronic format.
M.-R.C acknowledges support from the European Research Council (ERC) under the European
Union’s Horizon 2020 research and innovation programme (grant agreement no. 682115).
This research was supported by the Munich Institute for Astro- and Particle Physics (MIAPP) of the DFG cluster of 
excellence "Origin and Structure of the Universe", in connection with the inspiring workshop
"The Extragalactic Distance Scale in the Gaia era" organised by Lucas Macri, Rolf Kudritzki, Sherry Suyu, and Wolfgang Gieren.
This research has made use of the SIMBAD database and the VizieR catalogue access tool, operated at CDS, Strasbourg, France. 
The original description of the VizieR service was published in A\&AS 143, 23.
\end{acknowledgements}


\bibliographystyle{aa.bst}
	\bibliography{references.bib}





\begin{appendix}

\section{Simulations}
\label{App-Simul}

In this Appendix the simulations are described which were used to investigate any biases in the determination of the TRGB.

The simulations are carried out for a galaxy at a distance ($D$) of 50 kpc, where the TRGB is roughly at $K \sim 12.3$ mag.
The choice of the simulated galaxy is arbitrary, but some of the magnitude intervals listed below are tuned to this choice.
As an illustration the results of the simulations are compared with the analysis of the actual VMC data for the field 
around the dEB OGLE-LMC-ECL-09660.

The number of stars on the RGB, and the number of AGB and foreground contaminants, are described by a power law, 
$\log N \sim \alpha (m- m_{\rm o})$. For the latter, $\alpha = -0.05$, $m_{\rm o}= 10.0$ mag for magnitudes between 
10.0 and 14.5 mag, roughly corresponding to the brightest AGB stars and the start of the early-AGB in such a galaxy. 
For the RGB stars the slope is $\alpha = +0.3$, between $m_{\rm o}= K@$TRGB= $f(J-K)$ and 15.0 mag.
The two slopes are based on a comparison of the $K$-band luminosity function (LF) with real VMC data.
The probability of a star being an AGB or foreground contaminant is $f_{\rm c}$.

The simulation proceeds as follows. The total number of simulated stars is $N_{\rm sim}$.
A random number between 0 and 1 is drawn. If this number is $<$$f_{\rm c}$, a $K$-mag is drawn from the LF of AGB and foreground contaminants.
Otherwise the star is considered an RGB star.
We considered contaminations of $f_{\rm c} = 0.01, 0.1, 0.20, 0.38, 0.55, 0.75$.
For the field around LMC-ECL-09660 $f_{\rm c} = 0.20$ is appropriate.

In case of an RGB star, a random number is drawn to generate a $(J-K)$@TRGB according to a Gaussian distribution.
Here a mean of 1.0 mag and a dispersion of 0.05 mag are assumed, typical of the LMC (see Figure~\ref{Fig-Col}).

Assuming an absolute calibration $M_{K} = \alpha + \beta \; (J-K)$, with $\alpha= -4.196$ mag
and $\beta = -2.013$ (see Sect.~\ref{S-Calib}) the expected $K$-mag @TRGB in the simulated galaxy, $M_{K} + 5 \log (D) -5$,
is known, and an RGB $K$-mag is drawn from the LF mentioned above.

The $J$ magnitude is calculated from the $K$ mag and a $(J-K)$ colour, which is based on the generated $(J-K)$@TRGB and a 
mean $K - (J-K)$ relation based on real VMC data (see Figure~\ref{Fig-CMD-App}).

Gaussian distributed photometric errors in $J$ and $K$, based on real VMC data of the mean photometric error 
and dispersion as a function of $K$, are added to the simulated data 

Finally, the depth of the galaxy is simulated, by considering an exponential function $(\sim \exp(-d/H))$ along the los.
We have considered $H= 10$ pc (i.e. almost no effect), 800 pc (used in the examples shown here) and 2000 pc, 
representative for the LMC and SMC, respectively, according to \citet{HaschkeLMC, HaschkeSMC}\footnote{It is acknowledged that the scale height 
may depend on population, \citet{HaschkeSMC} find 2.0 $\pm$ 0.4 kpc for RRL [the value used here] and 2.7 $\pm$ 0.3 kpc for CCs, 
and recently even larger values have been reported, for example 4.3 $\pm$ 1.0 kpc for RRL in the SMC \citep{Muraveva2018}.}.
Finally the $T$ mag is calculated, $K - \beta \; (J-K)$.

\begin{figure} 
\centering
\resizebox{\hsize}{!}{\includegraphics{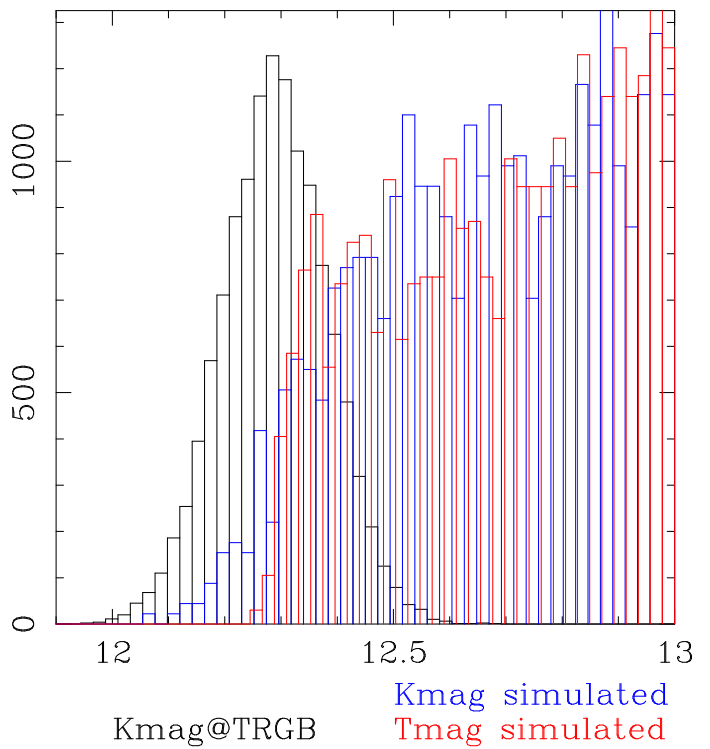}}
\caption[]{ 
For a Gaussian distribution of the $(J-K)$ colours at the TRGB with a width of 0.05 mag, 
and the relation between $M_{K}$ and $(J-K)$ discussed in the text, the black histogram is the theoretical distribution 
of the $K$-mag of stars at the TRGB for a Galaxy at the distance of the LMC. 
Since the LF is sampled, the actual distribution of all RGB stars in $K$ is the blue histogram. 
The cut-off is not sharp and samples neither the true brightest RGB stars, nor the peak in the true $K$-mag distribution.
The red histogram shows the simulated distribution in $T$-mag (shifted by $-2.013$ $\times$ the adopted 
mean $(J-K)$ colour at the TRGB). The cut-off is much sharper.
} 
\label{Fig-MKsim} 
\end{figure}

\begin{figure} 
\centering

\begin{minipage}{0.24\textwidth}
\resizebox{\hsize}{!}{\includegraphics{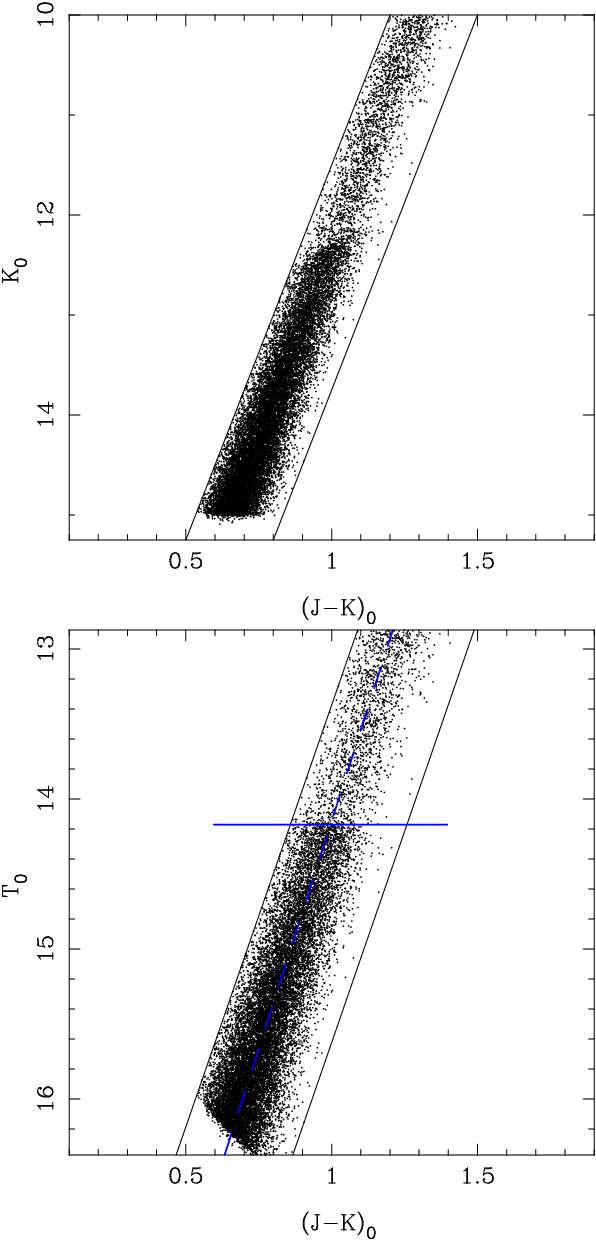}}
\end{minipage}
\begin{minipage}{0.24\textwidth}
\resizebox{\hsize}{!}{\includegraphics{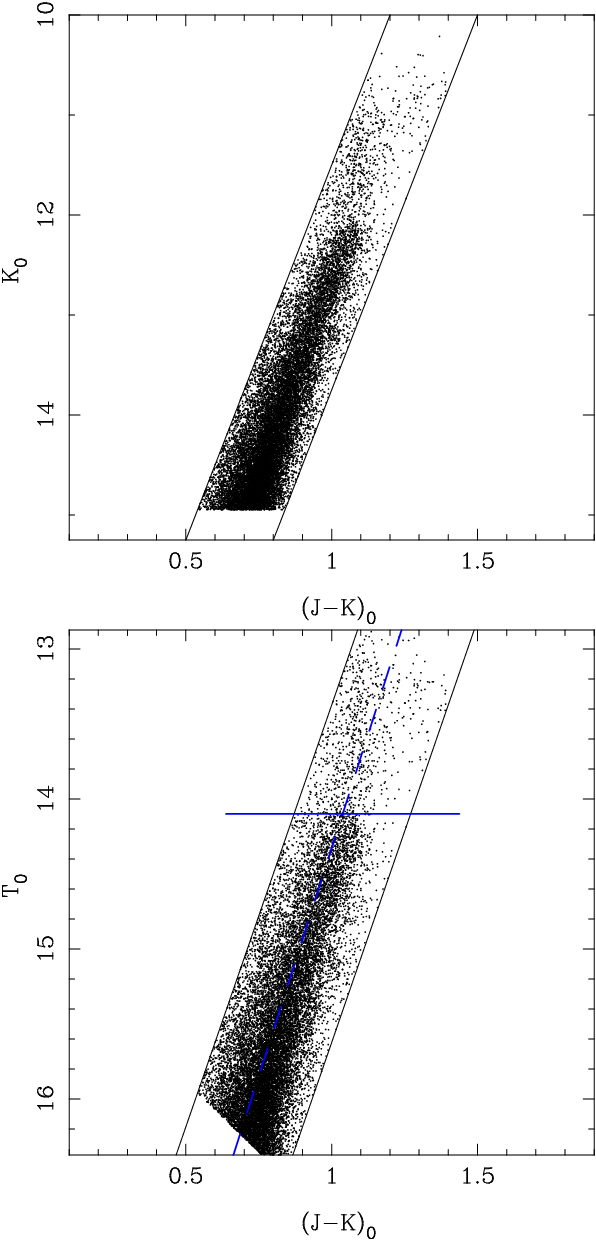}}
\end{minipage}

\caption[]{ 
Simulation (two left-hand panels) and real VMC data around LMC-ECL-09660 (two right-hand panels).
In the simulation 20~000 stars are generated. The top panels show a classic $K, (J-K)$ CMD.
In the left bottom panel the 18~730 stars are plotted that are within the colour selection box, in the  $T, (J-K)$ CMD.
The blue solid lines indicate the derived location of the TRGB, and the blue dashed line is the 
mean $T, (J-K)$ relation derived in the interval from the TRGB to one magnitude fainter, but shown for all magnitudes.
} 
\label{Fig-CMD-App} 
\end{figure}

The advantage of using the $T$ mag is illustrated in Fig.~\ref{Fig-MKsim}.
Assume a Gaussian distribution of the $(J-K)$ colour at the TRGB.
Since the absolute $K$ magnitude depends on colour, the theoretical $K$ magnitude at the TRGB is also Gaussian distributed, 
shown as the black histogram.
As discussed above, the RGB LF is sampled assuming a power-law distribution, and the blue histogram
indicates the LF of RGB stars. There is no clear cut-off.
The blue histogram shows the distribution in $T$ mag, that is $K -\beta \; (J-K)$,  shifted by the expected mean colour term.
The edge is defined much more clearly.

\begin{figure} 
\centering

\begin{minipage}{0.4\textwidth}
\resizebox{\hsize}{!}{\includegraphics{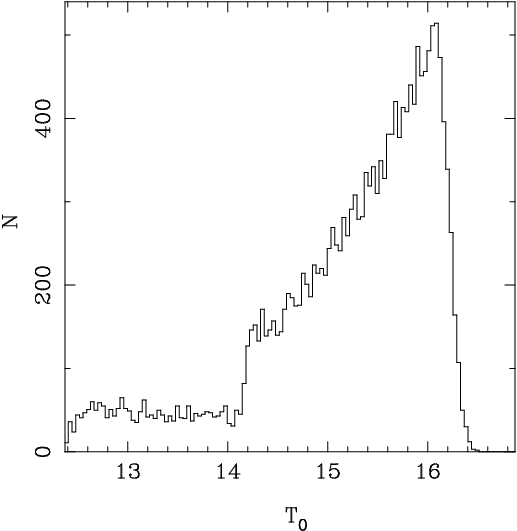}}
\end{minipage}
\begin{minipage}{0.4\textwidth}
\resizebox{\hsize}{!}{\includegraphics{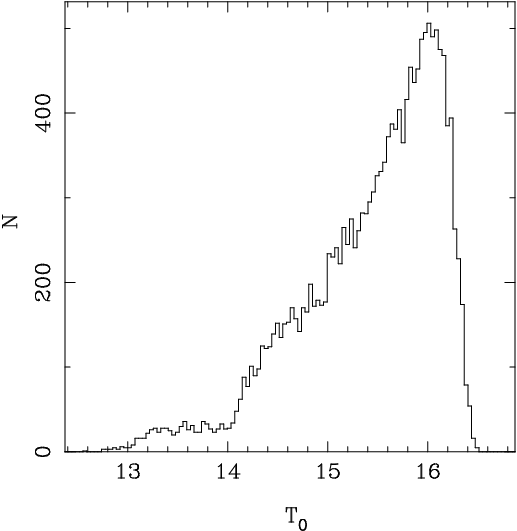}}
\end{minipage}

\caption[]{ 
The $T$ magnitude LF in the simulation (top) and in the field around LMC-ECL-09660 (bottom).
} 
\label{Fig-Bin} 
\end{figure}

\begin{figure} 
\centering

\begin{minipage}{0.4\textwidth}
\resizebox{\hsize}{!}{\includegraphics{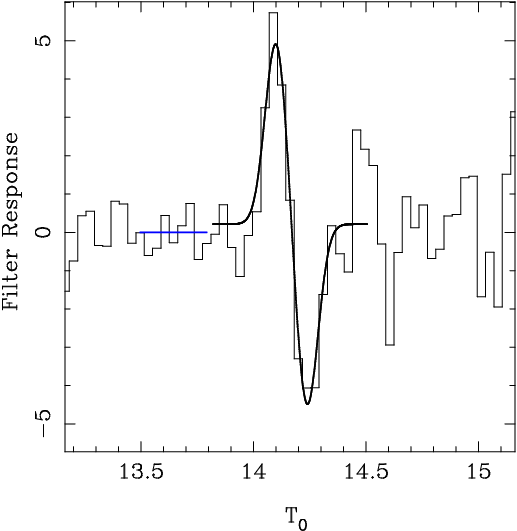}}
\end{minipage}

\begin{minipage}{0.4\textwidth}
\resizebox{\hsize}{!}{\includegraphics{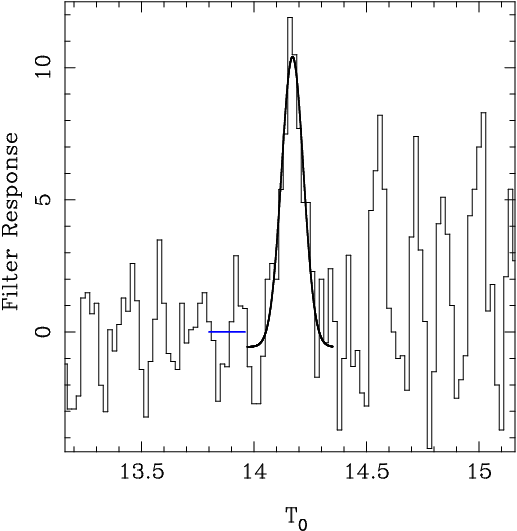}}
\end{minipage}

\caption[]{ 
Results of the simulation.
Response to the two filters used, one that derives the second-order derivative (and which is fitted with two Gaussians) in the top panel, 
and in the bottom panel the classic Sobel-like filter that finds the first-order derivative (and which is fitted with a single Gaussian).
The blue line indicates the region used to estimate the rms level in the response function.
In the top panel the bin width is 0.037 mag, the peak is detected with a S/N of 12, 
and there are 152 stars per bin between the TRGB and 0.5 mag fainter in the LF.  
In the bottom panel these numbers are 0.020 mag, 11, and 82 stars/bin. 
The derived DM are virtually identical: $18.5009 \pm 0.0027$ ($\chi^2_{\rm r}= 5.6$) and 
$18.5022 \pm 0.0046$ mag ($\chi^2_{\rm r}= 1.3$), respectively, and very close to the input value of 18.50 mag.
} 
\label{Fig-Fil-App} 
\end{figure}

\begin{figure} 
\centering

\begin{minipage}{0.4\textwidth}
\resizebox{\hsize}{!}{\includegraphics{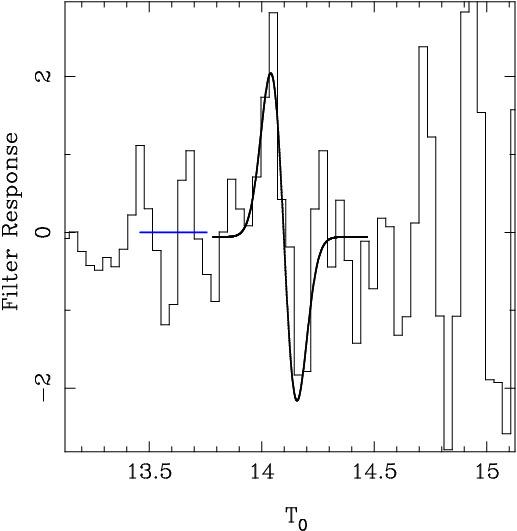}}
\end{minipage}

\begin{minipage}{0.4\textwidth}
\resizebox{\hsize}{!}{\includegraphics{TRGBFilter_exa_SG.ps}}
\end{minipage}

\caption[]{ 
Fit of single and double Gaussians to the filtered LF for LMC-ECL-09660.
The blue line indicates the range in magnitude used to estimate the noise level in the LF.
} 
\label{Fig-Fil} 
\end{figure}

A bin width ($w$) is chosen and the binned $T$ mag LF is then analysed using the first- and second-order derivative kernels using  
Savitzky-Golay coefficients as explained in the main text.
The response to the first-order derivative is fitted with a single Gaussian plus a constant (SG),
\begin{equation}
F(m)=  a_1 + a_2 \; \exp( -(m - a_3)^2 / (2 a_4^2)).
\end{equation}
The response to the second-order derivative is fitted with a double Gaussian plus a constant (DG),
\begin{eqnarray}
 F(m) &=&  a_1     + a_2 \; \exp( -(m - a_3 + a_5)^2 / (2 a_4^2) )  \nonumber \\
      & & \quad {} - a_2 \; \exp( -(m - a_3 - a_5)^2 / (2 a_4^2) ).
\end{eqnarray}
The fitting is done with the Levenberg-Marquardt algorithm (routine {\it  mrqmin} as implemented in 
Fortran in \citealt{Press1992}).

Initial guesses for the parameters are required; the constant $a_1$ is set to zero, 
the width of the Gaussian $a_4$ is set to the bin width, 
for the DG the difference between the two Gaussians $a_5$ is set to 1.5 times the bin width, and 
the location and height of the peak ($a_3$, $a_2$) are obtained from a rough analysis of the LF.
An "error" for the derived response function is determined by calculating the rms in a region brighter than the 
estimated location of the peak, as illustrated in Fig.~\ref{Fig-Fil-App}.

The fit parameters of interest are the mean magnitude and its error ($a_3$, the $T$ magnitude of the TRGB), 
the significance with which the peak is detected (SNpk= $a_2$/$\sigma_{\rm a_2}$), 
and the ratio of the error in the mean magnitude compared with the bin width ($\sigma_{\rm a_3}/w$).

Additional parameters are also derived: the number of stars within a 0.5 magnitude range brighter and fainter 
than the TRGB ($N_{\rm bright}$, $N_{\rm faint}$), from which one can calculate a contamination ratio
($N_{\rm bright} / (N_{\rm bright} + N_{\rm faint}$))
and the average number of RGB stars per bin ($N_{\rm bin}= N_{\rm faint}/ (0.5/{\rm bin~width})$).

\begin{figure} 
\centering

\begin{minipage}{0.4\textwidth}
\resizebox{\hsize}{!}{\includegraphics{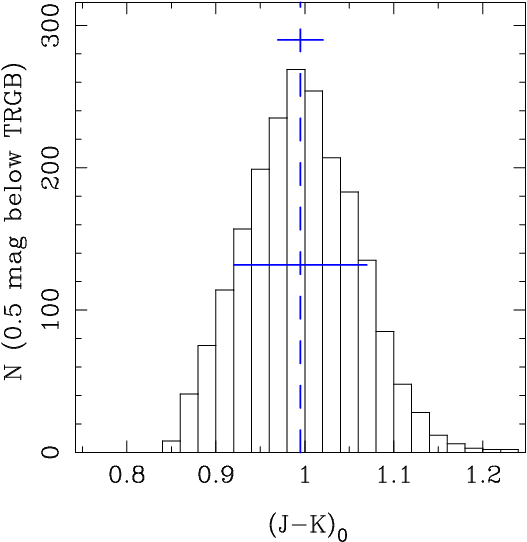}}
\end{minipage}
\begin{minipage}{0.4\textwidth}
\resizebox{\hsize}{!}{\includegraphics{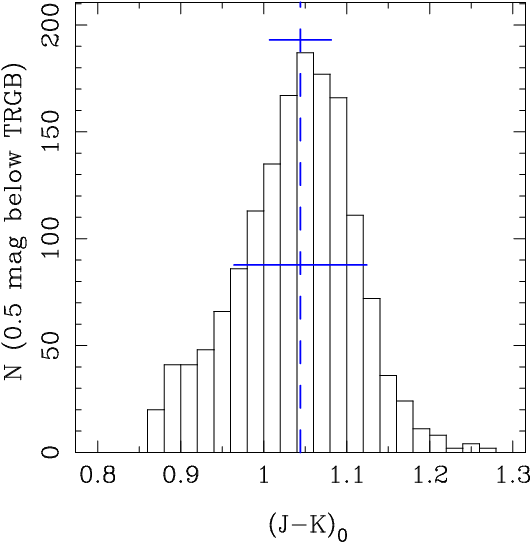}}
\end{minipage}

\caption[]{ 
The top panel shows the results for the simulation, the bottom panel the results for the field around LMC-ECL-09660.
The distribution of $(J-K)$ colour at the TRGB, estimated from the general $(J-K) - T$ relation 
(the blue line in Fig.~\ref{Fig-CMD-App}) using an 0.5 mag interval below the TRGB is shown.
The blue dashed and solid lines (roughly at half the maximum) indicate the mean and Gaussian dispersion, respectively.
The narrower blue line above the peak indicates the formal error in the $(J-K)$@TRGB estimate.
In the simulation the input was a Gaussian with mean 1.0 mag, and dispersion 0.05 mag.
The analysis of the simulated data  gives $(J-K)$@TRGB of 0.996 $\pm$ 0.025 mag, 
and a dispersion in the distribution of 0.063 mag.
} 
\label{Fig-Col} 
\end{figure}

The $(J-K)$ magnitude at the TRGB is also estimated.
Using the data in the one magnitude region below the tip a linear relation between $T$ and $(J-K)$ is determined.
From that the $(J-K)$@TRGB is determined from $a_3$, and its error based on $\sigma_{\rm a_3}$ and the errors
in the slope and zero point of the linear fitting relation.

The distribution in $(J-K)$ colour near the TRGB is also determined.
In the 0.5 mag region below the tip every ($(J-K)$, $T$) point is projected onto the mean $T - (J-K)$ relation.
This allows to estimate the $(J-K)$ as if this point were at the tip.

\begin{figure} 
\centering

\begin{minipage}{0.24\textwidth}
\resizebox{\hsize}{!}{\includegraphics{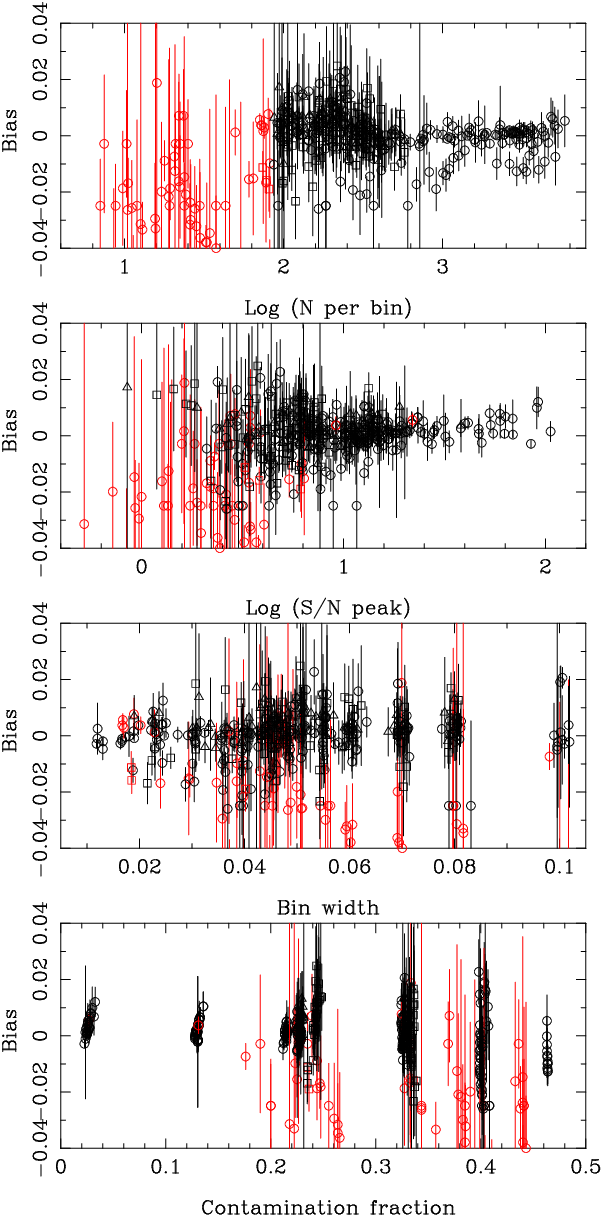}}
\end{minipage}
\begin{minipage}{0.24\textwidth}
\resizebox{\hsize}{!}{\includegraphics{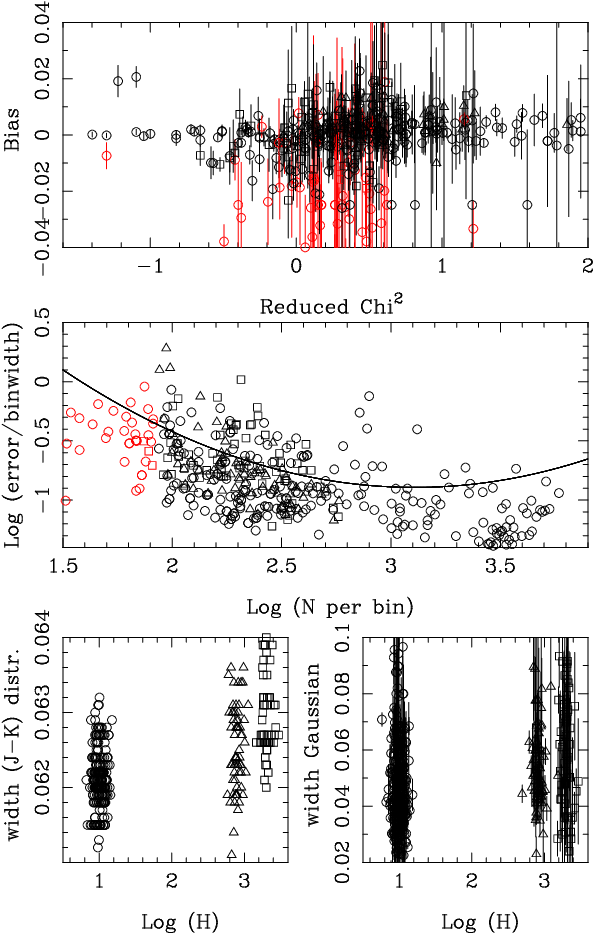}}
\end{minipage}

\caption[]{ 
Diagnostic plots for the DG-filter. Stars in red have fewer than 85 RGB stars per bin. 
The bias is defined as (derived DM $-$ true DM).
Scale height is coded as follows: $H= 10$ pc, circles; $H= 800$ pc, triangles; $H= 2000$ pc, squares. 
} 
\label{Fig-DG85} 
\end{figure}

\begin{figure} 
\centering

\begin{minipage}{0.24\textwidth}
\resizebox{\hsize}{!}{\includegraphics{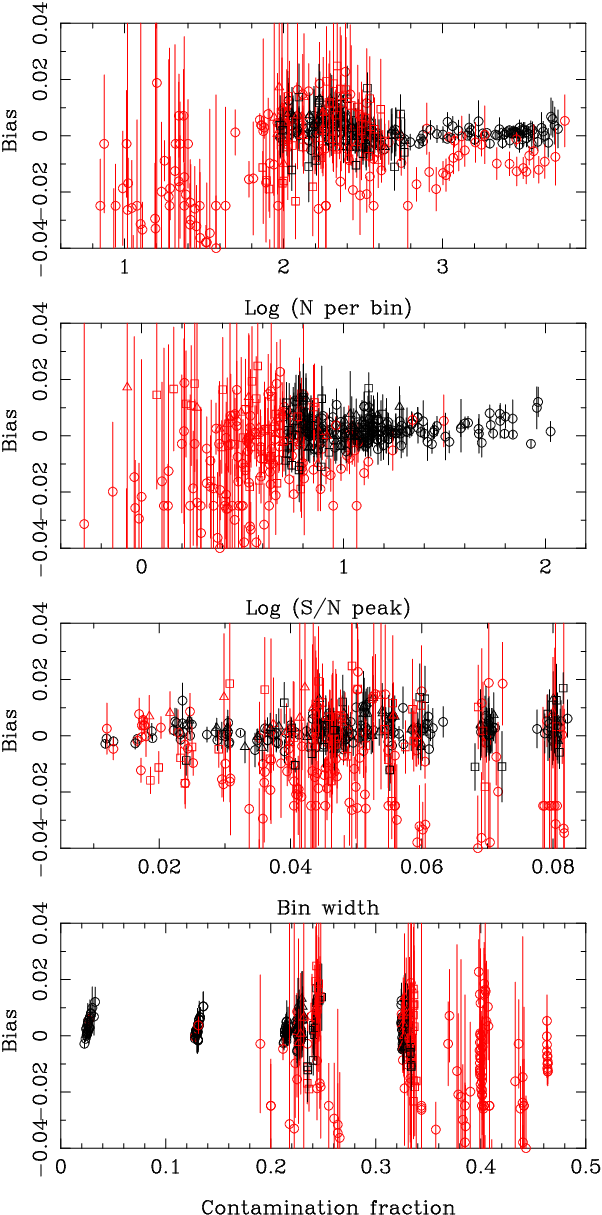}}
\end{minipage}
\begin{minipage}{0.24\textwidth}
\resizebox{\hsize}{!}{\includegraphics{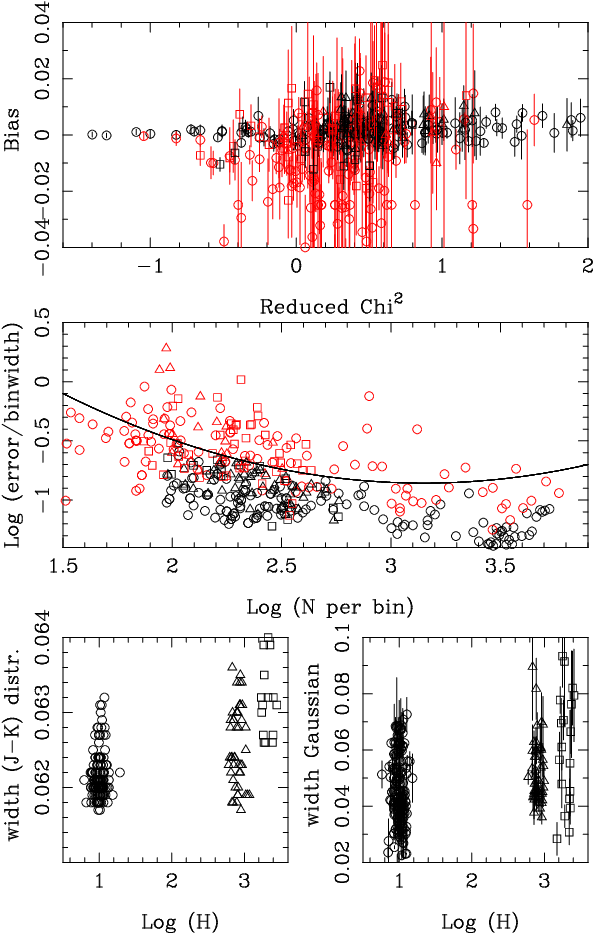}}
\end{minipage}

\caption[]{ 
Diagnostic plots for the DG-filter.
Final selection where red means excluded models with: Number of RGB stars per bin $< 85$, or S/N of the peak $<5$, 
or bin width $>0.085$, or a contamination fraction $>0.38$, or an (error / bin width) above the curve,
given by $y = 0.2778 \cdot x^2  -1.75 \cdot x +1.90$, where $x= \log$($N$ per bin) and $y= \log$ (error / bin width).
} 
\label{Fig-DGAll} 
\end{figure}

\begin{figure} 
\centering

\resizebox{\hsize}{!}{\includegraphics{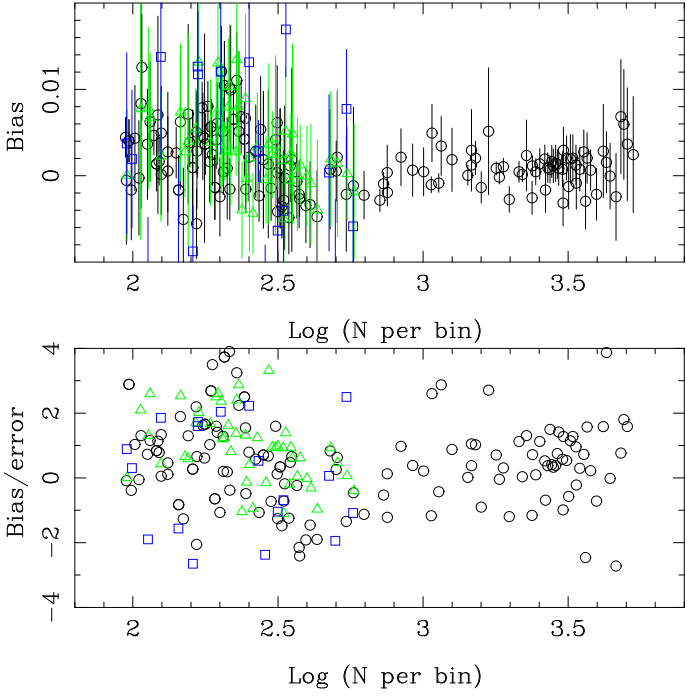}}

\caption[]{ 
Diagnostic plots for the DG-filter. 
Final selection, with all non-red points from Fig.~\ref{Fig-DGAll} plotted on a smaller scale ($-0.01$ to $+0.02$ mag).
Scale height also additionally colour coded ($H= 10$ pc, black circles; $H= 800$ pc, green triangles; $H= 2000$ pc, blue squares).
The bottom panel shows the bias divided by the error bar.
Larger simulations have only been run for the smallest scale height.
} 
\label{Fig-DGAll2} 
\end{figure}

\begin{figure} 
\centering

\begin{minipage}{0.23\textwidth}
\resizebox{\hsize}{!}{\includegraphics{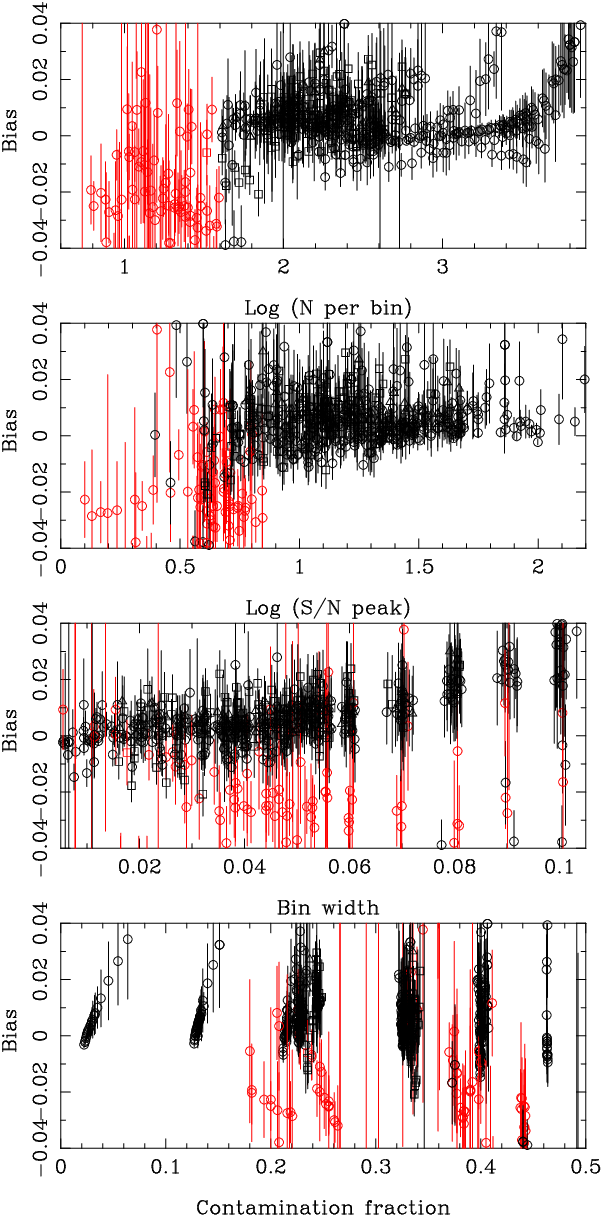}}
\end{minipage}
\begin{minipage}{0.23\textwidth}
\resizebox{\hsize}{!}{\includegraphics{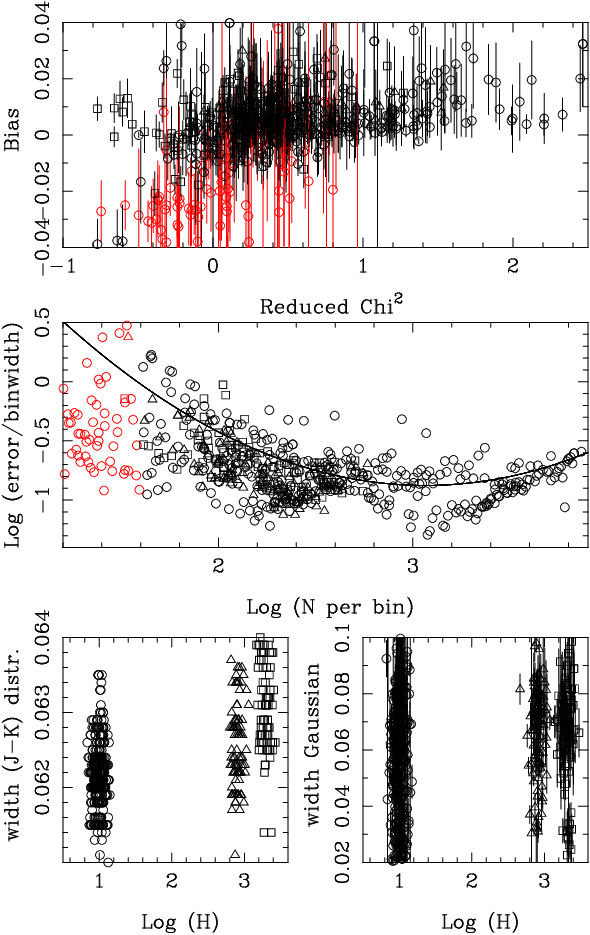}}
\end{minipage}

\caption[]{ 
Diagnostic plots for the SG-filter. Stars in red have fewer than 40 RGB stars per bin. 
%
} 
\label{Fig-Sob40} 
\end{figure}

\begin{figure} 
\centering

\begin{minipage}{0.23\textwidth}
\resizebox{\hsize}{!}{\includegraphics{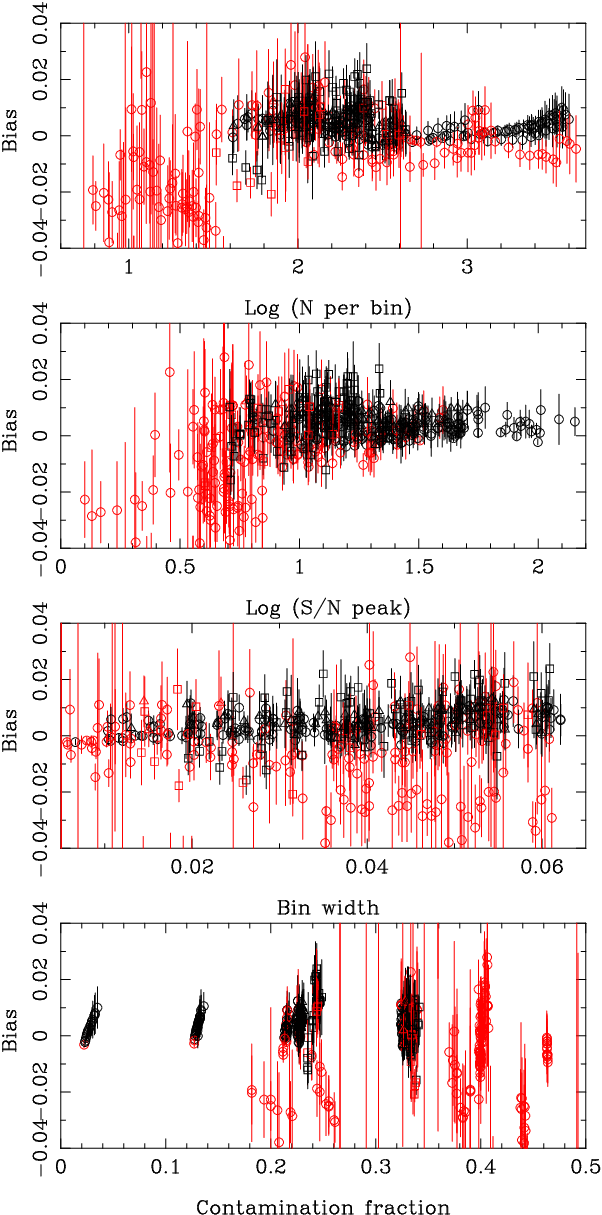}}
\end{minipage}
\begin{minipage}{0.23\textwidth}
\resizebox{\hsize}{!}{\includegraphics{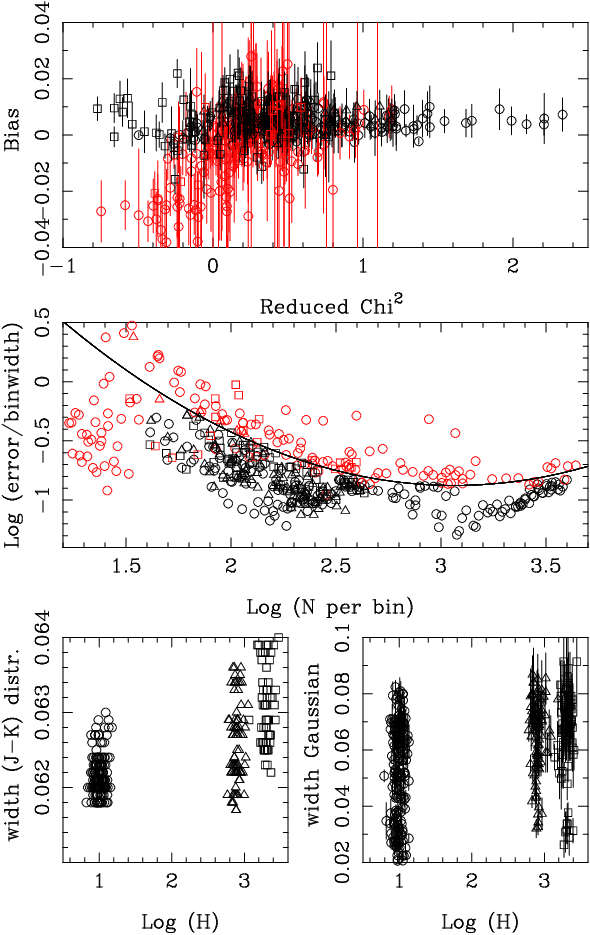}}
\end{minipage}

\caption[]{ 
Diagnostic plots for the SG-filter.
Final selection where red means excluded models with: Number of RGB stars per bin $< 40$, 
or S/N of the peak $<5$, or bin width $>0.065$, or a contamination fraction $>0.38$, 
or an (error / bin width) above the curve,
given by $y = 0.3990 \cdot x^2  -2.445 \cdot x +2.869$, where $x= \log$($N$ per bin) and $y= \log$ (error / bin width).
} 
\label{Fig-SobAll} 
\end{figure}

\begin{figure} 

\resizebox{\hsize}{!}{\includegraphics{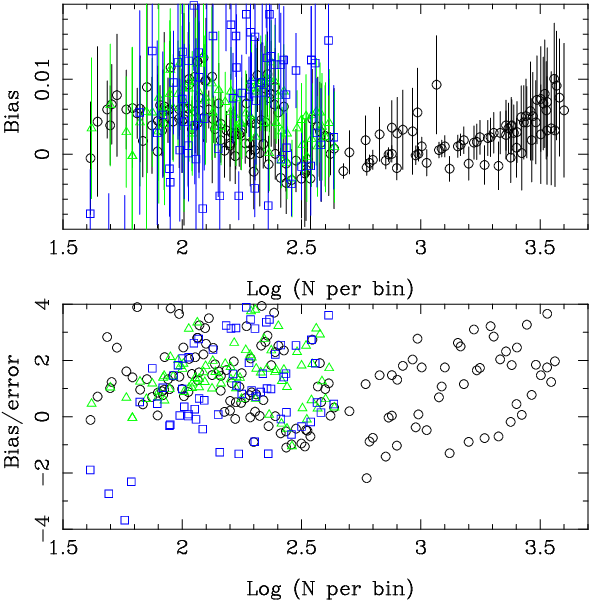}}

\caption[]{ 
Diagnostic plots for the SG-filter.
Final selection, with all non-red points from Fig.~\ref{Fig-SobAll} plotted on a smaller scale ($-0.01$ to $+0.02$ mag).
Scale height colour coded as in Fig.~\ref{Fig-DGAll2}.
The bottom panel shows the bias divided by the error bar.
Larger simulations have only been run for the smallest scale height.
} 
\label{Fig-SobAll2} 
\end{figure}

Almost 1200 simulations were run for different numbers of simulated stars, bin widths, 
fractions of AGB contaminants and three values of $H$. 
The figures below discuss the bias in DM, calculated as the fitted DM minus the true/input DM.
Figures~\ref{Fig-DG85} - \ref{Fig-DGAll2} are for the second-order derivative filter fitted
with the DG, and Figures~\ref{Fig-Sob40} - \ref{Fig-SobAll2} are for
the first-order derivative fitted with the SG.

Shown is the bias as a function of the quantities that are available from the fits:
the number of stars per bin, the signal-to-noise ratio (S/N) with with the peak in the response
function is derived, the bin width, the fraction of contaminants, the reduced $\chi^2$,
and the error in the magnitude of the peak compared with the width of the bin.

As expected qualitatively, if the RGB near the tip is well populated and the peak in the response
is well determined, the bias is essentially negligible (of order a few millimag), and smaller than 
the (systematic) errors due to uncertainties in reddening, 
transformation to the 2MASS system, or the absolute calibration of the TRGB method (see the main text).

The conditions that are used for the real data are a detection of the peak with a SNpk $>5$,
an average number of stars per bin in the 0.5 mag below the tip of $>85$ (second-order derivative),
or $>40$ (first-order derivative), and a ratio ($a_3/w$) that is small enough (see detailed relations 
in the captions of Fig.~\ref{Fig-DGAll} and Fig.~\ref{Fig-SobAll}).

Table~\ref{Tab-Bias} shows the bias and dispersion for the models that meet these conditions.
It shows that the bias and dispersion are $\sim$ 6 millimag or less, with the second-order derivative
filtering overall showing the tendency for slightly smaller values, for example inspect and compare 
Figs.~\ref{Fig-DGAll2} and \ref{Fig-SobAll2}.

\begin{table}

\caption{Median bias and dispersion in DM for models that meet the selection criteria.}
\label{Tab-Bias} 
\centering
  \begin{tabular}{rrrccccc}
  \hline\hline
  Number of RGB stars & Bias           & \#models  &   \\ 
  per bin             & (milli mag)    &           &   \\ \hline
\multicolumn{4}{c}{Double Gaussian $H= 10$ pc} \\ 
 (all)       & 1.35 $\pm$ 3.41 & 158     & \\ 
 $>3000$     & 1.25 $\pm$ 1.93 & 23 &  \\
1500--3000   & 0.85 $\pm$ 0.74 & 23 &  \\
 500--1500   & 0.35 $\pm$ 2.22 & 20 & \\
 300--500    & $-$2.15 $\pm$ 2.96 & 19 & \\
 200--300    &  4.24 $\pm$ 3.85 & 26 &  \\
 150--200    &  3.65 $\pm$ 2.96 & 22 &   \\
  85--150    &  2.64 $\pm$ 3.11 & 25 & \\
  85--500    &  2.44 $\pm$ 4.00 & 92 & \\
  85--200    &  3.65 $\pm$ 3.41 & 47 & \\

\multicolumn{4}{c}{Double Gaussian $H= 800$ pc} \\ 
 300--500    &  1.25 $\pm$ 1.93 & 15 &   \\
 200--300    &  2.85 $\pm$ 4.15 & 17 & \\
 150--200    &  3.95 $\pm$ 1.78 & 7 & \\
  85--150    &  6.75 $\pm$ 1.48 & 6 & \\
  85--500    &  2.85 $\pm$ 4.00 & 45 & \\
  85--200    &  5.45 $\pm$ 2.81 & 13 & \\

\multicolumn{4}{c}{Double Gaussian $H= 2000$ pc} \\ 
 300--500    & $-$6.35 $\pm$ 3.41  & 5 &  \\
 200--300    &  2.85 $\pm$ 13.6  & 4 & \\
 150--200    & $-$8.75 $\pm$ (0.0) & 3 & \\
  85--150    & $-$11.0 $\pm$ 1.78  & 5 & \\
  85--500    &  0.35 $\pm$ 11.6 & 17 & \\
  85--200    &  1.94 $\pm$ 15.8 &  8 & \\

\\

\multicolumn{4}{c}{Single Gaussian $H= 10$ pc} \\ 
 (all)      & 3.35 $\pm$ 3.70 & 174  & \\ 
 $>2500$    & 4.95 $\pm$ 2.66 &  25  &  \\
1400--2500   & 2.15 $\pm$ 1.03 &  20 &  \\
 600--1400   & 0.05 $\pm$ 1.48 &  21 & \\
 270--600   & $-$0.85 $\pm$ 2.51 &  21 & \\
 200--270    & 4.55 $\pm$ 5.19 &  19 &  \\
 140--200    & 2.65 $\pm$ 1.77 &  21 &   \\
 100--140    & 6.14 $\pm$ 4.15 &  20 & \\
  40--100    & 5.45 $\pm$ 1.48 &  27 & \\
 270--400   & $-$1.25 $\pm$ 2.96 &  14 & \\
  40--150    & 5.95 $\pm$ 2.37 &  50 & \\

\multicolumn{4}{c}{Single Gaussian $H= 800$ pc} \\ 
 270--400   &  2.25 $\pm$ 2.08 &  15 & \\
 200--270    & 8.35 $\pm$ 2.07 &  17 &  \\
 140--200    & 5.45 $\pm$ 2.22 &  14 &   \\
 100--140    & 7.15 $\pm$ 2.52 &  17 & \\
  40--100    & 5.55 $\pm$ 2.96 &  18 & \\
  40--150    & 6.15 $\pm$ 3.11 &  37 & \\

\multicolumn{4}{c}{Single Gaussian $H= 2000$ pc} \\ 
 270--400   &  1.25 $\pm$ 4.30 &  11 & \\
 200--270    & 11.9 $\pm$ 5.49 &  16 &  \\
 140--200    & 8.15 $\pm$ 11.0 &  13 &   \\
 100--140    & 4.74 $\pm$ 8.01 &  15 & \\
  40--100   & $-$1.95 $\pm$ 10.7 &  13 & \\
  40--110   &  1.15 $\pm$ 7.26 &  19 & \\
  40--150    & 3.84 $\pm$ 9.49 &  30 & \\

\hline
\hline
\end{tabular}
\end{table}

\end{appendix}

\end{document}